\documentclass[prb,reprint,superscriptaddress]{revtex4-1}

\usepackage[dvipsnames]{xcolor}
\usepackage{amsmath}
\usepackage{amssymb}
\usepackage{stackrel}
\usepackage{esint}
\usepackage{graphicx}
\usepackage{floatrow}
\usepackage[percent]{overpic}
\usepackage{colortbl}

\begin{document}

\title{The statistical physics of multi-component alloys using KKR-CPA}
\date{\today}
\author{Suffian N. Khan}
\email{khansn@ornl.gov}
\affiliation{Materials Science and Technology Division, Oak Ridge National Laboratory, Oak Ridge, TN 37831-6114}
\author{J. B. Staunton}
\email{J.B.Staunton@warwick.ac.uk}
\affiliation{Department of Physics, University of Warwick, Coventry, CV4 7AL}
\author{G. M. Stocks}
\email{stocksgm@ornl.gov}
\affiliation{Materials Science and Technology Division, Oak Ridge National Laboratory, Oak Ridge, TN 37831-6114}

\begin{abstract}
We apply variational principles from statistical physics and the Landau theory of phase transitions to multicomponent
alloys using the multiple-scattering theory of Korringa-Kohn-Rostoker
(KKR) and the coherent potential approximation (CPA). This theory is a multicomponent 
generalization of the $S^{(2)}$ theory of binary alloys developed by G. M. Stocks, J. B. Staunton, D. D. Johnson and others. It is
highly relevant to the chemical phase stability of high-entropy alloys as
it predicts the kind and size of finite-temperature chemical fluctuations. In doing so it includes effects of
rearranging charge and other electronics due to changing site
occupancies. When chemical fluctuations grow without bound an absolute
instability occurs and a second-order order-disorder phase transition may
be inferred. The $S^{(2)}$ theory is predicated on the fluctuation-dissipation theorem; thus we derive the linear response of the CPA medium to perturbations
in site-dependent chemical potentials in great detail.
The theory lends itself to a natural interpretation in terms
of competing effects: entropy driving disorder and favorable pair
interactions driving atomic ordering. To further clarify interpretation
we present results for representative ternary alloys CuAgAu, NiPdPt,
RhPdAg, and CoNiCu within a frozen charge (or \emph{band-only}) approximation. These results include the so-called Onsager mean field correction that extends the temperature range for which the theory is valid.
\end{abstract}
  
\maketitle

\section{Introduction}

Conventional alloys, like steel and aluminum-based alloys, are composed of one or two base metals and trace additions to stabilize the structure and tune the material properties. In contrast, high-entropy alloys (HEAs) are disordered alloys with five or more base metals.\cite{Zhang20141,hea-critical,e16094749,Guo201396} Examples include first-row transition metals in simple FCC or BCC phases, e.g. CrMnFeCoNi.  HEAs have been found with specific strength, corrosion resistance, and wear resistance that is comparable to, or exceeds that of, conventional alloys. From a scientific standpoint
they represent a vast uncharted space of possible alloys. To date there is limited phase data available for ternary alloys and almost none for quaternary or higher. Computation opens the potential for rapidly exploring this material space for capturing trends in properties. In particular we would like to know whether a possible HEA is stable at room-temperature.
In this article we examine the stability of multi-component alloys to chemical fluctuations. To do so we properly generalize and
interpret the $S^{(2)}$ theory developed for binary alloys.\cite{PhysRevLett.50.374,0305-4608-15-6-018,PhysRevB.50.1450} 
This theory addresses the stability of multicomponent alloys by calculating the free energy change as a result of an infinitesimal change in the site average occupancies of the components. The free-energy change includes not only entropic effects but also electronic effects from rearranging charge and changing electronic structure. The inclusion of all charge effects in the multicomponent case goes beyond what has been presented in the past\cite{althoff-duane,PhysRevB.53.10610} and more recently.\cite{PhysRevB.91.224204,johnson-review} We show a reciprocal connection between the free energy change and the derived short-range atomic order. We interpret our results results as a competition of entropy terms driving disorder and favorable pair energetics driving atomic ordering.

\begin{figure*}[t]
\hspace*{25pt}{\begin{overpic}[scale=0.8]{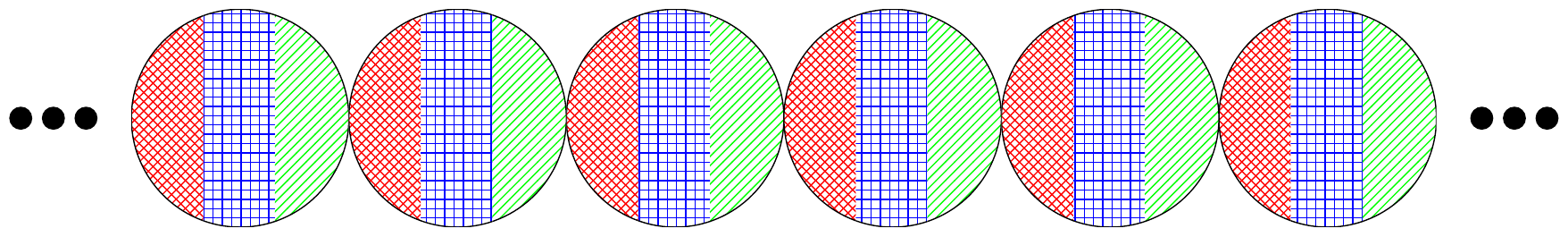}
\put (-5,12) { $(a)$}
\end{overpic}}
\hspace*{25pt}{\begin{overpic}[scale=0.8]{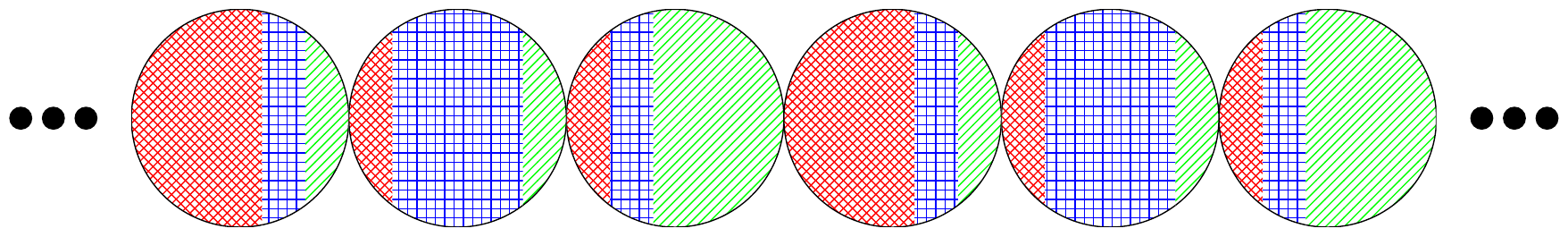}
\put (-5,12) { $(b)$}
\end{overpic}}
\caption{(Color online) (a) High-temperature fully disordered state of a hypothetical one-dimensional equiatomic alloy `ABC' with unspecified atomic interactions. The filling at each site represents an ensemble average of the occupancy by atoms of type A (red cross hatch), B (blue grid), and C (green north east lines). We are concerned with the free-energy change on imposing an infinitesimal, site-dependent variation in these average occupancies (i.e. concentrations) (b)~A finite variation in site concentrations establishing partial to full ordering. This variation may be viewed as the sum of two \emph{concentration waves}: $c(x) = [1/3,1/3,1/3] +  { \eta ( [-1/2-\sqrt{3}i/2, -1/2+\sqrt{3}i/2, 1 ] e^{i 2\pi x/3a} + c.c. )} $ for components A, B, C respectively and lattice constant $a$.}
\end{figure*}

Before proceeding, we contrast the $S^{(2)}$ theory with two well-known methods for predicting metallic phase transitions: cluster expansions and CALPHAD. Cluster expansions\cite{walle,PhysRevB.81.224202} are based on expanding the energy of an alloy configuration using nearest-neighbor lattice clusters. Each term consists of an unspecified prefactor and the product of ``spin variables" for sites within a cluster. The spin variable at a site reflects the atomic species occupying that site. The final energy is the sum of such terms over all permitted clusters. As is evident, this method has many free parameters that must be fit to either experimental data or the density-functional theory (DFT) energetics of specific configurations. Anywhere from 30-50 DFT energies of ordered compounds are needed to achieve a reliable fit. In addition, considerable care is required in choosing which clusters to permit and which ordered compounds to fit to. Otherwise over-fitting or poor reproduction of low-energy configurations occurs. But with a reliable fit the complete phase diagram may be assessed using Monte Carlo simulation. The other technique, CALPHAD,\cite{Hillert2001161,campbell} is based on large databases of experimental data available for ordered compounds. It predicts the Gibbs free energy of mixed phases using linear mixing (of Gibbs energy at end compounds), point entropy, and correction (or ``excess") terms. The correction terms are fit to be as consistent with the known experimental and/or DFT data as possible. By minimizing the Gibbs free energy it can also be used to predict a complete phase diagram.

In contrast to the above techniques we are here primarly considered with assessing the phase stability of very many HEAs. This is a single phase which presents itself only near the center of a multicomponent phase diagram. This is where the least experimental data is available and where extrapolation of data from binaries is of questionable validity. In addition, to enable high-throughput methods, a technique that requires limited guidance is needed. The $S^{(2)}$ theory is a self-contained DFT theory; requiring only lattice constant and choice of structure. Most HEAs, in fact, present themselves in simple close-packed structures: FCC, BCC, and HCP. The Korringa-Kohn-Rostoker (KKR)\cite{0034-4885-74-9-096501} method along with the coherent potential approximation (CPA)\cite{0034-4885-74-9-096501,PhysRev.156.809} is ideally suited to this case.

In the first four sections we give an overview of Landau theory, KKR-CPA, and mean-field theory within the context of multicomponent alloys. In the next four we discuss the details of the $S^{(2)}$ theory; including mapping to an effective pair interaction model, calculating the kind and size of chemical fluctuations, interpreting the possible modes of chemical polarization, and discussing the equations that give the complete linear response of the disordered alloy. After this we discuss the Onsager mean-field correction that restores certain sum rules of the short-range order parameter. We also describe a theoretical simplification that freezes all charge effects (the \emph{band-only} approximation). Lastly, as an example, we apply the theory within the band-only approximation to equiatomic alloys CuAgAu, NiPdPt, RhPdAg, and CoNiCu.

\section{Landau Theory}

The phase of an alloy is specified once the temperature, pressure,
and concentration of each component metal is known. Alternatively, we may choose to fix the alloy lattice constant and hence volume. We take the latter view throughout. In \emph{substitutional
alloys} the lattice structure is fixed and only atomic occupancies vary. In
\emph{interstitial alloys} additional atoms may occupy the interstices. These may also be treated as substitutional
alloys if interstice positions are included in the lattice and if vacancies
are considered as if a component atom. At high temperatures entropy dictates
component atoms have no site preference. As the temperature is lowered
this site symmetry is broken and either partial or full site ordering
is established. The Landau theory seeks to predict these site preferences
by minimizing the Helmholtz free energy.

For definiteness consider a crystal with Bravais lattice $\{R_{i}\}$,
basis $\{h_{1},h_{2},\ldots,h_{p}\}$, and atomic components $\{\alpha_{1},\alpha_{2},\ldots,\alpha_{n}\}$
at $N$ Bravais sites. We restrict ourselves to the case where all
site positions $\{R_{i}+h_{\text{a}}\}$ are crystallographically
equivalent. Later we will further restrict this to a Bravais lattice without basis. Let $\xi_{\mathfrak{i}\alpha}\in\{0,1\}$ indicate the
occupancy of an $\alpha$ atom at composite site index $\mathfrak{i}=(i,\text{a)}$.
Then $\{\xi_{\mathfrak{i}\alpha}\}$ briefly represents a specific
configuration. Now imagine an ensemble of configurations in which
we restrict $\langle\xi_{\mathfrak{i}\alpha}\rangle=c_{\mathfrak{i}\alpha}$
and $\sum_{\mathfrak{i}}c_{\mathfrak{i}\alpha}/N=c_{\alpha}$. Here
$\langle\cdot\rangle$ refers to an ensemble average. These constraints
permit one to continuously vary site occupancies $\{c_{\mathfrak{i}\alpha}\}$ while preserving
the known, total concentrations $\{c_{\alpha_{1}},c_{\alpha_{2}},\ldots,c_{\alpha_{n}}\}$.
The probability distribution $P[\{\xi_{\mathfrak{i}\alpha}\}]$ for
this ensemble is determined by minimizing the Helmholtz free energy
$F=\langle U\rangle-T\langle S\rangle$ subject to the aforementioned
constraints. It is not known a priori and will in general permit second
and higher order correlations among site occupancies $\xi_{\mathfrak{i}\alpha}$.
Relaxing the constraint $\langle\xi_{\mathfrak{i}\alpha}\rangle=c_{\mathfrak{i}\alpha}$
gives the physically realized Boltzmann distribution. By definition
$1=\langle\sum_{\alpha}\xi_{\mathfrak{i}\alpha}\rangle=\sum_{\alpha}c_{\mathfrak{i}\alpha}$.
This allows us to restrict the independent variables to the subset
$\{c_{\mathfrak{i}\alpha_1,}c_{\mathfrak{i}\alpha_2,}\ldots,c_{\mathfrak{i}\alpha_{(n-1)}}\}$.
We then speak of the $\alpha_{n}$ atom as a host species. Results
cannot, of course, depend on the choice of host atom. As mentioned,
at high $T$ the site concentrations $c_{\mathfrak{i}\alpha}=c_{\alpha}$
are site-independent and known. But, at some reduced $T_{c}$, partial
or full ordering is established. If $c_{\mathfrak{i}\alpha}(T)$ varies
smoothly through $T_{c}$ then the transition is second order. In
a first-order transition a discontinuity occurs in $c_{\mathfrak{i}\alpha}(T)$. 

The Landau theory is a series expansion of the free energy as an analytic
function of order parameters that characterize the phase transition.
In this case $F$ is being considered a functional of site-concentrations
$\{c_{\mathfrak{i}\alpha}=c_{\alpha}+\delta c_{\mathfrak{i}\alpha}\}$.\cite{kach_secondorder}
The perturbation amplitudes $\{\delta c_{\mathfrak{\mathfrak{i}\alpha}}\}$
vanish in the high $T$ phase. Thus they are long-range order parameters.
Performing a Taylor expansion of this $F$ about the high $T$ reference
state gives 
\begin{widetext}
\begin{align*}
F[\{c_{\mathfrak{i}1,}c_{\mathfrak{i}2,}\ldots,c_{\mathfrak{i}(n-1)}\}]= & F[\{c_{\alpha}\}]+\sideset{}{'}\sum_{\mathfrak{i\alpha}}\frac{\partial F}{\partial c_{\mathfrak{i}\alpha}}\bigg|_{\{c_{\alpha}\}}\delta c_{\mathfrak{i}\alpha}+\frac{1}{2}\sideset{}{'}\sum_{\mathfrak{i}\alpha;\mathfrak{j}\beta}\frac{\partial F}{\partial c_{\mathfrak{i}\alpha}\partial c_{\mathfrak{j\beta}}}\bigg|_{\{c_{\alpha}\}}\delta c_{\mathfrak{i}\alpha}\delta c_{\mathfrak{j}\beta}+\cdots
\end{align*}
\end{widetext}
where the prime on summations means the $\alpha_{n}$ (host) index
should be omitted. As all sites are equivalent in the reference state,
$\partial F/\partial c_{\mathfrak{i}\alpha}|_{\{c_{\alpha}\}}$ must
be independent of site position $\mathfrak{i}$. And clearly $\sum_{\mathfrak{i}}\delta c_{\mathfrak{i}\alpha}=0$
to conserve total concentrations. Taken together this implies the
first order term vanishes. Because the reference state has translational
symmetry, it is preferable to use Fourier transformed components $\delta c_{\text{a}\alpha}(k)$
(see Appendix \ref{sub:Lattice-Fourier-transform} for definition of lattice Fourier transforms). The wave vector
$k$ is confined to the first Brillouin zone in all such transforms. Then
\begin{multline}
\delta F[\{c_{\text{a}\alpha}(k)\}]= \\ \frac{1}{2}\sideset{}{'}\sum_{\text{a}\alpha;\text{b}\beta}\sum_{k}\delta c_{\text{a}\alpha}(k)^{*}F^{(2)}(k)_{\text{a\ensuremath{\alpha};\text{b\ensuremath{\beta}}}}\delta c_{\text{b}\beta}(k)+\cdots\label{eq:free-energy-expansion}
\end{multline}
for suitably defined $F^{(2)}(k)_{\text{a\ensuremath{\alpha};\text{b}\ensuremath{\beta}}}\rightarrow\mathbb{F}^{(2)}(k)$.
The diagonalization in $k$-space of the second order term is a consequence
of translational symmetry. As long as $\mathbb{F}^{(2)}(k)$ is a
positive definite matrix (i.e. all positive eigenvalues), the system
is stable to infinitesimal fluctuations from the high $T$ reference
state. A second order phase transition occurs when the minimum in
free energy expanded about the homogenuous reference bifurcates along
some mode $k_{0}$ and its star. This can only occur if both the second
\emph{and} third order terms for this star of wave vectors vanishes
at $T_{c}$. The mode $k_{0}$ and temperature $T_{c}$ is fixed by
zeroing the lowest eigenvalue: $\text{min}_{k,\sigma}\,\lambda_{\sigma}[\mathbb{F}^{(2)}(k,T_{c})]=0$.
Here $\lambda_{\sigma}(M)$ stands for the $\sigma^{\text{th}}$ eigenvalue
of matrix $M$. \emph{This determines the partial ordering established} ($k_{0}$)
\emph{and temperature at which it occurs} ($T_{c}$). To ensure the transition
is indeed second order a selection rule is needed for the third order
term
\begin{multline*}
\frac{1}{3!}\sideset{}{'}\sum_{\text{a}\alpha;\text{b}\beta;c\gamma}\sum_{k_{1}k_{2}k_{3}}F^{(3)}(k_{1},k_{2},k_{3})_{\text{a\ensuremath{\alpha};\text{b\ensuremath{\beta}};c\ensuremath{\gamma}}}\times \\ \delta c_{\text{a}\alpha}(k_{1})\delta c_{\text{b}\beta}(k_{2})\delta c_{c\gamma}(k_{3}).
\end{multline*}
By translational symmetry $F^{(3)}(k_{1},k_{2},k_{3})$ vanishes unless
$k_{1}+k_{2}+k_{3}=K$ is a reciprocal lattice vector. Thus a second
order transition generally only occurs when $k_{1}+k_{2}+k_{3}\neq K$
for any vectors $\{k_{i}\}$ within the star of $k_{0}$.\cite{kach_secondorder,landau,lifshitz} If the third-order term does not vanish then a first-order transition may take place at some \emph{higher} temperature. In either case, the vanishing of the second-order term marks an \emph{absolute instability point}. 

We will also take advantage of the grand canonical
ensemble throughout much of this article. In this case the relevant thermodynamic potential is the
grand potential as a function of chemical potentials $\nu_{\mathfrak{i}\alpha}$
for atoms of type $\alpha$ at site index~$\mathfrak{i}$. We can
write the grand potential as 
\begin{multline}
\Omega(T,\{\nu_{\mathfrak{i}\alpha}\})= -\frac{1}{\beta}\log\sum_{\{\xi_{\mathfrak{i}\alpha}\}}e^{-\beta(\Omega_{el}[\{\xi_{\mathfrak{i}\alpha}\}]-\sum_{\mathfrak{i}\alpha}\nu_{\mathfrak{i}\alpha}\xi_{\mathfrak{i}\alpha})} \\ \label{eq:omega-boltzmann-sum}
\end{multline}
where the electronic grand potential $\Omega_{el}[\{\xi_{\mathfrak{i}\alpha}\}]$
isolates the electronic degrees of freedom and $\beta$ is inverse
temperature. The above expression should make clear the site-dependent chemical potentials may undergo a
gauge transform $\nu_{\mathfrak{i}\alpha}\rightarrow\nu_{\mathfrak{i}\alpha}+\gamma_{\mathfrak{i}}$
without changing the probabilities $P[\{\xi_{\mathfrak{i}\alpha}\}]$.
We use this freedom to set $\nu_{\mathfrak{i}n}=0$ (for brevity $\nu_{\mathfrak{i}\alpha_n} \rightarrow \nu_{\mathfrak{i}n}$). Note that there
is a reciprocal relationship between site-concentrations $\{c_{\mathfrak{i}1,}c_{\mathfrak{i}2,}\ldots,c_{\mathfrak{i}(n-1)}\}$
and chemical potentials $\{\nu_{\mathfrak{i}1},\nu_{\mathfrak{i}2},\ldots,\nu_{\mathfrak{i}(n-1)}\}$
via $\langle\xi_{\mathfrak{i}\alpha}\rangle[\{\nu_{\mathfrak{i}1},\nu_{\mathfrak{i}2},\ldots,\nu_{\mathfrak{i}(n-1)}\}]=c_{\mathfrak{i}\alpha}$.
Thus we may alternatively seek to minimize $\Omega(T,\{c_{\mathfrak{i}\alpha}\})$ relative to $\{c_{\mathfrak{i}\alpha}\}$ with unspecified $\{\nu_{\mathfrak{i}\alpha}\}$
subject to the constraint $\sum_{\mathfrak{i}}c_{\mathfrak{i}\alpha}/N=c_{\alpha}$.
We may then perform the same perturbative expansion in site concentrations
as for the Helmholtz $F$.

\section{Variational Grand Potential}

It remains to determine an explicit form for the grand potential $\Omega(T,\{c_{\mathfrak{i}\alpha}\})$.
In principle $\Omega_{el}[\{\xi_{\mathfrak{i}\alpha}\}]$ of Eq.~(\ref{eq:omega-boltzmann-sum})
can be computed for a supercell within the context of DFT. However this is near the limit of computational tractability.
The first simplification that can be made is to consider the distribution
$P[\{\xi_{\mathfrak{i}\alpha}\}]$ to be a perturbation from an uncorrelated
distribution $P_{0}[\{\xi_{\mathfrak{i}\alpha}\}]=\prod_{\mathfrak{i}}\mathcal{P}_{0}(\xi_{\mathfrak{i}\alpha})$.
Here $\mathcal{P}_{0}[\xi_{\mathfrak{i1}},\xi_{\mathfrak{i2}},\ldots,\xi_{\mathfrak{i}n}]=\bar{c}_{\mathfrak{i}\alpha}$
if $\xi_{\mathfrak{i}\alpha}=1$. The bar notation $\bar{c}_{\mathfrak{i}\alpha}$
is a reminder that the uncorrelated distribution is arbitrary at this
stage. If $H_{0}$
is the mean-field Hamiltonian that gives rise to the uncorrelated
distribution $P_{0}[\{\xi_{\mathfrak{i}\alpha}\}]$, then a first-order
expansion of Eq.~(\ref{eq:omega-boltzmann-sum}) from this reference
state is
\begin{align}
\Omega(T,\{\nu_{\mathfrak{i}\alpha}\}) \mkern-50mu \nonumber \\ 
=& -\frac{1}{\beta}\log\sum_{\{\xi_{\mathfrak{i}\alpha}\}}e^{-\beta(\Omega_{el}-H_{0})-\beta(H_{0}-\sum_{\mathfrak{i}\alpha}\nu_{\mathfrak{i}\alpha}\xi_{\mathfrak{i}\alpha})} \nonumber  \\
\approx & -\frac{1}{\beta}\log\sum_{\{\xi_{\mathfrak{i}\alpha}\}}\left(1-\beta(\Omega_{el}-H_{0})\right)e^{-\beta(H_{0}-\sum_{\mathfrak{i}\alpha}\nu_{\mathfrak{i}\alpha}\xi_{\mathfrak{i}\alpha})} \nonumber  \\
\approx & \,\,\Omega_{0}+\langle\Omega_{el}-H_{0}\rangle_{0}\equiv\Omega^{(1)} \label{eq:omega-expansion}
\end{align}
where the logarithm has been expanded to first order in $\beta(\Omega_{el}-H_{0})$
and $\langle\cdot\rangle_{0}$ means ensemble average with respect
to the uncorrelated distribution. We emphasize that this expansion is most valid 
for small $\beta$ and/or weakly correlated systems. 
The entropy of the uncorrelated
reference state is easily known and we explicitly write
\begin{align}
\Omega^{(1)}[\{\nu_{\mathfrak{i}\alpha}\},\{\bar{c}_{\mathfrak{i}\alpha}\}] \mkern-110mu \nonumber \\
&= -TS_{0}-\sideset{}{'}\sum_{\mathfrak{i}\alpha}\nu_{\mathfrak{i}\alpha}N_{0,\mathfrak{i}\alpha}+\langle\Omega_{el}\rangle_{0}[\{\bar{c}_{\mathfrak{i}\alpha}\}] \nonumber \\
&=  \beta^{-1}\sum_{\mathfrak{i}\alpha}\bar{c}_{\mathfrak{i}\alpha}\log\bar{c}_{\mathfrak{i}\alpha}-\sideset{}{'}\sum_{\mathfrak{i}\alpha}\nu_{\mathfrak{i}\alpha}\bar{c}_{\mathfrak{i}\alpha}+\langle\Omega_{el}\rangle_{0}[\{\bar{c}_{\mathfrak{i}\alpha}\}],
\label{eq:omega1-final}
\end{align}
where as before $\nu_{\mathfrak{i}n}=0$. By the Gibbs-Bogoliubov-Feynman
inequality,\cite{feynman-statphy} $\Omega^{(1)}$ is in fact a variational upper bound on $\Omega$.
Minimizing with respect to $\{\bar{c}_{\mathfrak{i}1},\ldots,\bar{c}_{\mathfrak{i(}n-1)}\}$
gives the optimal uncorrelated reference system. That is
\begin{align}
0=\frac{\partial\Omega^{(1)}}{\partial\bar{c}_{\mathfrak{i}\alpha}}=\beta^{-1}\log\frac{\bar{c}_{\mathfrak{i}\alpha}}{\bar{c}_{\mathfrak{i}n}}-\nu_{\mathfrak{i}\alpha}+\frac{\partial\langle\Omega_{el}\rangle_{0}}{\partial\bar{c}_{\mathfrak{i}\alpha}}.\label{eq:optimal-site-concentrations}
\end{align}
This equation establishes a reciprocal relationship between $\{\bar{c}_{\mathfrak{i}1},\ldots,\bar{c}_{\mathfrak{i(}n-1)}\}$
and $\{\nu_{\mathfrak{i}1},\ldots,\nu_{\mathfrak{i}(n-1)}\}$. It
effectively pins each uncorrelated reference system to a corresponding
physical system and vice-versa depending on $\langle\Omega_{el}\rangle_{0}$.
A perturbative Landau analysis on $\Omega^{(1)}[\{\bar{c}_{\mathfrak{i}\alpha}\}]$
precedes as before. Also note that $\bar{c}_{\mathfrak{i}\alpha}=\langle\xi_{\mathfrak{i}\alpha}\rangle_{0}$
and $c_{\mathfrak{i}\alpha}=\langle\xi_{\mathfrak{i\alpha}}\rangle$
need not coincide for given $\{\nu_{\mathfrak{i}\alpha}\}$. While
the above relation for $\Omega\approx\Omega^{(1)}$ is more explicit
than before, it remains to determine 

\begin{align*}
\langle\Omega_{el}\rangle_{0}= & \sum_{\{\xi_{\mathfrak{i}\alpha}\}}\prod_{\mathfrak{i}}\mathcal{P}_{0}(\xi_{\mathfrak{i}\alpha})\Omega_{el}[\{\xi_{\mathfrak{i}\alpha}\}].
\end{align*}
Note $\langle\Omega_{el}\rangle_{0}$ has no explicit dependence on chemical
potentials $\{\nu_{\mathfrak{i}\alpha}\}$. While the ensemble average
is now uncorrelated, it still contains the intractable factor $\Omega_{el}[\{\xi_{\mathfrak{i}\alpha}\}]$.
We now consider the computation of this term from first-principles
electronic structure theory.

\begin{figure}
\begin{overpic}[scale=0.45]{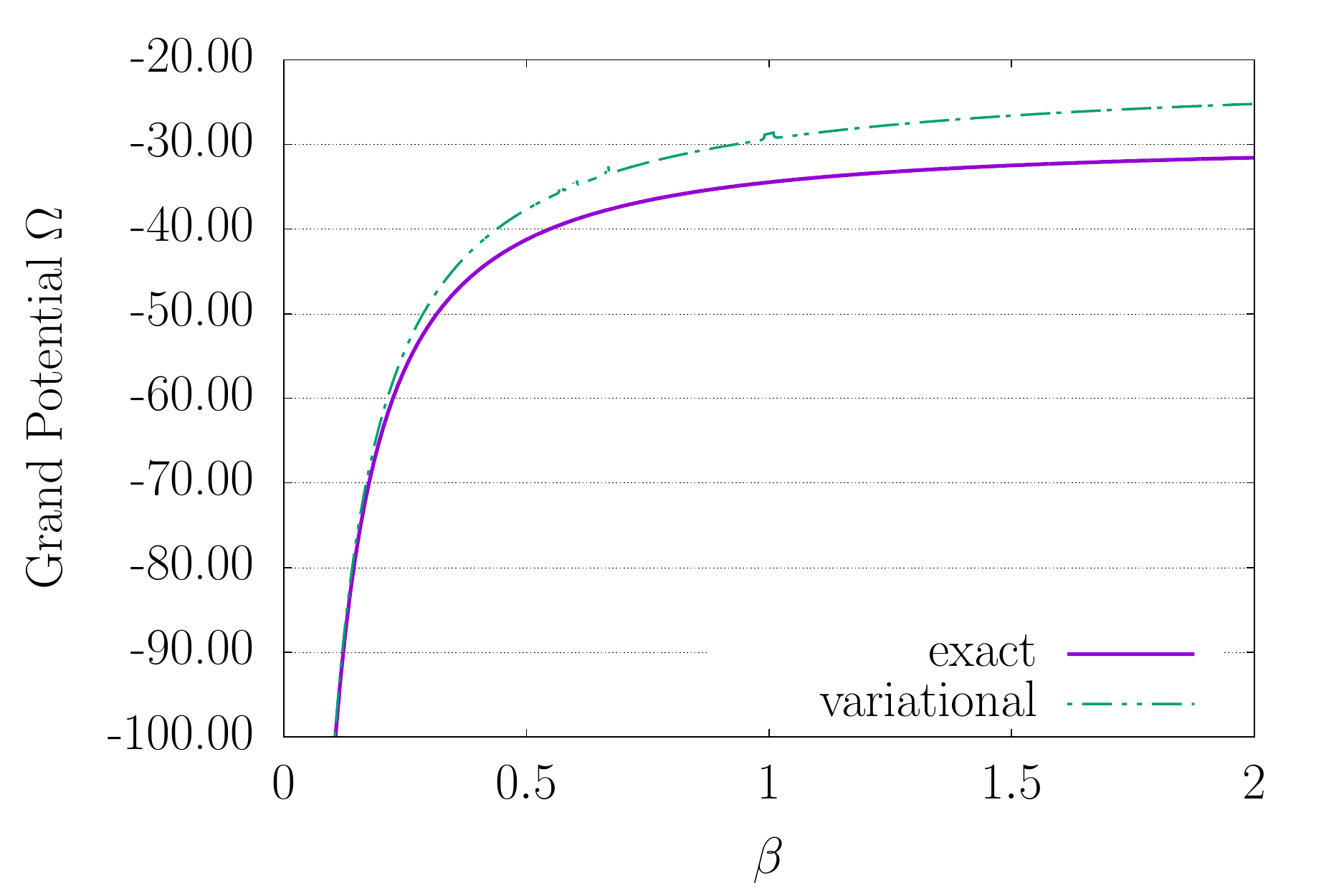}
\put(0,60) {$(a)$}
\end{overpic}
\begin{overpic}[scale=0.45]{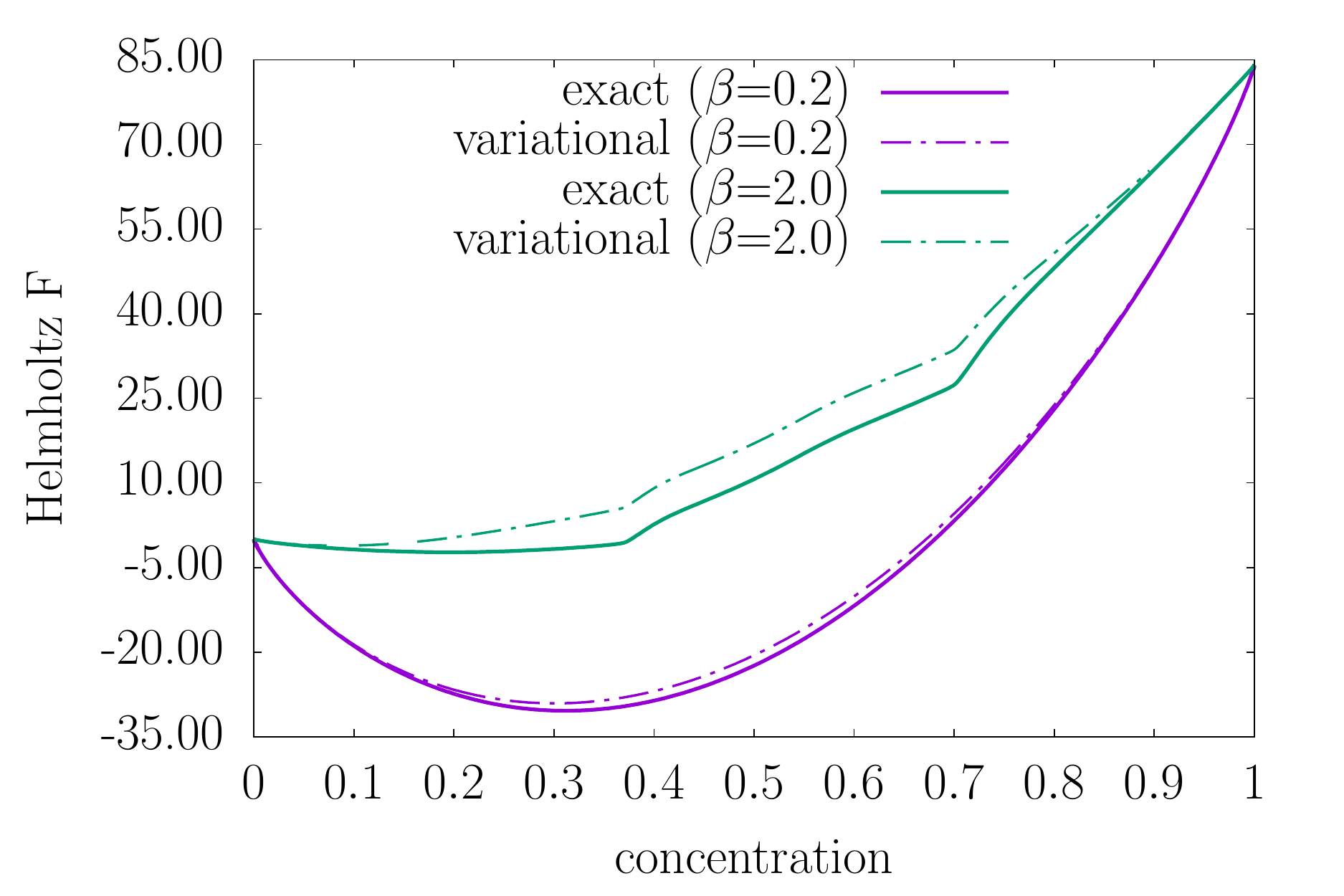}
\put(0,60) {$(b)$}
\end{overpic}
\caption{(Color online) Results for a toy one-dimensional binary alloy consisting of twelve sites with periodic boundary conditions. Atoms are of kind A or B and the units of energy are arbitrary. Nearest neighbor A-A pair energy is 5.0 and second nearest-neighbor A-A pair energy is 2.0. All other pair interactions are zero. (a) Exact and variational grand potential for fixed concentration 0.5. The exact answer is calculated from a grand canonical weighted sum over all 2$^{12}$ configurations. Note that the variational upper bound becomes loose at large $\beta$. (b) Exact and variational Helmholtz free energy versus concentration of A atoms for $\beta=0.2$ (lower curves) and $\beta=2.0$ (upper curves). The variational bound becomes tight in the ordered limits $c \rightarrow 0$ and $c \rightarrow 1$.}
\label{fig:toy-model-potentials}
\end{figure}

\section{Multiple Scattering Theory\label{sec:MST}}

To evaluate $\langle\Omega_{el}\rangle_{0}$ a framework is needed
to solve the electronic structure problem and to effectively perform
the ensemble average. Here the intention is to solve the electronic
structure using DFT and the multiple scattering
technique. The advantage of the multiple-scattering (or KKR) technique\cite{zabloudil,0022-3719-20-16-010}
is that it provides a generalization for approximating the ensemble
averages. This is based on the CPA and described in the next section.
We briefly mention the key notions and equations of multiple-scattering
without derivation. This will provide a starting
point for the linear response theory outlined later.

Density functional theory maps the many-electron problem to that of
a single electron traveling in a effective crystal potential $V(r)$.
The $V(r)$ is the average Coloumb field of the nuclei and electrons
plus an additional tem $V_{xc}(r)$ that compensates for exchange
and correlation effects. It is nominally a full functional of the
electron-density. In the local-density approximation this dependence
is reduced to $V_{xc}(r)=f(\rho(r))$ where $f(\rho)$ is a univariate
function. Many choices are available for $V_{xc}(r)$ and any of them
is equally suitable for our purposes.

Multiple-scattering theory solves the reduced one-electron Schr\"{o}dinger
equation by giving a procedure for calculating the Greens function
$G(E;r,r')\equiv\langle r|(E- H)^{-1}|r'\rangle$. It is based on a
partitioning of real space into volumes $V_{\mathfrak{i}}$ about
each site. This naturally defines a set of non-overlapping potentials
$V_{\mathfrak{i}}(r)=V(r)$ for $r\in V_{\mathfrak{i}}$ and $V_{\mathfrak{i}}(r)=0$
otherwise. The procedure for $G$ then proceeds in two steps:

\textbf{\emph{Step 1}}. For each site $\mathfrak{i}$ and composite
angular momentum index $L=(\ell,m)$ the Schr\"{o}dinger equation $(-\nabla^{2}+V_{\mathfrak{i}}(r))\psi=E\psi$
is solved for two linearly independent solutions $\phi_{\mathfrak{i}L}(E;r)$
and $J_{\mathfrak{i}L}(E;r$). These are defined by boundary conditions
\begin{align*}
\lim_{r\rightarrow0}\phi_{\mathfrak{i}L}(E;r)\rightarrow j_{\ell}(\sqrt{E}r)Y_{\ell m}(r) \\
\lim_{r\notin V_{\mathfrak{i}}}J_{\mathfrak{i}L}(E;r)\rightarrow j_{\ell}(\sqrt{E}r)Y_{\ell m}(r)
\end{align*}
for spherical bessel $j_{l}(r)$ and spherical harmonic $Y_{\ell m}(r)$.
The Jost function $\phi_{\mathfrak{i}L}(E;r)$ is transformed to the more useful $Z_{\mathfrak{i}L}(E;r)=\sum_{L'} \phi_{\mathfrak{i}L'}(E;r) (\alpha_{\mathfrak{i}} t_\mathfrak{i}^{-1})_{L'L}$ using matrices $\alpha_{\mathfrak{i};LL'}(E)$ and $t_{\mathfrak{i};LL'}(E)$ to be defined presently. Both $Z_{\mathfrak{i}L}(E;r)$ and $J_{\mathfrak{i}L}(E;r)$ play a key role in the theory.
Occasionally we also have need for the regular scattering solution; self-consistently
defined as 
\begin{multline}
R_{\mathfrak{i}L}(E;r)=j_{\ell}(\sqrt{E}r)Y_{\ell m}(r)+\\ \int dr'G_{0}(E,r,r')V_{\mathfrak{i}}(r')R_{\mathfrak{i}L}(E;r') \label{eq:reg-sol-definition}
\end{multline}
where $G_{0}(E,r,r')$ is the well-known free-particle Green function.
From this we can also define a so-called alpha matrix $\alpha_{\mathfrak{i};LL'}$
via $R_{\mathfrak{i}L}(E;r) \rightarrow \sum_{L'}j_{\ell'}(\sqrt{E}r)Y_{L'}(r)\alpha_{\mathfrak{i};L'L}$ as $r\rightarrow0$.
The alpha matrix will be used in \emph{Lloyd's formula}, to be described later. 

In addition to these wave solutions, the on-shell scattering $T(E)$
operator for each potential $V_{\mathfrak{i}}(r)$ is needed. The definition
and computation of the $T$ operator follows from conventional scattering
theory.\cite{taylor} We calculate this operator in a basis of $j_{\ell}(\sqrt{E}r)Y_{\ell m}(r)$,
writing $t_{\mathfrak{i};LL'}(E)$. When $V_{\mathfrak{i}}(r)$ is
a spherical scatterer and the site scattering phase shifts $\delta_{\mathfrak{i}\ell}(E)$
are known, then $t_{\mathfrak{i};LL'}(E)=-\delta_{\ell\ell'}e^{i\delta_{\mathfrak{i}\ell}(E)}\sin\delta_{\mathfrak{i}\ell}(E)/\sqrt{E}$
for Kronecker delta $\delta_{\ell\ell'}$. It is not however necessary
that $V_{\mathfrak{i}}(r)$ be spherical. In general, 
\begin{align}
t_{\mathfrak{i};LL'}(E)=\int drj_{\ell}(\sqrt{E}r)Y_{\ell m}(r)V_{\mathfrak{i}}(r)R_{\mathfrak{i}L'}(E;r). \label{eq:tmatrix-definition}
\end{align}
Lastly the $t_{\mathfrak{i}}$ matrices are concatenated along the
diagonal of a supermatrix $\mathbf{t}_{\mathfrak{i}L;\mathfrak{j}L'}(E)=\delta_{\mathfrak{i}\mathfrak{j}}t_{\mathfrak{i};LL'}(E).$
This supermatrix has combined row (column) index $(\mathfrak{i},L)$.
$\blacklozenge$

\textbf{\emph{Step 2}}. The independent, site-centered solutions are
stitched together by calculating the so-called \emph{scattering path
operator} (SPO) supermatrix 
\begin{align}
\mathbf{\tau}_{\mathfrak{\mathfrak{i}}L;\mathfrak{j}L'}\mathbf{=[}\mathbf{t^{-1}-G_{0}]}_{\mathfrak{\mathfrak{i}}L;\mathfrak{j}L'}^{-1}. \label{eq:spo-definition}
\end{align}
The\emph{ structure constants }$\mathbf{G_{0}}\vphantom{G}_{;\mathfrak{i}L;\mathfrak{j}L'}(E)$
are a priori known given lattice site positions $\{R_{\mathfrak{i}}=R_{i}+h_{\text{a}}\}$.\cite{zabloudil}
They are independent of the crystal potential $V(r)$. Since we consider
the lattice fixed we may take the structure constants for granted.
The interpretation of the SPO element $\tau_{\mathfrak{i}\mathfrak{j}}$
is it gives the analog of the $T$ matrix that connects incoming waves
on site $\mathfrak{j}$ to outgoing waves on site $\mathfrak{i}$.
Finally,
\begin{multline}
G(E;r,r')= \sum_{LL'}Z_{\mathfrak{i}L}(E;r)\tau_{\mathfrak{i}L;\mathfrak{j}L'}(E)Z_{\mathfrak{j}L'}(E;r')\\- 
\delta_{\mathfrak{i}\mathfrak{j}}\sum_{L}Z_{\mathfrak{i}L}(E;r_{<})J_{\mathfrak{i}L}(E;r_{>})\label{eq:green-function}
\end{multline}
for $r\in V_{\mathfrak{i}}$ and $r'\in V_{\mathfrak{j}}$ and $r_{<}=\text{min}(r-R_{\mathfrak{i}},r'-R_{\mathfrak{i}})$
and $r_{>}=\text{max}(r-R_{\mathfrak{i}},r'-R_{\mathfrak{i}})$. $\blacklozenge$

Using the Greens function it is easy to compute the electron density
$\rho(r)$ and density of states $n(E)$ as a post-processing step.
These are
\begin{alignat}{1}
\rho(r)= & -\frac{1}{\pi}\lim_{\epsilon\rightarrow0}\text{Im}\int f(E-\mu)G(E+i\epsilon;r,r)dE\label{eq:charge-from-green}\\
n(E)= & -\frac{1}{\pi}\lim_{\epsilon\rightarrow0}\text{Im}\int f(E-\mu)G(E+i\epsilon;r,r)dr\nonumber 
\end{alignat}
for Fermi-Dirac function $f(E-\mu$). The electronic potential $\mu$
is fixed to ensure an overall charge-neutral system. It is at this
stage that finite-temperatures enter the electronic formalism. The
choice of a numerical grid of energies $\{E_{i}\}$ for evaluating
the above densities dictates the energies that need to be considered
in the above process. If a potential $V_{\text{Hart.}}(r)$ is solved
via the Poisson equation $\nabla^{2}V_{\text{Hart.}}(r)=-e\rho(r)$;
then the previous procedure can be repeated until $V_{\text{out}}(r):=V_\text{Hart.}(r)+V_\text{xc.}(r)=V_{\text{in}}(r)$.
This establishes a self-consistent potential. Using $\rho(r)$ and
$V(r)$ it is possible to write an expression for the grand
potential $\Omega_{el}$. We do this in the next section when we simultaneously
consider how to simulate the $\langle\cdot\rangle_{0}$ ensemble averaging.

\section{Coherent Potential Approximation\label{sec:cpa}}

The coherent potential approximation (CPA)\cite{PhysRev.156.809} is a mean-field technique
for addressing the ensemble average in $\langle\Omega_{el}\rangle_{0}$.
To accomodate disorder, the single potential $V_{\mathfrak{i}}(r)$
at each site $\mathfrak{i}$ is replaced by the set of potentials
$\{V_{\mathfrak{i}\alpha}(r)\}$. This in turn leads to a series of
associated $T$ matrices $\{t_{\mathfrak{i}\alpha}(E)\}$. To continue,
the CPA seeks an optimal mean-field medium of scatterers $\{t_{\mathfrak{i}c}(E)\}$
($c$ for CPA) that coherently accounts for the average scattering
properties of $\{t_{\mathfrak{i}\alpha}\}$. As per multiple-scattering
theory, this optimal mean-field medium has corresponding SPO $\mathbf{\tau}_{\mathfrak{i}L;\mathfrak{j}L'}^{c}\mathbf{=[}\mathbf{t_{c}^{-1}-G_{0}]}_{\mathfrak{\mathfrak{i}}L;\mathfrak{j}L'}^{-1}$.
Now consider the same mean-field medium $\{t_{\mathfrak{i}c}\}$ but
with embedded impurity atom $\alpha$ at site $\mathfrak{i}_{0}$.
In this case we make the site substitution $\mathbf{t}_{\mathfrak{i}L;\mathfrak{j}L'}^{\mathfrak{i}_{0}\alpha}=\delta_{\mathfrak{i}\mathfrak{j}}\delta_{\mathfrak{i}\mathfrak{i}_{0}}t_{\mathfrak{i}\alpha;LL'}+\delta_{\mathfrak{i}\mathfrak{j}}(1-\delta_{\mathfrak{i}\mathfrak{i}_{0}})t_{\mathfrak{i}c;LL'}$.
Its corresponding SPO is $\tau^{\mathfrak{i}_{0}\alpha}=[(\mathbf{t}^{\mathfrak{i}_{0}\alpha})^{-1}-\mathbf{G_{0}}]^{-1}$.
Using Eq.~(\ref{eq:green-function}) we can also construct an associated
Greens function $G^{\mathfrak{i}_{0}\alpha}(E,r,r')$. To fix the
medium $\{t_{\mathfrak{i}c}(E)\}$ the CPA makes the physically sensible
constraint that 
\begin{align}
\tau_{\mathfrak{i}\mathfrak{i}}^{c}= \sum_{\alpha}\bar{c}_{\mathfrak{i}\alpha}\tau_{\mathfrak{i}\mathfrak{i}}^{\mathfrak{i}\alpha}\label{eq:CPA-condition}
\end{align}
at every site $\mathfrak{i}$ for ensemble provided site concentrations
$\bar{c}_{\mathfrak{i}\alpha}=\langle\xi_{\mathfrak{i}\alpha}\rangle_{0}$.
This condition states that performing an SPO averaging over impurities
at a given site restores the mean-field SPO. It could also be reformulated
as an averaging over Greens functions if desired. Given $\tau^{\mathfrak{i}\alpha}$
we can define site-dependent electron densities $\rho_{\mathfrak{i}\alpha}(r)$
and density of states $n_{\mathfrak{i}\alpha}(E)$ via Eq.~(\ref{eq:charge-from-green})
with $G=G^{\mathfrak{i}\alpha}$. It remains
how to determine $V_{\mathfrak{i}\alpha,\text{out}}(r)$. This has
been considered in detail by Johnson \emph{et
al.}\cite{PhysRevB.41.9701} and is given by
\begin{multline}
V_{\mathfrak{i}\alpha}(r)=V_{\text{xc}}(\rho_{\mathfrak{i}\alpha}(r))+ 
 e^{2}\int_{V_{\mathfrak{i}}}dr'\frac{\rho_{\mathfrak{i}\alpha}(r')-Z_{\alpha}\delta(r')}{|r-r'|} \\
 +e^{2}\sum_{\mathfrak{j}\neq\mathfrak{i}}\int_{V_{\mathfrak{j}}}dr'\frac{\overline{\rho}_{\mathfrak{j}}(r')-\bar{Z}_{\mathfrak{j}}\delta(r')}{|r+R_{\mathfrak{i}}-R_{\mathfrak{j}}-r'|}\label{eq:cpa-potential}
\end{multline}
where $Z_{\alpha}$ is the atomic number of atom $\alpha$, and $\overline{\rho}_{\mathfrak{i}}(r)=\sum_{\alpha}\bar{c}_{\mathfrak{i}\alpha}\rho_{\mathfrak{i}\alpha}(r)$
and $\bar{Z}_{\mathfrak{i}}=\sum_{\alpha}\bar{c}_{\mathfrak{i}\alpha}Z_{\mathfrak{\alpha}}$
are site averages. The second and third terms represent the intra-
and intersite Coloumb interactions respectively. Using this prescription
one can take $V_{\mathfrak{i}\alpha,\text{out}}(r)\rightarrow V_{\mathfrak{i}\alpha,\text{in}}(r)$
until self-consistency is achieved. 

For convenience we here define CPA related quantities that are used extensively in expressions to follow, 
\begin{align}
\Delta_{\mathfrak{i}\alpha}=t_{\mathfrak{i}\alpha}^{-1}-t_{\mathfrak{i}c}^{-1} \label{eq:cpa-definitions-delta} \\
D_{\mathfrak{i}\alpha}=[1+\tau_{\mathfrak{i}\mathfrak{i}}^{c}\Delta_{\mathfrak{i}\alpha}]^{-1}=\Delta_{\mathfrak{i}\alpha}^{-1}X_{\mathfrak{i}\alpha} \label{} \label{eq:cpa-definitions-Dbar} \\
\bar{D}_{\mathfrak{i}\alpha}=[1+\Delta_{\mathfrak{i}\alpha}\tau_{\mathfrak{i}\mathfrak{i}}^{c}]^{-1}=X_{\mathfrak{i}\alpha}\Delta_{\mathfrak{i}\alpha}^{-1}.\label{eq:cpa-definitions-D}
\end{align}

The electronic grand potential is related to the total number of electrons by the thermodynamic 
relation  $\partial\Omega/\partial\mu=-N$. The average integrated density of states $\langle N(E)\rangle_{0}$ is approximated
within the CPA by \emph{Lloyd's formula}\cite{0953-8984-16-36-011}
\begin{multline}
N_{c}(E)=N_{0}(E)\,+ \\ \frac{1}{\pi}\text{Im}\bigg\{ \text{log}||\mathbf{\tau}^{c}||+ 
\sum_{\mathfrak{i}\mu}\bar{c}_{\mathfrak{i}\mu}( 
\log||\alpha_{\mathfrak{i}\mu} t_{\mathfrak{i}\mu}^{-1}||-\log||\bar{D}_{\mathfrak{i}\mu}^{-1}||)\bigg\} \label{eq:lloyd-formula}
\end{multline}
where $N_{0}(E)$ is the free-electron integrated density of states
and the $\alpha_{\mathfrak{i}\mu}$ matrix is defined in Section~\ref{sec:MST}. 
The determinant $||\tau^{c}||$ is over composite
indices $(\mathfrak{i},L)$ while remaining determinants are over
indices $L$ only. The Lloyd formula obeys a variational property
$\delta N_{c}/\delta t_{\mathfrak{i}c}^{-1}=0$ when varying mean-field
medium $\{t_{\mathfrak{i}c}\}$ away from the CPA solution while holding
potentials $\{V_{\mathfrak{i}\alpha}(r)\}$ fixed.  Notably, this formula is the multiple scattering generalization of the Friedel sum rule.

Performing a series of integrations and non-trivial
substitutions on $\partial\Omega/\partial\mu=-N,$ one obtains an expression for the
grand potential. Johnson \emph{et~al.}\cite{PhysRevB.41.9701} have derived

\begin{widetext}
\vspace*{-10pt}
\begin{multline}
\Omega_{c}= \bigg\{ -\int dE\,f(E-\mu)N_{c}(E)-
\sum_{\mathfrak{i}\alpha}\bar{c}_{\mathfrak{i}\alpha} \int_{V_{\mathfrak{i}}}dr\rho_{\mathfrak{i}\alpha}(r)V_{\mathfrak{i}\alpha}(r)\bigg\} \\
+\bigg\{\sum_{\mathfrak{i}\alpha}\bar{c}_{\mathfrak{i}\alpha} \int_{V_{\mathfrak{i}}}dr\,\rho_{\mathfrak{i}\alpha}(r)\epsilon_{\text{xc}}(\rho_{\mathfrak{i}\alpha}(r)) 
+\frac{e^{2}}{2}\sum_{\mathfrak{i}\alpha}\bar{c}_{\mathfrak{i}\alpha} \int_{V_{\mathfrak{i}}}dr\int_{V_{\mathfrak{i}}}dr'\frac{1}{|r-r'|}\left[\rho_{\mathfrak{i}\alpha}(r)\rho_{\mathfrak{i}\alpha}(r')-2Z_{\alpha}\delta(r)\rho_{\mathfrak{i}\alpha}(r')\right]\\
+\frac{e^{2}}{2}\sum_{\mathfrak{i}\neq\mathfrak{j}}\sum_{\alpha\beta}\bar{c}_{\mathfrak{i}\alpha}\bar{c}_{\mathfrak{j}\beta}\int_{V_{\mathfrak{i}}}dr\int_{V_{\mathfrak{j}}}dr'\frac{1}{|r+R_{\mathfrak{i}}-R_{\mathfrak{j}}-r'|}\left[\rho_{\mathfrak{i}\alpha}(r)\rho_{\mathfrak{j}\beta}(r')-2Z_{\alpha}\delta(r)\rho_{\mathfrak{j}\beta}(r')+Z_{\alpha}\delta(r)Z_{\beta}\delta(r')\right]\bigg\}.\label{eq:cpa-grand-potential}
\end{multline}
\end{widetext}

The univariate function $\epsilon_{xc}(\rho)$ will depend on the
choice of exchange-correlation functional. It can be shown that the
first term in braces is the band contribution and the remaining term is double-counting
corrections. An important property $\Omega_{c}$ satisfies is $\delta\Omega_{c}/\delta\rho_{\mathfrak{i}\alpha}(r)=0$
for all $\rho_{\mathfrak{i}\alpha}(r)$ at fixed $\{\bar{c}_{\mathfrak{i}\alpha}\}$.
Therefore it satisfies a variational principle much in the spirit
of finite-temperature DFT as described by Mermin for ordered systems.\cite{PhysRev.137.A1441} The above $\Omega_{c}$ provides the explicit
description for $\langle\Omega_{el}\rangle_{0}$ needed to evaluate
$\Omega^{(1)}$.

\section{Effective Pair Interaction\label{sec:effective-params}}

Considerable effort must be expended to evaluate $\langle \Omega_\text{el.} \rangle$ in a first-principles framework. We see in this section how the resulting theory can be mapped to an effective pair interaction model. These effective pair potentials are ideally suited for Monte Carlo simulation. This circumvents the need for Landau theory and in-principle enables us to anticipate both first and second-order transitions. Recall the Landau theory as we have applied it only computes an absolute instability of the high-temperature state. Therefore the Landau based theory is best suited for second-order transitions.

Key to this section is that the expansion in Eq.~(\ref{eq:omega1-final}) will be unaffected if we substitute some $\langle H_\text{eff} \rangle_0$ that \emph{mimics} $\langle \Omega_\text{el.} \rangle_0$. In particular we desire $\delta \langle H_\text{eff} \rangle_0[\{\bar{c}_{\mathfrak{i}\alpha}\}] = \delta \langle \Omega_\text{el.} \rangle_0[\{\bar{c}_{\mathfrak{i}\alpha}\}]$ for allowed $\{\bar{c}_{\mathfrak{i}\alpha}\}$.
In this case Eq.~(\ref{eq:omega1-final}) may be identified as the grand potential of a system with uncorrelated probability distribution $P_0[\{\xi_{\mathfrak{i}\alpha}\}]$ and total energy $U = \langle H_\text{eff} \rangle_0$. Suppose we make the ansatz that a given configuration $\{ \xi_{\mathfrak{i}\alpha} \}$ has effective energies
\begin{align}
\tilde{H}_\text{eff}[\{\xi_{\mathfrak{i}\alpha}\}] &= \sum_{\mathfrak{i}\alpha;\mathfrak{j}\beta} \tilde{V}_{\mathfrak{i}\alpha;\mathfrak{j}\beta} \xi_{\mathfrak{i}\alpha} \xi_{\mathfrak{j}\beta} \label{eq:pair-pot-all} \\
H_\text{eff}[\{\xi_{\mathfrak{i}\alpha}\}] &= \sideset{}{'}\sum_{\mathfrak{i}\alpha;\mathfrak{j}\beta} V_{\mathfrak{i}\alpha;\mathfrak{j}\beta} \xi_{\mathfrak{i}\alpha} \xi_{\mathfrak{j}\beta}. \label{eq:pair-pot}
\end{align}
Recall a prime on a summation omits the $\alpha_n$ index. We take the above pair interaction parameters to be symmetric, that is $V_{\mathfrak{i}\alpha;\mathfrak{j}\beta} = V_{\mathfrak{i}\beta;\mathfrak{j}\alpha}$ etc. Eq.~(\ref{eq:pair-pot-all}) assumes a host-invariant picture and assigns pair energy $\tilde{V}_{\mathfrak{i}\alpha;\mathfrak{j}\beta}$ between atom $\alpha$ at site $\mathfrak{i}$ and atom $\beta$ at $\mathfrak{j}$. $\tilde{V}_{\mathfrak{i}\alpha;\mathfrak{j}\beta}$ is an $n \times n$ matrix in component indices. On the other hand Eq.~(\ref{eq:pair-pot}) considers $V_{\mathfrak{i}\alpha;\mathfrak{j}\beta}$ as the energy of exciting pairs from a host medium of $\alpha_n$ atoms. In this case $V_{\mathfrak{i}\alpha;\mathfrak{j}\beta}$ is an \mbox{($n$-1)$\times$ ($n$-1)} matrix. Again, our key requirement is for Eqs.~(\ref{eq:pair-pot-all})-(\ref{eq:pair-pot}) to be valid substitutions in Eq.~(\ref{eq:omega1-final}). Therefore we demand $\delta \langle H_\text{eff} \rangle_0 = \delta \langle \tilde{H}_\text{eff} \rangle_0 = \delta \langle \Omega_\text{el.} \rangle_0$ for allowed site-concentration variations. Thus 
\begin{align}
\delta \langle H_\text{eff} \rangle_0 
&= \sideset{}{'}\sum_{\mathfrak{i}\alpha;\mathfrak{j}\beta} V_{\mathfrak{i}\alpha;\mathfrak{j}\beta}  ( \delta \bar{c}_{\mathfrak{i}\alpha} \bar{c}_{\beta} 
+  \bar{c}_{\alpha} \delta \bar{c}_{\mathfrak{j}\beta} + \delta \bar{c}_{\mathfrak{i}\alpha} \delta \bar{c}_{\mathfrak{j}\beta}) \nonumber \\
&= \sideset{}{'}\sum_{\mathfrak{i}\alpha;\mathfrak{j}\beta} V_{\mathfrak{i}\alpha;\mathfrak{j}\beta} \delta \bar{c}_{\mathfrak{i}\alpha} \delta \bar{c}_{\mathfrak{j}\beta} = \sum_{\mathfrak{i}\alpha;\mathfrak{j}\beta} \tilde{V}_{\mathfrak{i}\alpha;\mathfrak{j}\beta} \delta \bar{c}_{\mathfrak{i}\alpha} \delta \bar{c}_{\mathfrak{j}\beta} \label{eq:delta-eff-hamil-exp}
\end{align}
when expanding about the high-temperature disordered state. The first-order terms vanish due to translational invariance and
$\sum_\mathfrak{i} \delta \bar{c}_{\mathfrak{i}\alpha} = 0$ for allowed variations. Eq.~(\ref{eq:delta-eff-hamil-exp}) relates the two pair parameters by
\begin{align}
V_{\mathfrak{i}\alpha;\mathfrak{j}\beta} = \tilde{V}_{\mathfrak{i}\alpha;\mathfrak{j}\beta} + \tilde{V}_{\mathfrak{i}n;\mathfrak{j}n}
-\tilde{V}_{\mathfrak{i}n;\mathfrak{j}\beta}-\tilde{V}_{\mathfrak{i}\alpha;\mathfrak{j}n}. \label{eq:host-indep-convert}
\end{align}
The reverse transform from $V_{\mathfrak{i}\alpha;\mathfrak{j}\beta} \rightarrow \tilde{V}_{\mathfrak{i}\alpha;\mathfrak{j}\beta}$ is not unambiguously defined. In fact, we may gauge transform $\tilde{V}_{\mathfrak{i}\alpha;\mathfrak{j}\beta} \rightarrow \tilde{V}_{\mathfrak{i}\alpha;\mathfrak{j}\beta} + \phi_{\alpha} \phi_{\beta}$ for any mean-field term $\phi_{\alpha}$ without affecting the expansion in Eq.~(\ref{eq:delta-eff-hamil-exp}). We fix this gauge momentarily. By comparison to Eq.~(\ref{eq:omega1-final}) we can make the convenient identification 
\begin{align}
V_{\mathfrak{i}\alpha;\mathfrak{j}\beta} =& \,\partial^{2}\langle\Omega_{el}\rangle_{0}/\partial\bar{c}_{\mathfrak{j}\beta}\partial\bar{c}_{\mathfrak{i}\alpha}\bigg|_{\bar{c}_{\mathfrak{i}n}\text{ dependent}} =: -S_{\mathfrak{i\alpha;\mathfrak{j}\beta}}^{(2)}  \label{eq:smat-definition} \\
\tilde{V}_{\mathfrak{i}\alpha;\mathfrak{j}\beta} =& \,\partial^{2}\langle\Omega_{el}\rangle_{0}/\partial\bar{c}_{\mathfrak{j}\beta}\partial\bar{c}_{\mathfrak{i}\alpha}\bigg|_{\bar{c}_{\mathfrak{i}n}\text{ independent}} =: -\tilde{S}_{\mathfrak{i\alpha;\mathfrak{j}\beta}}^{(2)}
\label{eq:smat-inv-definition}
\end{align}
In Eq.~(\ref{eq:smat-definition}) the last concentration $\bar{c}_{\mathfrak{i}n}$ is considered dependent on the others via $\sum_{\alpha} \bar{c}_{\mathfrak{i}\alpha}=1$. In Eq.~(\ref{eq:smat-inv-definition}) this constraint is dropped and the derivative is only defined in a formal sense. The superscript ``(2)" is conventional and denotes a second derivative. We shall see in Section~$\ref{sec:band-reduction}$ that $\tilde{S}_{\mathfrak{i\alpha;\mathfrak{j}\beta}}^{(2)}$ obeys the sum rule 
\begin{align}
\sum_{\alpha} \bar{c}_{\mathfrak{i}\alpha} \tilde{S}_{\mathfrak{i\alpha;\mathfrak{j}\beta}}^{(2)}=0 \label{eq:smat-sumrule}
\end{align}
for all $\mathfrak{i};\mathfrak{j}\beta$. This permits us to fix the gauge on $\tilde{V}_{\mathfrak{i}\alpha;\mathfrak{j}\beta}$ and define a reverse map $V_{\mathfrak{i}\alpha;\mathfrak{j}\beta} \rightarrow \tilde{V}_{\mathfrak{i}\alpha;\mathfrak{j}\beta}$. This is  
\begin{align*}
\tilde{V}_{\mathfrak{i}n;\mathfrak{j}n} &= \sideset{}{'}\sum_{\alpha\beta} \bar{c}_{\mathfrak{i}\alpha}  \bar{c}_{\mathfrak{j}\alpha} V_{\mathfrak{i}\alpha;\mathfrak{j}\beta} \\
\tilde{V}_{\mathfrak{i}n;\mathfrak{j}\beta} &= \tilde{V}_{\mathfrak{i}n;\mathfrak{j}n} - \sideset{}{'}\sum_{\alpha} \bar{c}_{\mathfrak{i}\alpha} V_{\mathfrak{i}\alpha;\mathfrak{j}\beta} \\
\tilde{V}_{\mathfrak{i}\alpha;\mathfrak{j}\beta} &= V_{\mathfrak{i}\alpha;\mathfrak{j}\beta} - \tilde{V}_{\mathfrak{i}n;\mathfrak{j}n}
+ \tilde{V}_{\mathfrak{i}n;\mathfrak{j}\beta} + \tilde{V}_{\mathfrak{i}n;\mathfrak{j}\alpha}.
\end{align*}
It will be convenient to convert between host-dependent and host-invariant interaction pictures as needed. 

\section{Chemical Fluctuations\label{sec:atomic-fluctuations}}

The diffuse scattering intensity in alloy diffraction experiments
is directly proportional to a sum over second-order correlations among site occupancies.
We define short-range order 
\begin{align}
\Psi_{\mathfrak{i}\alpha;\mathfrak{j}\beta}:=\langle\xi_{\mathfrak{i}\alpha}\xi_{\mathfrak{j}\beta}\rangle-\langle\xi_{\mathfrak{i}\alpha}\rangle\langle\xi_{\mathfrak{j}\beta}\rangle. \label{eq:sro-exact}
\end{align}
From Eq.~(\ref{eq:omega-boltzmann-sum}) it is easy to see $-\partial\Omega/\partial\nu_{\mathfrak{i}\alpha}=\langle\xi_{\mathfrak{i}\alpha}\rangle=c_{\mathfrak{i}\alpha}$
and $-\partial^{2}\Omega/\partial\nu_{\mathfrak{j}\beta}\partial\nu_{\mathfrak{i}\alpha}=\partial c_{\mathfrak{i}\alpha}/\partial\nu_{\mathfrak{j}\beta}=\beta\,\Psi_{\mathfrak{i}\alpha;\mathfrak{j}\beta}$. It is also easy to see $\sum_\alpha \Psi_{\mathfrak{i}\alpha;\mathfrak{j}\beta}=0$. Therefore this is a singular $n \times n$ matrix for given $\mathfrak{i}, \mathfrak{j}$.
Now the relation between site concentrations $\{c_{\mathfrak{i}\alpha}\}$
and site chemical potentials $\{\nu_{\mathfrak{i}\alpha}\}$ is unknown. Instead we can relate
optimal variational parameters $\{\bar{c}_{\mathfrak{i}\alpha}\}$
to $\{\nu_{\mathfrak{i}\alpha}\}$ via Eq.~(\ref{eq:optimal-site-concentrations}).
This allows us to estimate $\Psi_{\mathfrak{i}\alpha;\mathfrak{j}\beta}$
via 
\begin{align}
\bar{\Psi}_{\mathfrak{i}\alpha;\mathfrak{j}\beta}:=\beta^{-1}\partial\bar{c}_{\mathfrak{i}\alpha}/\partial\nu_{\mathfrak{j}\beta}.
\label{eq:sro-variational}
\end{align}
The bar notation is a reminder that this is an approximation. Because $\sum_\alpha \bar{c}_{\mathfrak{i}\alpha}=1$, it also satisfies $\sum_{\alpha} \bar{\Psi}_{\mathfrak{i}\alpha;\mathfrak{j}\beta} = 0$. Nevertheless, $\bar{\Psi}_{\mathfrak{i}\alpha;\mathfrak{j}\beta}$ is not guaranteed to satisfy all the sum rules $\Psi_{\mathfrak{i}\alpha;\mathfrak{j}\beta}$ does. For instance the site-diagonal piece $\Psi_{\mathfrak{i}\alpha;\mathfrak{\mathfrak{i}}\beta}=\langle\xi_{\mathfrak{i}\alpha}\delta_{\alpha\beta}\rangle-\langle\xi_{\mathfrak{i}\alpha}\rangle\langle\xi_{\mathfrak{i}\alpha}\rangle=c_{\mathfrak{i}\alpha}\delta_{\alpha\beta}-c_{\mathfrak{i}\alpha}c_{\mathfrak{i}\beta}$.
This need not be true for $\bar{\Psi}_{\mathfrak{i}\alpha;\mathfrak{j}\beta}$.
We discuss how to restore this site-diagonal sum rule in Section~\ref{sec:onsager}. By differentiating Eq.~(\ref{eq:optimal-site-concentrations})
with respect to $c_{\mathfrak{j}\beta}$ while holding remaining $\{c_{\mathfrak{i}1},\ldots c_{\mathfrak{i}(n-1)}\}$ fixed, we find
\begin{align}
0= & \,\beta^{-1}\delta_{\mathfrak{i}\mathfrak{j}}(\frac{\delta_{\alpha\beta}}{\bar{c}_{\mathfrak{i}\alpha}}+\frac{1}{\bar{c}_{\mathfrak{i}n}})-\frac{\partial\nu_{\mathfrak{i}\alpha}}{\partial\bar{c}_{\mathfrak{j}\beta}}+\frac{\partial^{2}\langle\Omega_{el}\rangle_{0}}{\partial\bar{c}_{\mathfrak{j}\beta}\partial\bar{c}_{\mathfrak{i}\alpha}}\nonumber \\
0= & \,\beta^{-1}(C{}_{\alpha\beta}^{-1}\delta_{\mathfrak{i}\mathfrak{j}}-\bar{\Psi}_{\mathfrak{i}\alpha;\mathfrak{j}\beta}^{-1})-S_{\mathfrak{i}\alpha;\mathfrak{j}\beta}^{(2)}.\label{eq:short-range-order}
\end{align}
where we used the definition in Eq.~(\ref{eq:smat-definition}) and also define 
\begin{align}
C_{\alpha\beta} :=& \,\bar{c}_{\alpha}(\delta_{\alpha\beta}-\bar{c}_{\beta}) \label{eq:cmat-definition} \\
C_{\alpha\beta}^{-1} :=& \,(\delta_{\alpha\beta}/\bar{c}_{\alpha}+1/\bar{c}_{n}) \nonumber \\
\tilde{C}_{\alpha\beta}^{-1} :=& \,\delta_{\alpha\beta}/\bar{c}_\alpha. \nonumber 
\end{align}
The host terms arise because $c_{\mathfrak{i}n} = c_{\mathfrak{i}n}[c_{\mathfrak{i}1},\ldots,c_{\mathfrak{i}(n-1)}]$ is a function of the other $n-1$ on-site concentrations. Note that $\bar{\Psi}_{\mathfrak{i}\alpha;\mathfrak{j}\beta}^{-1}$ is defined as the inverse of the upper-left \mbox{($n$-1)$\times$($n$-1)} block (in component indices) of $\bar{\Psi}_{\mathfrak{i}\alpha;\mathfrak{j}\beta}$. Eq.~(\ref{eq:short-range-order}) relates the approximate short-range order $\bar{\Psi}$ to electronics
of the CPA medium through matrix  $S_{\mathfrak{i}\alpha;\mathfrak{j}\beta}^{(2)}$. This
relationship is formally similar to the short-range order expression
derived in a Gorsky-Bragg-Williams\cite{Bragg699} model
with pair interactions $V_{\mathfrak{i}\alpha;\mathfrak{j}\beta}$
substituted by $S_{\mathfrak{i}\alpha;\mathfrak{j}\beta}^{(2)}$.
Again, we see it is possible to interpret $S_{\mathfrak{i}\alpha;\mathfrak{j}\beta}^{(2)}$
as an effective pairwise interaction.

If we use Eq.~(\ref{eq:optimal-site-concentrations}) we can set
$\Omega^{(1)}[\{\nu_{\mathfrak{i}\alpha}\},\{\bar{c}_{\mathfrak{i}\alpha}\}]\rightarrow\Omega^{(1)}[\{\bar{c}_{\mathfrak{i}\alpha}\}]$
as a function of site concentrations $\{\bar{c}_{\mathfrak{i}\alpha}\}$
only. Performing a second order expansion then gives 
\begin{multline}
\delta\Omega^{(1)}= \\ \frac{1}{2}\sideset{}{'}\sum_{\text{\ensuremath{\mathfrak{i}\alpha};\text{\ensuremath{\mathfrak{j}}}\ensuremath{\beta}}}\delta\bar{c}_{\text{\ensuremath{\mathfrak{i}\alpha}}}\left[\beta^{-1}\bar{\Psi}_{\mathfrak{i}\alpha;\mathfrak{j}\beta}^{-1}-\sideset{}{'}\sum_{\mathfrak{k}\gamma}\bar{c}_{\mathfrak{k}\gamma}\frac{\partial^{2}\nu_{\mathfrak{k}\gamma}}{\partial\bar{c}_{\mathfrak{j}\beta}\partial\bar{c}_{\mathfrak{i}\alpha}}\right]\delta\bar{c}_{\mathfrak{j}\beta}+\cdots. \label{eq:omega1-initial-expansion}
\end{multline}
This expression gives the change in grand potential by indirectly
varying the physical system through a variation of the corresponding,
pinned uncorrelated reference medium (c.f. Eq.~(\ref{eq:optimal-site-concentrations})). The second term in brackets
accounts for changing chemical potentials $\{\nu_{\mathfrak{i}\alpha}\}$
as $\{\bar{c}_{\mathfrak{i}\alpha}\}$ varies. This term would be absent if we instead held $\{\nu_{\mathfrak{i}\alpha}\}$
fixed and independent of $\{\bar{c}_{\mathfrak{i}\alpha}\}$. By independently setting $\{ \bar{c}_{\mathfrak{i}\alpha} \}$ and allowing $\nu_{\mathfrak{i}\alpha} = \nu_{\mathfrak{i}\alpha}[ \{\bar{c}_{\mathfrak{i}\alpha}\} ]$ to vary, we are in effect working in the canonical ensemble. The canonical ensemble fixes $\{ c_{\mathfrak{i}\alpha} \}$ and allows fluctuations in $\{ \nu_{\mathfrak{i}\alpha} \}$. The reverse is true in the grand canonical ensemble. In the thermodynamic limit these fluctuations are assumed not to play an important role. \emph{Based on these expectations we ignore the fluctuations in $\partial \nu_{\mathfrak{k}\gamma} / \partial \bar{c}_{\mathfrak{i}\alpha} \partial \bar{c}_{\mathfrak{j}\beta}$ as insignificant to the relevant physics.} Thus, we drop the second term in Eq.~(\ref{eq:omega1-initial-expansion}) and identify $\delta \Omega^{(1)} = \delta F^{(1)}$. In that case we find the
physical system is unstable to infinitesimal fluctuations when $\bar{\Psi}_{\mathfrak{i}\alpha;\mathfrak{j}\beta}^{-1}$
is no longer positive definite. If we Fourier transform we have to second order
\begin{align}
\delta F^{(1)}  \mkern-50mu &\nonumber \\ 
&=\frac{1}{2}\sum_{k}\sideset{}{'}\sum_{\text{a}\text{\ensuremath{\alpha};\text{b}\ensuremath{\beta}}}\delta\bar{c}_{\text{\ensuremath{\mathfrak{\text{a}}\alpha}}}(k)^{*}\left[\beta^{-1}\bar{\Psi}_{\text{a}\alpha;\text{b}\beta}^{-1}(k)\right]\delta\bar{c}_{\text{b}\beta}(k) \nonumber \\
&= \frac{1}{2}\sum_{k}\sideset{}{'}\sum_{\text{a}\text{\ensuremath{\alpha};\text{b}\ensuremath{\beta}}}\delta\bar{c}_{\text{\ensuremath{\mathfrak{\text{a}}\alpha}}}(k)^{*}\left[(\beta C){}_{\alpha\beta}^{-1}\delta_\text{ab} -S_{\mathfrak{\text{a}}\alpha;\text{b}\beta}^{(2)}(k) \right]\delta\bar{c}_{\text{b}\beta}(k) \nonumber \\
&= \frac{1}{2}\sum_{k}\sum_{\text{a}\text{\ensuremath{\alpha};\text{b}\ensuremath{\beta}}}\delta\bar{c}_{\text{\ensuremath{\mathfrak{\text{a}}\alpha}}}(k)^{*}\left[(\beta \tilde{C}){}_{\alpha\beta}^{-1}\delta_\text{ab} -\tilde{S}_{\mathfrak{\text{a}}\alpha;\text{b}\beta}^{(2)}(k) \right]\delta\bar{c}_{\text{b}\beta}(k). \label{eq:free-energy-final-result}
\end{align}
Note that the third line implies a sum over all components and uses only host-invariant parameters. Similar to Eq.~(\ref{eq:omega1-final}) the variational free energy is 
\begin{align}
F^{(1)} = -TS + U = \beta^{-1} \sum_{\mathfrak{i}\alpha} \bar{c}_{\mathfrak{i}\alpha} \log \bar{c}_{\mathfrak{i}\alpha} + \langle \Omega_\text{el.} \rangle_0. \label{eq:variational-helmholtz}
\end{align} 
From Eq.~(\ref{eq:variational-helmholtz}) we identify $(\beta \tilde{C})^{-1}_{\alpha\beta} = \delta_{\alpha\beta}/\bar{c}_\alpha$ as the \emph{entropy cost} of a variation. Similarly, we identify $\tilde{S}_{\mathfrak{\text{a}}\alpha;\text{b}\beta}^{(2)}(k)$ as the \emph{energy cost of pair creation}. Further, Eq.~(\ref{eq:free-energy-final-result}) implies \emph{the cost of a fluctuation along mode $k$ is inversely proportional to short-range order parameter $\bar{\Psi}_{\text{a}\alpha;\text{b}\beta}(k)$}. This is intuitively satisfying as the short-range order parameter is a measure of the tendency of atoms to cluster. Lastly, we infer an absolute instability point at mode $k_{0}$ when matrix $\beta^{-1}\bar{\Psi}_{\text{a}\alpha;\text{b}\beta}^{-1}(k_0)$ has its lowest eigenvalue pass through zero. We discuss the interpretation of these eigenvalues and eigenvectors for the multicomponent case in the next section.

\section{Chemical Polarizations\label{sec:chemical-pol}}

\begin{figure}
\setbox1=\hbox{\includegraphics[height=9cm]{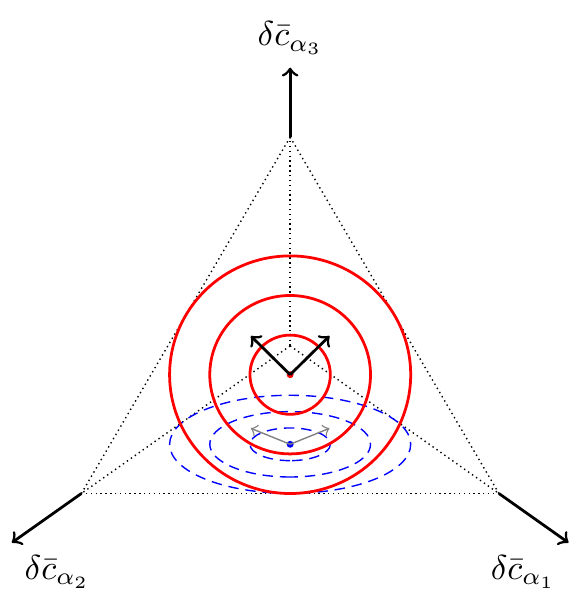}}
\includegraphics[height=9cm]{eigen_projection1.pdf}\llap{\makebox[\wd1][l]{\raisebox{3.5cm}{\includegraphics[height=3.5cm]{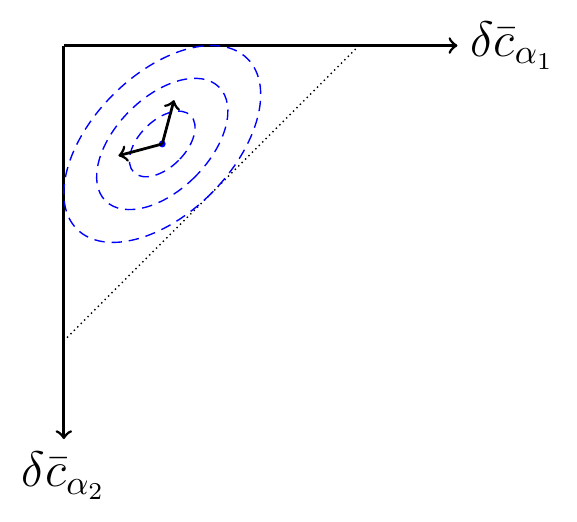}}}}
\caption{(Color online) Variations in concentration space of a three-component alloy with hypothetical $\delta F^{(1)}$ contours (solid red) centered about the origin $\delta \bar{c} = (0,0,0)$ (red dot).   The contours have a contrived symmetry so that every vector within the subspace $\delta \bar{c}_{\alpha_1} + \delta \bar{c}_{\alpha_2} + \delta \bar{c}_{\alpha_3}=0$ is an eigenvector. One orthogonal pair of eigenvectors is shown (centered arrows). The projection of contours and this pair of eigenvectors onto the subspace spanned by $(\delta \bar{c}_{\alpha_1},\delta \bar{c}_{\alpha_2},0)$ is shown below (dashed blue). The inset shows a top-down view of projected contours and eigenvectors. Key here is that \emph{projected eigenvectors do not align with projected contours.} Therefore care must be taken to define eigenvectors in an unambiguous and consistent manner.}
\label{fig:eigen-projection}
\end{figure}

The set of variables $(\delta \bar{c}_{\text{a}1}(k),\ldots, \delta \bar{c}_{\text{a}n}(k))$ for given wave-vector $k$ and unit cell basis position `a' form coordinates in a \emph{concentration space}. For simplicity we consider a monatomic basis and drop latin index `a'. The origin \mbox{$\delta \bar{c}_\alpha (k) = 0$} corresponds to the fully disordered high-temperature state. From this reference state we only allows coordinate moves that preserve \mbox{$\sum_\alpha \delta \bar{c}_{\alpha}(k) = 0$}. This confines us to a subspace that preserves the total component concentrations. Throughout this article we have frequently chosen to work with the $n$-1 independent variables $\{\delta \bar{c}_{1}(k), \ldots, \delta \bar{c}_{(n-1)}(k)\}$. In this framework, $(\beta C)^{-1}_{\alpha\beta}$ and $S^{(2)}_{\alpha\beta}(k)$ of Eq.~(\ref{eq:free-energy-final-result}) are \mbox{($n$-1)$\times$($n$-1)} matrices with respect to component indices. One difficulty with this point of view is that diagonalizing such quantities in the subspace $(\delta \bar{c}_{1}(k), \ldots, \delta \bar{c}_{(n-1)}(k))$ assumes the metric $||\delta \bar{c} ||^2 = \sum_\alpha' \delta \bar{c}_{\alpha}(k)^2$ (n.b. prime). This is a host-dependent metric and leads to eigenvalues and eigenvectors that are only meaningful in this frame of reference (c.f. Fig. \ref{fig:eigen-projection}). On the other hand, the most canonical metric over concentration space is $||\delta \bar{c}||^2 =  \sum_\alpha \delta \bar{c}_{\alpha}(k)^2$ (\emph{no} prime) as it is host-invariant and a good gauge of the total size of a fluctuation. (As a point of contrast we note that Singh et al.\cite{PhysRevB.91.224204} choose a host-invariant metric by considering $\{\delta \bar{c}_1(k),\ldots,\delta \bar{c}_n(k)\}$ to be the $n$~barycentric coordinates of an ($n$-1)-simplex embedded in a ($n$-1)-dimensional Cartesian space. This is motivated by a preference to work in the coordinate space of the Gibbs triangle or its higher dimensional variants.) Thus our scheme is to diagonalize the $n\times n$ matrices $(\beta \tilde{C})^{-1}_{\alpha\beta} $ and $\tilde{S}^{(2)}_{\alpha\beta}(k)$ (n.b. tilde) over the complete space $(\delta \bar{c}_{1}(k),\ldots, \delta \bar{c}_{n}(k))$. This uses host-invariant coefficients and metric. However it permits eigenvectors that do \emph{not} preserve \mbox{$\sum_\alpha \delta \bar{c}_{\alpha}(k)=0$} because of the unconstrained diagonalization. To constrain the diagonalization we first perform a \emph{norm-conserving} change of variables to isolate the non-physical degree of freedom. Thus we define a new set of variables $\delta \eta_{\alpha}(k) = \sum_\beta O_{\alpha\beta} \delta \bar{c}_{\beta}(k)$ where
\begin{align*}
O = \left[ \begin{tabular}{rrrrrrrc} 
1/$\sqrt{2}$ &(1 & -1 & 0 & 0 & 0 & $\cdots$ & 0) \\
1/$\sqrt{6}$ & (1 & 1 & -2 & 0 & 0 & $\cdots$ & 0) \\
1/$\sqrt{12}$ & (1 & 1 & 1 & -3 & 0 & $\cdots$ & 0) \\
\rule{0pt}{4ex} 
 & \vdots& & \vdots&   &\vdots & & \\
 \rule{0pt}{4ex} 
1/$\sqrt{n(n\!-\! 1)}$ & (1 & 1 & 1 & 1 & 1 & $\cdots$ & $1\!-\! n$) \\
1/$\sqrt{n}$ & (1 & 1 & 1 & 1 & 1 & $\cdots$ & 1)
\end{tabular} \right]
\end{align*}
is an orthogonal transform (i.e. $O O^T = 1$). It is easy to see by inspection that the rows of $O$ form an orthonormal set. The last row isolates the frozen degree of freedom $\delta\eta_{n}(k) = \sum_\alpha \delta \bar{c}_{\alpha}(k)/\sqrt{n}=0$. In this new system of variables 
\begin{align*}
\tilde{S}^{(2)}_{\alpha\beta}(k) \rightarrow & \, \sum_{\mu\nu} O_{\alpha\mu} \tilde{S}^{(2)}_{\mu\nu}(k) O^T_{\nu\beta} =: \breve{S}^{(2)}_{\alpha\beta}(k) \\
(\beta \tilde{C})^{-1}_{\alpha\beta} \rightarrow & \,  \sum_{\mu\nu} O_{\alpha\mu} (\beta \tilde{C})^{-1}_{\mu\nu} O^T_{\nu\beta} =: (\beta \breve{C})^{-1}_{\alpha\beta}.
\end{align*}
In terms of which the free-energy of Eq.~(\ref{eq:free-energy-final-result}) is
\begin{multline}
\delta F^{(1)} = \\ \frac{1}{2}\sum_{k}\sum_{\alpha\beta} \delta\eta_{\alpha}(k)^{*} \left[(\beta \breve{C}){}_{\alpha\beta}^{-1} -\breve{S}_{\alpha\beta}^{(2)}(k) \right]\delta\eta_{\beta}(k). \label{eq:free-energy-eta-variables}
\end{multline} 
To restrict the diagonalization to the relevant subspace we replace $\breve{S}^{(2)}_{\alpha n}(k) = \breve{S}^{(2)}_{n \alpha}(k) =  (\beta \breve{C})^{-1}_{n\alpha} = (\beta \breve{C})^{-1}_{\alpha n} = 0$ for all $\alpha$. These matrix coefficients are irrelevant since $\delta\eta_{n}(k) = 0$ always. On finding the eigenvectors and eigenvalues in the $\delta\eta$ variables we may always transform back using $\delta \bar{c}_{\alpha}(k) = \sum_\beta O^T_{\alpha\beta} \delta \eta_{\beta}(k)$. There are $n$-1 eigenvectors but each eigenvector has $n$ components on including $\delta \eta_n(k) = 0$. We may then write
\begin{align}
\delta F^{(1)} = \frac{1}{2} \sum_{ks} \lambda_{s}(k) \langle \delta e_s(k) | \delta \bar{c}(k)  \rangle ^2 \label{eq:free-energy-diagonal}
\end{align}
for eigenvalues $\lambda_s(k)$, eigenvectors $\delta e_s(k)$, and concentration space inner product $\langle v | w \rangle := \sum_\alpha v_\alpha^* w_\alpha$. \emph{Eigenvalue $\lambda_s(k)$ is the energy cost for a concentration wave $\delta e_s(k)$ with magnitude $\sum_\alpha \delta e_{s;\alpha}(k)^2 = \langle \delta e_{s}(k) | \delta e_s(k) \rangle = 1$}. All eigenvectors satisfy $\sum_\alpha \delta e_{s;\alpha}(k) = 0$. This reflects the sum of concentrations being preserved for each mode. Finally, eigenvectors are ``orthogonal" to each other, i.e. $ \langle \delta e_s(k) | \delta e_t(k) \rangle = \delta_{st}$. At high temperatures $(\beta \breve{C})^{-1}_{\alpha\beta}$ dominants Eq.~(\ref{eq:free-energy-eta-variables}). In this case eigenvectors point in directions of maximum entropy increase and the electronics of the alloy are not relevant. At low temperatures $\breve{S}^{(2)}_{\alpha\beta}$ dominates. In this case eigenvectors point in directions of favorable atomic ordering as based on the electronics.


\section{Linear Response\label{sec:linear-response}}

One way to compute approximate atomic correlations $\bar{\Psi}$ is by working out
the linear response and then computing the ratio $\bar{\Psi}_{\mathfrak{i}\alpha;\mathfrak{j}\beta}=\beta^{-1} \delta\bar{c}_{\mathfrak{i}\alpha}/\delta\nu_{\mathfrak{j}\beta}$.
Therefore in this section we seek to determine the linear response
of the homogenous CPA medium on applying infinitesimal variations
$\{\delta\nu_{\mathfrak{i}\alpha}\}$. We also find $\mathcal{\mathbb{S}}^{(2)}$
as byproduct of this procedure via Eq.~(\ref{eq:short-range-order}). 

Before proceeding we note the CPA solution is self-consistently constructed
out of many interconnected quantities; including site chemical potentials
$\{\mathfrak{\nu_{\mathfrak{i}\alpha}}\}$, site concentrations $\{\bar{c}_{\mathfrak{i}\alpha}\}$,
site charge densities $\{\rho_{\mathfrak{i}\alpha}\}$, site potentials
$\{V_{\mathfrak{i}\alpha}\}$, site scattering matrices $\{t_{\mathfrak{i}\alpha}\}$,
site CPA scattering matrices $\{t_{\mathfrak{i}c}\}$, and CPA scattering path
operator $\tau_{c}$. All these quantities are ultimately determined
by external site chemical potentials $\{\nu_{\mathfrak{i}\alpha}\}$.
However it is simpler to find only the variational relationship between those quantities that are directly coupled. This leads to a ring of coupled equations that together determine the total variation of the CPA medium. This staged approach also helps to organize and interpret the mathematics. 

The key variations needed are Eq. (\ref{eq:CPA-condition}) to establish
variation of CPA medium $t_{\mathfrak{i}c}$; Eq. (\ref{eq:cpa-potential})
for variation of site potential $V_{\mathfrak{i}\alpha}$; Eq. (\ref{eq:charge-from-green})
for variation of charge density $\rho_{\mathfrak{i\alpha}}$; and Eq. (\ref{eq:cpa-grand-potential})
for variation of the electronic grand potential. The
variation of each of these requires a concerted effort and is relegated
to appendices. Here we define needed quantities and give the final coupled equations. 

There are a few simplifications made in the course of solving the mathematics. First, we only consider a Bravais lattice without basis. Therefore $\mathfrak{i} \rightarrow i$ in what follows. Indeed, many high-entropy alloys are either of the FCC or BCC type. Second, the charge-density response $\delta\rho_{\mathfrak{i}\alpha}(r)$ is expanded in terms of an orthonormal basis $f_{n}(r)$ of functions. These satisfy $\int f_{n}(r)dr=\delta_{n1}$ and $f_1(r)=1$. Legendre polynomials may be used to fit this requirement. This basis expansion reduces the related degrees of freedom from the number of points along a grid ($\sim1000$) to a small number of basis coefficients ($\sim5$). It also discretizes the charge density associated volume integrals. Any superscripts $n,m$ will refer to indices in this basis. Context will distinguish these indices from number of components $n$. Explicitly, we define the charge response 
\begin{align}
\Phi_{i\alpha;j\beta}^{n}=\frac{\delta}{\delta\nu_{j\beta}}\int dr\rho_{i\alpha}(r)f_{n}(r)dr. \label{eq:charge-response-definition}
\end{align}
Take care to note this is a rectangular matrix of dimensions $n_c \times (n_c-1)$ for an $n_c$ component alloy. Third, we make the approximation $|r+R_i-R_j-r'|\rightarrow |R_i-R_j|$ when appropriate (c.f. Eq.~(\ref{eq:cpa-grand-potential})). This is equivalent to keeping only leading monopole terms for Coulomb interactions between pairs of cells. It permits us to work in terms of site charges
\begin{align}
Q_{i\alpha}:=\int_{V_{0}}dr[\rho_{i\alpha}(r)-Z_{\alpha}\delta(r)], \label{eq:site-charge-definition}
\end{align}
polarization $P_i := \sum_\alpha \bar{c}_{i\alpha} Q_{i\alpha}$, and Fourier transform 
\begin{align}
M(k):=\sum_{i\neq0}\frac{e^{2}}{R_{i}}e^{-k\cdot R_{i}} \label{eq:M-definition}
\end{align}
of the lattice electrostatic pair interaction.

We now define site relevant quantities using Fermi-Dirac function $f(\epsilon-\mu)$, site impurity Green
function $G_{\alpha}$, site regular solution $Z_{\alpha;L}(r)$, and basis functions $f_n(r)$:
\begin{widetext}
\begin{align}
A_{\alpha}^{mn}:=&-\frac{1}{\pi}\int d\epsilon\,f(\epsilon-\mu)\text{ Im }\int_{V_{0}}dr\int_{V_{0}}dr'f_{m}(r)G_{\alpha}(\epsilon;r,r') \, \times \nonumber \\
& \mkern300mu \left[\frac{dV_{\text{xc}}}{d\rho}(\rho_{\alpha}(r'))f_{n}(r')+\left(\int_{V_{0}}dr''\frac{e^{2}}{|r'-r''|}f_{n}(r'')\right)\right]G_{\alpha}(\epsilon;r',r)\label{eq:A-definition} \\
B_{\alpha}^{m}:=&-\frac{1}{\pi}\int d\epsilon\,f(\epsilon-\mu)\text{ Im }\int_{V_{0}}dr\int_{V_{0}}dr'f_{m}(r)G_{\alpha}(\epsilon;r,r')G_{\alpha}(\epsilon;r',r) \label{eq:B-definition} \\
F_{\alpha;LL'}^{n}:=&-\int_{V_{0}}drZ_{\alpha;L}(r)Z_{\alpha;L'}(r)f_{n}(r) \label{eq:F-definition} \\
U_{\alpha;LL'}^{n}:=&-\int_{V_{0}}drZ_{\alpha;L}(r)Z_{\alpha;L'}(r)\left\{ \frac{dV_{\text{xc}}}{d\rho}(\rho_{\alpha}(r))f_{n}(r)+\left(\int_{V_{0}}dr'\frac{e^{2}}{|r-r'|}f_{n}(r')\right)\right\}. \label{eq:U-definition}
\end{align}
\end{widetext}
The lack of site indices $i$ follows from the equivalence of all sites in the homogenous reference. 

In addition, it turns out that it simplifies the expressions
to work with enlarged supermatrices with row (column) indices
given by composite index $(L_{1},L_{2})$ for $L_{1},L_{2}$ independent
angular momentum indices. Understanding this we can define CPA related
supermatrices
\begin{align}
\mathbb{\bar{D}}_{\alpha;L_{1}L_{2};L_{3}L_{4}}:= & \,\bar{D}_{\alpha;L_{1}L_{3}}D_{\alpha;L_{4}L_{2}} \label{eq:superD-definition}
\\ \mathbb{\mathcal{\mathbb{D}}}_{\alpha;L_{1}L_{2};L_{3}L_{4}}:= & \,D_{\alpha;L_{1}L_{3}}\bar{D}_{\alpha;L_{4}L_{2}} \label{eq:superDbar-definition}
\\[5pt] \mathbb{X}_{L_{1}L_{2};L_{3}L_{4}}:= & \sum\limits _{\alpha}\bar{c}_{\alpha}X_{\alpha;L_{1}L_{3}}X_{\alpha;L_{4}L_{2}}\label{eq:superX-definition}
\\[2pt] \mathbb{C}_{L_{1}L_{2};L_{3}L_{4}}(k):= & \frac{1}{V_{\text{BZ}}}\int dq\:\Delta\tau^c(q){}_{L_{1}L_{3}}\Delta\tau^c(q-k){}_{L_{4}L_{2}} \label{eq:superC-definition}
\end{align}
where the matrix $\Delta\tau^c(q)=\tau^c(q)-\tau^c_{00}$ and $\tau^c(q)$
is the Fourier transform of the CPA SPO $\tau^c_{ij}$. These may be thought of as linear operators on the vector space $L\times\nobreak L$.
They are used to describe the response of $t_\mu$ and $t_c$ matrices.
The computation of $\mathcal{\mathbb{C}}(k)$ is expensive as it requires
a convolution integral of the SPO $\tau^c(k)$ over the Brillouin zone.

We can now state a set of coupled equations for Fourier transformed short-range order parameter $\bar{\Psi}_{\alpha\beta}(k)$:

\begin{align}
\frac{\delta P}{\delta\nu_{\gamma}}(k):= & \sum_{\sigma}(Q_{\sigma}-Q_{n}) \beta \bar{\Psi}_{\sigma\gamma}(k)+\sum_{\sigma}\bar{c}_{\sigma}\Phi_{\sigma\gamma}^{1}(k)\label{eq:master-pol-variation}
\end{align}
\begin{align}
\frac{\delta t_{\mu}^{-1}}{\delta\nu_{\gamma}}(k)= & \sum_{n}U_{\mu}^{n}\Phi_{\mu\gamma}^{n}(k)+F_{\mu}^{1}M(k)\frac{\delta P}{\delta\nu_{\gamma}}(k)\label{eq:master-pot-variation}
\end{align}
\begin{widetext}
\begin{align}
\left[\mathbb{I}-\mathbb{X}\mathbb{C}(k)\right]\frac{\delta t_{c}^{-1}}{\delta\nu_{\gamma}}(k)= \stackrel[\mu]{}{\sum}(X_{\mu}-X_{n})\beta \bar{\Psi}_{\mu\gamma}(k)+\sum\limits _{\mu}\bar{c}_{\mu}\mathbb{\bar{D}_{\mu}}\frac{\delta t_{\mu}^{-1}}{\delta\nu_{\gamma}}(k)\label{eq:master-cpa-variation}
\end{align}
\begin{align}
\Phi_{\mu\gamma}^{m}(k)= & \sum_{n}A_{\mu}^{mn}\Phi_{\mu\gamma}^{n}(k)+B_{\mu}^{m}M(k)\frac{\delta P}{\delta\nu_{\gamma}}(k)-\frac{1}{\pi}\int d\epsilon\,f(\epsilon-\mu)\text{ Im}\sum_{LL'}F_{\mu;LL'}^{m}(\mathcal{\mathbb{D}}_{\mu}\mathbb{C}(k)\frac{\delta t_{c}^{-1}}{\delta\nu_{\gamma}}(k))_{LL'}\label{eq:master-charge-variation}
\end{align}
\begin{align}
\beta \bar{\Psi}_{\mu\gamma}(k)= & \beta C_{\mu\gamma}+\sum_{\sigma}\beta C_{\mu\sigma}\left\{ \frac{1}{\pi}\int d\epsilon f(\epsilon-\mu)\text{Im}\text{Tr}\left[\,(X_{\sigma}-X_{n})\mathbb{C}(k)\frac{\delta t_{c}^{-1}}{\delta\nu_{\gamma}}(k)\right]-(Q_{\sigma}-Q_{n})M(k)\frac{\delta P}{\delta\nu_{\gamma}}(k)\right\}. \label{eq:master-conc-variation}
\end{align}
\end{widetext}

Here $P_i := \sum_\mu \bar{c}_{i\mu} Q_{i\mu}$
describes how charge rearrangements polarize the inhomogenous medium. Charge neutrality requires
$\sum_i P_i = 0$. Eq.~(\ref{eq:master-pol-variation}) simply describes the changing polarization in terms
of changing site charges and concentrations. Eq.~(\ref{eq:master-pot-variation}) describes how 
 $t_{i\mu}$ varies as charge rearrangements influence the on-site potential $V_{i\mu}$ via Eq.~(\ref{eq:cpa-potential}). The variation of Eq.~(\ref{eq:cpa-potential}) gives rise to Eqs.~(\ref{eq:master-pol-variation})-(\ref{eq:master-pot-variation}) and is derived in detail in  Appendix~\ref{sub:Potential-Variation}. The coherent medium response $\delta t_{ic}^{-1}$ is determined in terms
of changing site-scattering matrices $\delta t_{i\mu}$ and their occupancies $\delta \bar{c}_{i\mu}$. Both of these feed into Eq.~(\ref{eq:master-cpa-variation}) and arise from a variation to Eq.~(\ref{eq:CPA-condition}). It is derived
in Appendix~\ref{sub:CPA-Variation}. Eq.~(\ref{eq:master-charge-variation}) encodes the charge response $\delta \rho_{i\mu}$. The first two terms give the response from a direct variation of on-site $V_{i\mu}$. The remaining term gives the response due to the off-site, average CPA medium. Eq.~(\ref{eq:master-charge-variation}) arises from a variation to Eq.~(\ref{eq:charge-from-green}) and is
derived in Appendix~\ref{sub:Charge-Variation}. Lastly, Eq.~(\ref{eq:master-conc-variation})
relates the atomic correlations to the changing energetics. The first term in braces gives the band-energy contribution and the second the Madelung energy. Eq.~(\ref{eq:master-conc-variation}) arises from a variation of Eq.~(\ref{eq:optimal-site-concentrations}) and is derived in Appendix~\ref{sub:Concentration-Variation}. Note
that the above equations do not couple different $k$ vectors. This
enables for different $k$ values to be solved simultaneously.

\section{Onsager Reaction Field\label{sec:onsager}}

\begin{figure}
\begin{overpic}[scale=0.45]{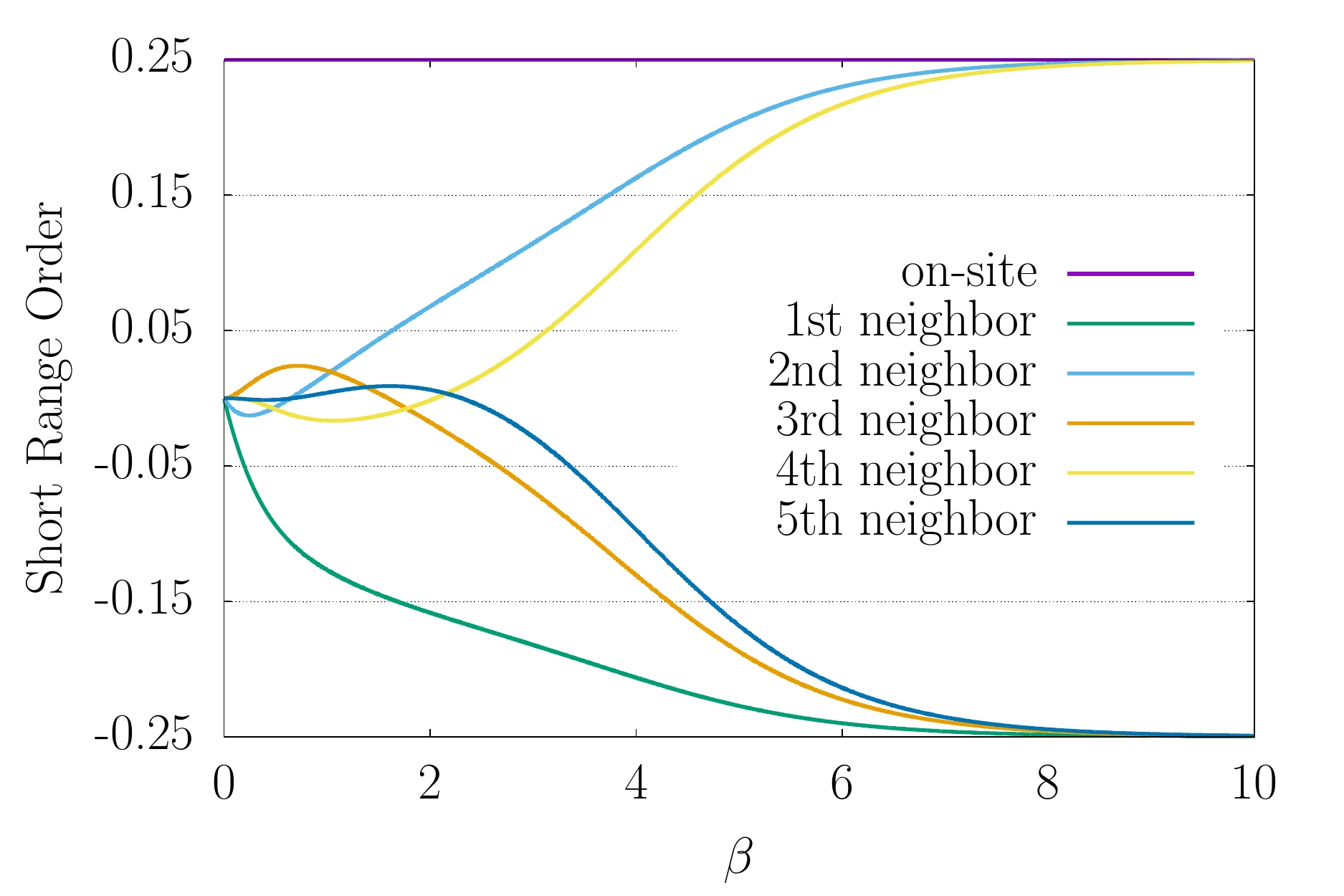}
\put(0,60) {$(a)$}
\end{overpic}
\begin{overpic}[scale=0.45]{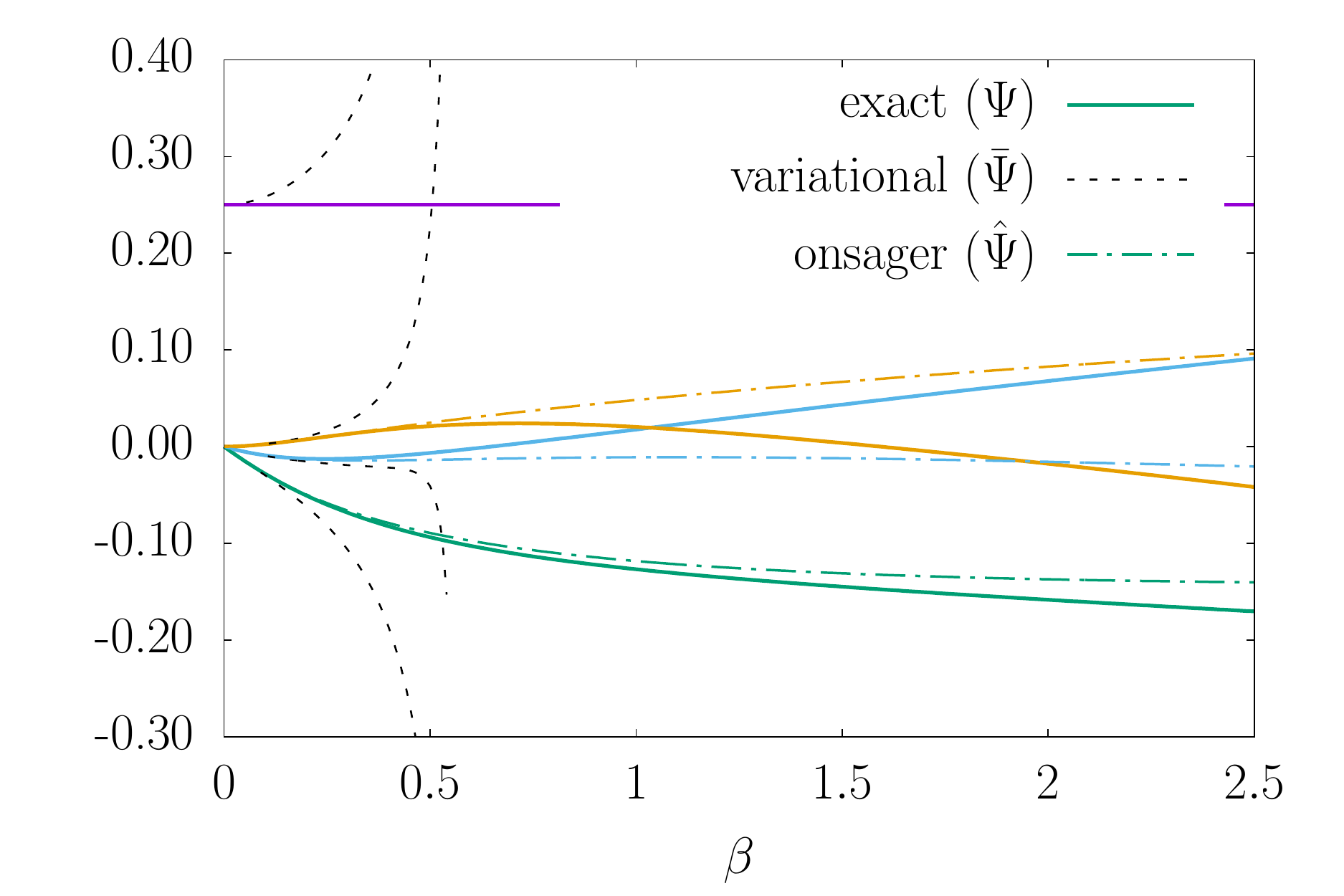}
\put(0,60) {$(b)$}
\end{overpic}
\caption{(Color online) Short range order parameters for the same toy model as described in the caption of Fig.~\ref{fig:toy-model-potentials}. One-dimensional short-ranged models are a especially difficult case for mean-field theories. Nevertheless this toy model serve as a useful illustration of the utility of an Onsager reaction-field correction. (a)~The exact short range order of Eq. (\ref{eq:sro-exact}). From top to bottom (at large $\beta$) curves correspond to on-site, 2nd, 4th, 5th, 3rd, and 1st neighbor respectively. Note that the small $\beta$ ordering does not necessarily reflect the large $\beta$ ordering. At large $\beta$ the system establishes a -A-B-A-B-A-B-A-B-A-B-A-B- pattern. (b) The variational short range order $\bar{\Psi}$ of Eq.~(\ref{eq:sro-variational}) and Onsager corrected short range order $\hat{\Psi}$ of Section~\ref{sec:onsager} in the small $\beta$ limit. Note that inclusion of an Onsager correction suppresses a divergence in $\bar{\Psi}$ and shows a striking increase in range of validity.}
\label{fig:toy-model-sro}
\end{figure}

As mentioned in Section~\ref{sec:atomic-fluctuations}, the true short-range order parameters obey site-diagonal sum rule
$\Psi_{\mathfrak{i}\alpha;\mathfrak{i}\beta} = C_{\alpha\beta}$. In addition to this there is a sum rule obeyed by the exact charge response defined in Eq.~(\ref{eq:charge-response-definition}): $\Phi^{(m)}_{\mathfrak{i}\alpha;\mathfrak{i}\beta}$ = 0. This result follows from fully considering
\begin{multline}
\frac{\partial \rho_{\mathfrak{i}\alpha}(r) }{\partial \nu_{\mathfrak{i}\beta}} = \frac{\partial}{\partial \nu_{\mathfrak{i}\beta}} \frac{ \langle \rho_\mathfrak{i}(r) \xi_{\mathfrak{i}\alpha} \rangle }{ \langle \xi_{\mathfrak{i}\alpha} \rangle }
 \\ =  \frac{\partial}{\partial \nu_{\mathfrak{i}\beta}}  \frac{1}{\langle \xi_{\mathfrak{i}\alpha} \rangle } \sum_{ \{\xi_{\mathfrak{j}\mu } \} } \rho_\mathfrak{i}(r) \xi_{\mathfrak{i}\alpha} e^{-\beta( \Omega_{el} - \sum_{\mathfrak{j}\mu} \nu_{\mathfrak{j}\mu} \xi_{\mathfrak{j}\mu} )}.
\end{multline}
On the other hand the charge response $\Phi^{(m)}_{\mathfrak{i}\alpha;\mathfrak{j}\beta}$ inherent in Eqs.~(\ref{eq:master-pol-variation})-(\ref{eq:master-conc-variation}) need not obey this sum rule because of approximations used.

The Onsager reaction
field is a technique that can re-establish these sum rules in the approximate linear response.\cite{PhysRevB.50.1450} Consider first the sum rule for the short-range order.
Using Eq.~(\ref{eq:short-range-order})
we can write the short-range order in the self-consistent fashion
\begin{align}
\beta \bar{\Psi}_{\mathfrak{i}\alpha;\mathfrak{j}\beta} =& \left[(\beta C)^{-1}- S^{(2)}\right]^{-1}_{\mathfrak{i}\alpha;\mathfrak{j}\beta} \nonumber
\\ =& \left[(\beta C) + (\beta C)S^{(2)}(\beta C) + \cdots\right]_{\mathfrak{i}\alpha;\mathfrak{j}\beta} \nonumber
\\ =& \left[(\beta C) + (\beta C)S^{(2)}(\beta \bar{\Psi})\right]_{\mathfrak{i}\alpha;\mathfrak{j}\beta}. \label{eq:sro-expansion}
\end{align}
In terms of the explicit variations $\{\delta \bar{c}_{\mathfrak{i}\alpha}\}$ and $\{\delta \nu_{\mathfrak{i}\alpha}\}$;
\begin{align}
\delta \bar{c}_{\mathfrak{i}\alpha}  = \sideset{}{'} \sum_{\beta} (\beta C)_{\alpha\beta} \delta \nu_{\mathfrak{i}\beta}
 + \sideset{}{'}\sum_{\gamma;\mathfrak{k}\delta}(\beta C)_{\alpha\gamma} S^{(2)}_{\mathfrak{i}\gamma;\mathfrak{k}\delta} \delta \bar{c}_{\mathfrak{k}\delta}. \label{eq:pre-onsager}
\end{align}
As before, we know the true short-range order obeys $\delta c_{\mathfrak{i}\alpha} / \delta \nu_{\mathfrak{i}\beta} = \beta C_{\alpha\beta}$ when varying $\delta \nu_{\mathfrak{i}\beta}$ and setting $\delta \nu_{\mathfrak{j}\gamma} = 0$ otherwise; \emph{regardless of the correlations}. We see Eq.~(\ref{eq:pre-onsager}) violates the sum rule in such an instance due to the presence of the second term. Let us therefore define a self-reaction field $\delta c_{\mathfrak{j}\beta}^{(\delta \nu_\mathfrak{i})}$ to be the concentration variation at site $\mathfrak{j}$ when considering only variations of on-site chemical potentials $\{ \delta \nu_{\mathfrak{i}1}, \ldots, \delta\nu_{\mathfrak{i}(n-1)}\}$. To restore the sum rule we then consider the ansatz
 \begin{align}
 \delta \hat{c}_{\mathfrak{i}\alpha} -  
  \sideset{}{'} \sum_{\beta} (\beta C)_{\alpha\beta} \delta \nu_{\mathfrak{i}\beta} &= 
  \nonumber \\ & \mkern-150mu = \sideset{}{'}\sum_{\gamma;\mathfrak{k}\delta}(\beta C)_{\alpha\gamma} S^{(2)}_{\mathfrak{i}\gamma;\mathfrak{k}\delta} (\delta \hat{c}_{\mathfrak{k}\delta} - \delta \hat{c}_{\mathfrak{k}\delta}^{(\delta \nu_{\mathfrak{i}})})
\nonumber \\ & \mkern-150mu = \sideset{}{'}\sum_{\gamma;\mathfrak{k}\delta}(\beta C)_{\alpha\gamma} S^{(2)}_{\mathfrak{i}\gamma;\mathfrak{k}\delta} (\delta \hat{c}_{\mathfrak{k}\delta} - \sideset{}{'}\sum_{\mu\epsilon} \frac{\partial \hat{c}_{\mathfrak{k}\delta}}{\partial \nu_{\mathfrak{i}\epsilon}} (\beta C)^{-1}_{\epsilon\mu} \delta \hat{c}_{\mathfrak{i}\mu}). \label{eq:post-onsager}
\end{align}
The change in notation $\delta \bar{c}_{\mathfrak{i}\alpha} \rightarrow \delta \hat{c}_{\mathfrak{i}\alpha}$ reflects the changed definition of the concentration variation in terms of site chemical potentials. The above definition is consistent with $ \delta \hat{c}_{\mathfrak{i}\alpha}/\delta \nu_{\mathfrak{i}\beta} = (\beta C)_{\alpha\beta}$ when only on-site $\delta \nu_{\mathfrak{i}\beta}$ varies since in this case $\delta \hat{c}_{\mathfrak{i}\mu} = (\beta C)_{\mu\beta} \delta \nu_{\mathfrak{i}\beta}$ and the second term in Eq.~(\ref{eq:post-onsager}) cancels. This restores the on-site sum rule at all temperatures. By reorganizing Eq.~(\ref{eq:post-onsager}) we can also interpret the effect of the Onsager reaction field as shifting the pair parameters via
\begin{multline}
S^{(2)}_{\mathfrak{i}\gamma;\mathfrak{k}\delta} \rightarrow S^{(2)}_{\mathfrak{i}\gamma;\mathfrak{k}\delta} - \sideset{}{'}\sum_{\mathfrak{l}\nu;\epsilon} S^{(2)}_{\mathfrak{i}\gamma;\mathfrak{l}\nu} (\beta \hat{\Psi}_{\mathfrak{l}\nu;\mathfrak{i}\epsilon}) (\beta C)^{-1}_{\epsilon\delta} \delta_{\mathfrak{i}\mathfrak{k}}  \\
=: S^{(2)}_{\mathfrak{i}\gamma;\mathfrak{k}\delta} - \Lambda_{\gamma\delta} \delta_{\mathfrak{i}\mathfrak{k}} \label{eq:onsager-shift}
\end{multline}
where we have relabeled indices and naturally defined the revised short range order $\beta \hat{\Psi}_{\mathfrak{l}\delta;\mathfrak{i}\epsilon} := \partial \hat{c}_{\mathfrak{l}\delta}/
\partial \delta \nu_{\mathfrak{i}\alpha}$.
After taking a lattice Fourier transform of Eq.~(\ref{eq:post-onsager}) we find the site-independent Onsager reaction field
\begin{align}
\Lambda_{\gamma\delta} = \frac{1}{V_\text{BZ}} \int dk \sideset{}{'}\sum_{\nu\epsilon}  S^{(2)}_{\gamma\nu}(k) \hat{\Psi}_{\nu\epsilon}(k) C^{-1}_{\epsilon\delta}. \label{eq:onsager-lambda}
\end{align}


We can also consider this result in the context of Eqs.~(\ref{eq:master-pol-variation})-(\ref{eq:master-conc-variation}). In this case the $S^{(2)}_{\alpha\beta}(k)$ parameters are not readily identifiable. However the same logic of subtracting a shift $\Lambda_{\alpha\beta}$ of the effective pair parameters can be applied directly to Eq.~(\ref{eq:master-conc-variation}). Thus, we replace Eq.~(\ref{eq:master-conc-variation}) with
\begin{widetext}
\begin{multline}
\beta \hat{\Psi}_{\mu\gamma}(k)=  \beta C_{\mu\gamma}+\sum_{\sigma}\beta C_{\mu\sigma}\bigg\{ \frac{1}{\pi}\int d\epsilon f(\epsilon-\mu)\text{Im}\text{Tr}\left[\,(X_{\sigma}-X_{n})\mathbb{C}(k)\frac{\delta t_{c}^{-1}}{\delta\nu_{\gamma}}(k)\right] \\
-(Q_{\sigma}-Q_{n})M(k)\frac{\delta P}{\delta\nu_{\gamma}}(k) - \sideset{}{'}\sum_\delta \Lambda_{\sigma\delta} \hat{\Psi}_{\delta\gamma}(k) \bigg\}. \label{eq:master-conc-variation-onsager}
\end{multline}
\begin{align}
\Lambda_{\sigma\delta} := \frac{1}{V_\text{BZ}} \int dk \sideset{}{'}\sum_\gamma \bigg\{ \frac{1}{\pi}\int d\epsilon f(\epsilon-\mu)\text{Im}\text{Tr}\left[\,(X_{\sigma}-X_{n})\mathbb{C}(k)\frac{\delta t_{c}^{-1}}{\delta\nu_{\gamma}}(k)\right] 
-(Q_{\sigma}-Q_{n})M(k)\frac{\delta P}{\delta\nu_{\gamma}}(k) \bigg\} C^{-1}_{\gamma\delta}. \label{eq:onsager-lambda-final}
\end{align}
\end{widetext}
It is easy to verify the short-range order sum rule is obeyed by integrating both sides of Eq.~(\ref{eq:master-conc-variation-onsager}) over the Brillouin zone. Similarly, we can restore the on-site charge response sum rule by replacing Eq.~(\ref{eq:master-charge-variation}) with
\begin{widetext}
\begin{multline}
\hat{\Phi}_{\mu\gamma}^{m}(k)=  \sum_{n}A_{\mu}^{mn}\hat{\Phi}_{\mu\gamma}^{n}(k)+B_{\mu}^{m}M(k)\frac{\delta P}{\delta\nu_{\gamma}}(k) \\ -\frac{1}{\pi}\int d\epsilon\,f(\epsilon-\mu)\text{ Im}\sum_{LL'}F_{\mu;LL'}^{m}(\mathcal{\mathbb{D}}_{\mu}\mathbb{C}(k)\frac{\delta t_{c}^{-1}}{\delta\nu_{\gamma}}(k))_{LL'} - \sideset{}{'} \sum_{\sigma} \Lambda^{m}_{\mu\delta} \hat{\Psi}_{\delta\gamma}(k) \label{eq:master-charge-variation-onsager}
\end{multline}
\begin{align}
 \Lambda^{m}_{\mu\delta} :=  \frac{1}{V_\text{BZ}} \int dk \sideset{}{'}\sum_\gamma \bigg\{ B_{\mu}^{m}M(k)\frac{\delta P}{\delta\nu_{\gamma}}(k) -\frac{1}{\pi}\int d\epsilon\,f(\epsilon-\mu)\text{ Im}\sum_{LL'}F_{\mu;LL'}^{m}(\mathcal{\mathbb{D}}_{\mu}\mathbb{C}(k)\frac{\delta t_{c}^{-1}}{\delta\nu_{\gamma}}(k))_{LL'} \bigg\} C^{-1}_{\gamma\delta}.  \label{eq:master-charge-onsager-final}
\end{align}
\end{widetext}
Again, the on-site charge response sum rule can be confirmed by applying $\int dk(\cdot)/V_\text{BZ}$ to both sides of Eq.~(\ref{eq:master-charge-variation-onsager}). The Onsager reaction field improves the linear response of Eqs.~(\ref{eq:master-pol-variation})-(\ref{eq:master-conc-variation}) by inclusion of reaction fields $\Lambda_{\sigma\delta}$ and $\Lambda_{\sigma\delta}^m$ (specified by Eq.~(\ref{eq:onsager-lambda-final}) and Eq.~(\ref{eq:master-charge-onsager-final}) respectively) to restore on-site sum rules.

\section{Band-only Reduction\label{sec:band-reduction}}

Due to the complexity of Eqs.~(\ref{eq:master-pol-variation})-(\ref{eq:master-conc-variation}), we present for the purposes of this paper a major simplification in which we demand there is \emph{no charge transfer} and \emph{no charge response}. In other words $Q_\alpha = Q_\beta$ and $\Phi^n_{\alpha\beta}(k) = 0$. In this case Eqs.~(\ref{eq:master-pol-variation})-(\ref{eq:master-conc-variation}) reduce to the single
\begin{multline}
\beta \bar{\Psi}_{\mu\gamma}(k)= \beta C_{\mu\gamma}+ 
\sideset{}{'}\sum_{\sigma\nu}\beta C_{\mu\sigma}\bigg\{ \frac{1}{\pi}\int d\epsilon f(\epsilon-\mu)\text{Im}\text{Tr}\bigg[  \\
\,(X_{\sigma}-X_{n}) \mathbb{C}(k) \left[\mathbb{I}-\mathbb{X}\mathbb{C}(k)\right]^{-1} (X_{\nu}-X_{n}) \bigg]\bigg\} \beta \bar{\Psi}_{\nu\gamma}(k). \label{eq:master-band-variation}
\end{multline}
By comparison with Eq.~(\ref{eq:sro-expansion}) we may identify the factor in braces as $S^{(2)}_{\sigma\nu}(k)$. And by comparison to Eq.~(\ref{eq:host-indep-convert}), we may identify 
\begin{align}
\tilde{S}^{(2)}_{\alpha\beta}(k)\! = \! \frac{1}{\pi}\int\! d\epsilon f(\epsilon\!-\!\mu)\text{Im}\text{Tr}\bigg[ 
 X_{\alpha} \mathbb{C}(k) \left[\mathbb{I}\!-\!\mathbb{X}\mathbb{C}(k)\right]^{-1} X_\beta \bigg].
\raisetag{-7pt} \label{eq:s2-band-only-host-invariant}
\end{align}
Despite freezing the charge, we still include the electronic response due to band-terms in the total energy. It will incorporate all band-related mechanisms, e.g. Fermi surface nesting and van Hove singularities. Eq.~(\ref{eq:master-band-variation}) retains the computationally most demanding piece of the calculation, which is the convolution integral $\mathbb{C}(k)$ and inversion $[\mathbb{I}-\mathbb{X}\mathbb{C}(k)]^{-1}$. From Eq.~(51) and the relation $\sum_\alpha \bar{c}_\alpha X_\alpha = 0$ derived in Appendix \ref{sub:CPA-Variation}  \emph{it is clear $\sum_\alpha \bar{c}_\alpha \tilde{S}^{(2)}_{\alpha\beta}(k) = 0$ as used in Section \ref{sec:effective-params}}. From the form of Eq.~(\ref{eq:master-conc-variation}) we expect this sum rule to hold in the general case as well.

\section{Band-only Results}

To solve the KKR-CPA equations we used the \emph{Hutsepot} code made available to us by 
M. Daene.\cite{daene} We used the atomic sphere approximation,\cite{0305-4608-8-12-008} a $20\times 20\times 20$ Monkhorst-Pack grid\cite{PhysRevB.13.5188} for Brillouin zone
integrals, $l_\text{max}$=3 for basis set expansions, and a 24 point semi-circular Gauss-Legendre grid in the complex plane for integrating over valence energies. All self-consistent potentials are in the disordered local moment (DLM) state.\cite{0305-4608-15-6-018} This simulates the high-temperature paramagnetic state. Calculations
of the convolution integral $\mathbb{C}$ and band-only, multicomponent $S^{(2)}_{\alpha\beta}(k)$ is based on in-house code. An adaptive scheme based on nested line integrals and Simpson's rule is used for Brillouin zone integrals of Eq.~(\ref{eq:superC-definition}). This code used $l_\text{max}=2$, 26 energy points along a rectangular contour for energy integration, and fixed $T=300$ K for evaluating $S^{(2)}_{\alpha\beta}(k)$ at $24\times24\times24$ k-points. The multicomponent Onsager field correction uses an internally developed code. A double Monkhorst-Pack grid scheme using a high-resolution $96\times96\times96$ mesh near the peak $S^{(2)}_{\alpha\beta}(k)$ eigenvalue and lower-resolution $24\times24\times24$ mesh otherwise is used for Brillouin zone integrals of Eq.~(\ref{eq:onsager-lambda}). The exchange-correlation functional is that of Perdew-Wang.\cite{PhysRevB.45.13244}

\begin{table}
\begin{tabular}{c|cc|ccc|ccc}
Alloy & $a$ & $\mu$ & $Q_1$  & $Q_2$ & $Q_3$ & $M_1$ & $M_2$ & $M_3$  \\
\hline
CuAgAu\cite{CuAgAu-latt} & 7.523 & 0.551 & -0.115 & 0.036 & 0.078 & 0.00 & 0.00 & 0.00  \\
NiPdPt\cite{NiPdPt-latt} &  6.901 & 0.801 & -0.175 & 0.024 & 0.151 & 0.38 & 0.00 & 0.00 \\
RhPdAg\cite{RhPdAg-latt} & 7.722 & 0.469 & 0.019 & -0.029 & 0.010 & 0.00 & 0.00 & 0.00 \\
CoNiCu\cite{CoNiCu-latt} & 6.832 & 0.634 & 0.035 & -0.023 & -0.012 & 1.49 & 0.00 & 0.00 
\end{tabular}
\caption{Self-consistent KKR-CPA solutions of representative equiatomic ternary alloys on an FCC lattice using the \emph{Hutsepot} code.
Columns are the lattice constant (Bohr), electronic chemical potential (Ryd), net site charge ($e$), and site magnetic moment ($\mu_B$) ordered by atomic number. Moments are disordered according to the DLM approximation and thus reflect a paramagnetic state.
References provide the source of lattice constant data.}
\label{tbl:alloy-reference}
\end{table}

\begin{table}
\begin{floatrow}
\begin{tabular}{c|cccc}
$\tilde{V}_{\alpha\beta}(1)$&  & Cu & Ag & Au \\
\hline
Cu & & 0.998 & 0.253 & -1.244\\
Ag & & 0.253 & 0.007 & -0.258\\
Au & & -1.244 & -0.258 & 1.495\\
\end{tabular}

\begin{tabular}{c|cccc}
$\tilde{V}_{\alpha\beta}(2)$&  & Cu & Ag & Au \\
\hline
Cu & & -0.033 & -0.002 & 0.035\\
Ag & & -0.002 & 0.034 & -0.032\\
Au & & 0.035 & -0.032 & -0.003\\
\end{tabular}
\end{floatrow}

\begin{floatrow}
\begin{tabular}{c|cccc}
$\tilde{V}_{\alpha\beta}(1)$&  & Ni & Pd & Pt \\
\hline
Ni & & 2.054 & 0.021 & -2.062\\
Pd & & 0.021 & 0.013 & -0.033\\
Pt & & -2.062 & -0.033 & 2.083\\
\end{tabular}

\begin{tabular}{c|cccc}
$\tilde{V}_{\alpha\beta}(2)$&  & Ni & Pd & Pt \\
\hline
Ni & & -0.297 & -0.032 & 0.327\\
Pd & & -0.032 & 0.044 & -0.013\\
Pt & & 0.327 & -0.013 & -0.312\\
\end{tabular}
\end{floatrow}

\begin{floatrow}
\begin{tabular}{c|cccc}
$\tilde{V}_{\alpha\beta}(1)$&  & Rh & Pd & Ag \\
\hline
Rh & & -3.123 & 0.526 & 2.578\\
Pd & & 0.526 & 0.197 & -0.720\\
Ag & & 2.578 & -0.720 & -1.843\\
\end{tabular}

\begin{tabular}{c|cccc}
$\tilde{V}_{\alpha\beta}(2)$&  & Rh & Pd & Ag \\
\hline
Rh & & -0.120 & 0.087 & 0.032\\
Pd & & 0.087 & 0.017 & -0.104\\
Ag & & 0.032 & -0.104 & 0.071\\
\end{tabular}
\end{floatrow}

\begin{floatrow}
\begin{tabular}{c|cccc}
$\tilde{V}_{\alpha\beta}(1)$&  & Co & Ni & Cu \\
\hline
Co & & -0.303 & 0.171 & 0.130\\
Ni & & 0.171 & 0.047 & -0.217\\
Cu & & 0.130 & -0.217 & 0.088\\
\end{tabular}

\begin{tabular}{c|cccc}
$\tilde{V}_{\alpha\beta}(2)$&  & Co & Ni & Cu \\
\hline
Co & & 0.110 & -0.035 & -0.074\\
Ni & & -0.035 & -0.009 & 0.044\\
Cu & & -0.074 & 0.044 & 0.030\\
\end{tabular}
\end{floatrow}

\caption{Effective pair parameters (mRyd) at \mbox{$R_1 = a/\sqrt{2}$} and \mbox{$R_2 = a$} 
calculated at T = 300 K. 
Pair energies are an order of magnitude 
reduced in the second shell. Note that rows and columns sum to approximately zero. In general
$\sum_\alpha \bar{c}_{\alpha} \tilde{S}^{(2)}_{\alpha\beta} = 0$.}
\label{tbl:pair-pot}
\end{table}

Before proceeding we note that our band-only results are in fair agreement with a number of past calculations. These past results have shown favorable comparison to experiment.\cite{PhysRevB.50.1450,PhysRevB.53.10610} For {PdRh} on FCC lattice past results find a concentration wave instability at $k$=(000) occuring at $T_c$= 1850 K (1580 K with Onsager correction).\cite{PhysRevB.50.1450} Using our codes and settings described we find 2300 K (1770 K). For {NiZn} previous results find instability for $k$=(100) at 1925 K (1430 K).\cite{PhysRevB.53.10610} We find 2140 K (1430 K). Past results for {CuZn} find incommensurate vector $k$=(0,0.15,1) at 425 K without Onsager correction and commensurate vector $k$=(100) at 230 K with Onsager correction.\cite{PhysRevB.53.10610} We find instability at $k$=(0,0.2,1) at 542 K (160 K with Onsager). Past results for {CuNi} find $k$=(000) at 680 K (560 K).\cite{PhysRevB.53.10610} We find 560 K (445 K). Finally, for ternary alloy {Cu$_{0.50}$Ni$_{0.25}$Zn$_{0.25}$} past results find $k$=(100) at 1243 K (985 K).\cite{PhysRevB.53.10610} We find 1210 K (885 K). We also note that for the Ising model on SC, BCC, and FCC lattices the ratio of the mean-field predicted transition $T_\text{MF}$ to Onsager predicted transition $T_\text{Ons}$ is precisely known to be 1.516, 1.393, and 1.345 respectively.\cite{joyce} We get 1.53, 1.38, and 1.33 respectively. Differences are likely due to the resolution of numerical grids in the solver.

We now present band-only results for CuAgAu, NiPdPt, RhPdAg, and CoNiCu on an FCC lattice. The first two alloys respectively are isoelectronic (same group) and the next two have adjacent atomic numbers (same period). In all cases we take equiatomic concentrations. In Table~\ref{tbl:alloy-reference} we present site charges and moments of the high temperature fully disordered paramagnetic reference state. There is greater charge-transfer for the isoelectronic alloys. In brief, we find: For CuAgAu the concentration wave instability occurs at $k$=(100) with $T_c$=580 K (210 K with Onsager correction). For NiPdPt at $k$=(100) with 980 K (270 K). For RhPdAg at $k$=(000) at 4660 K (3980 K). For CoNiCu at $k$=(100) at 280 K (210 K). 

\begin{figure}
\begin{overpic}[scale=0.5]{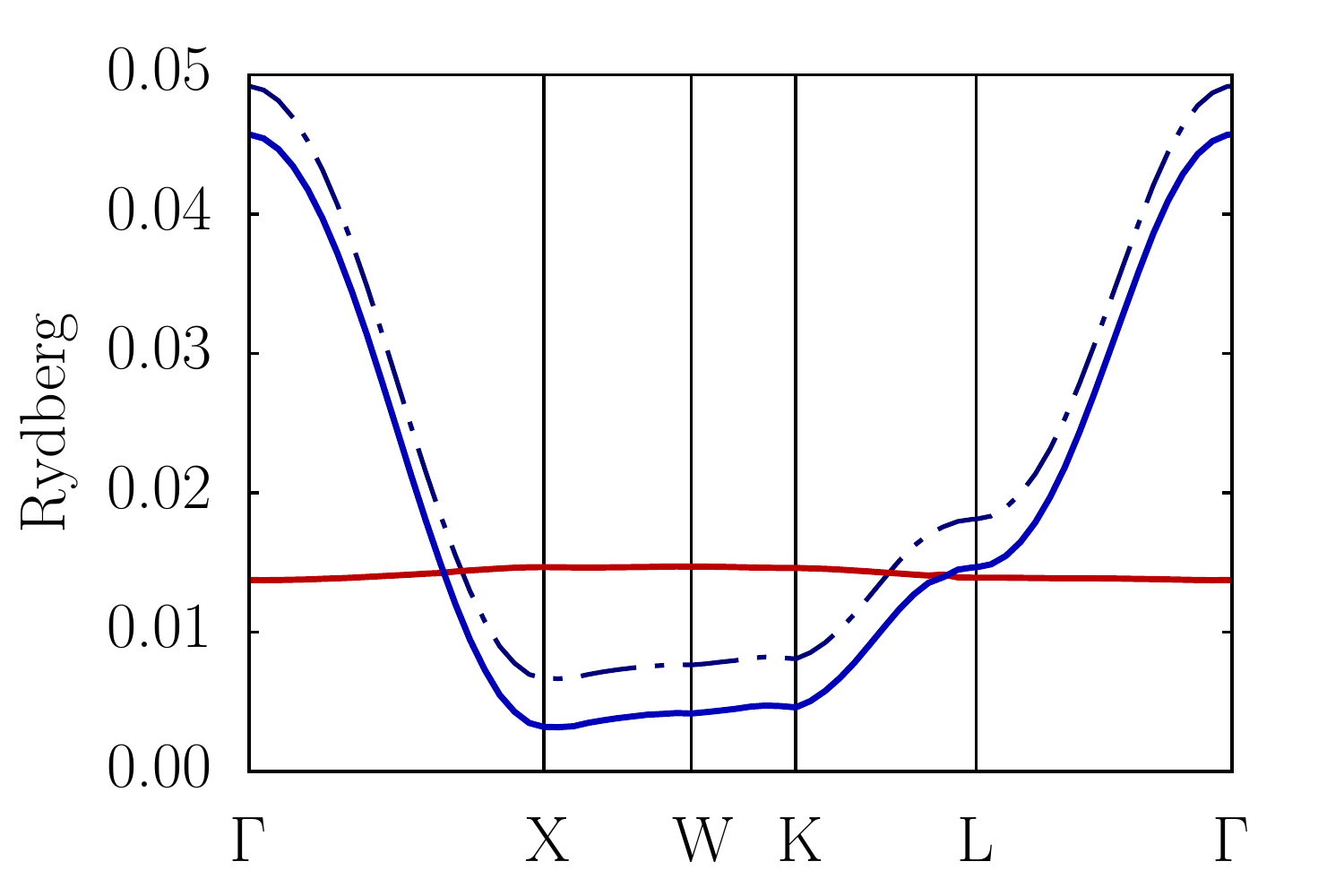}
\put(0,60) {$(a)$}
\end{overpic}
\begin{overpic}[scale=0.5]{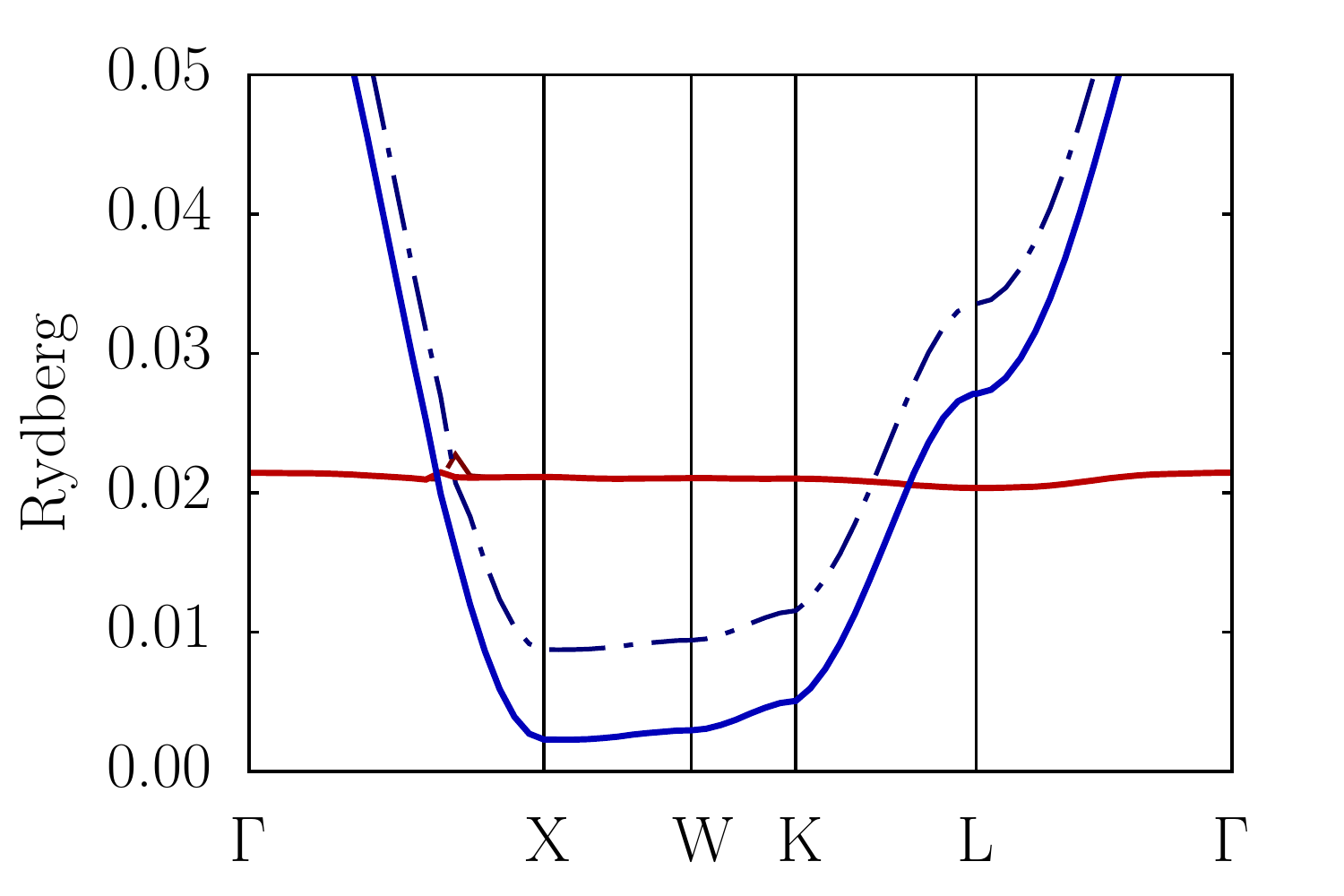}
\put(0,60) {$(b)$}
\end{overpic}
\begin{overpic}[scale=0.5]{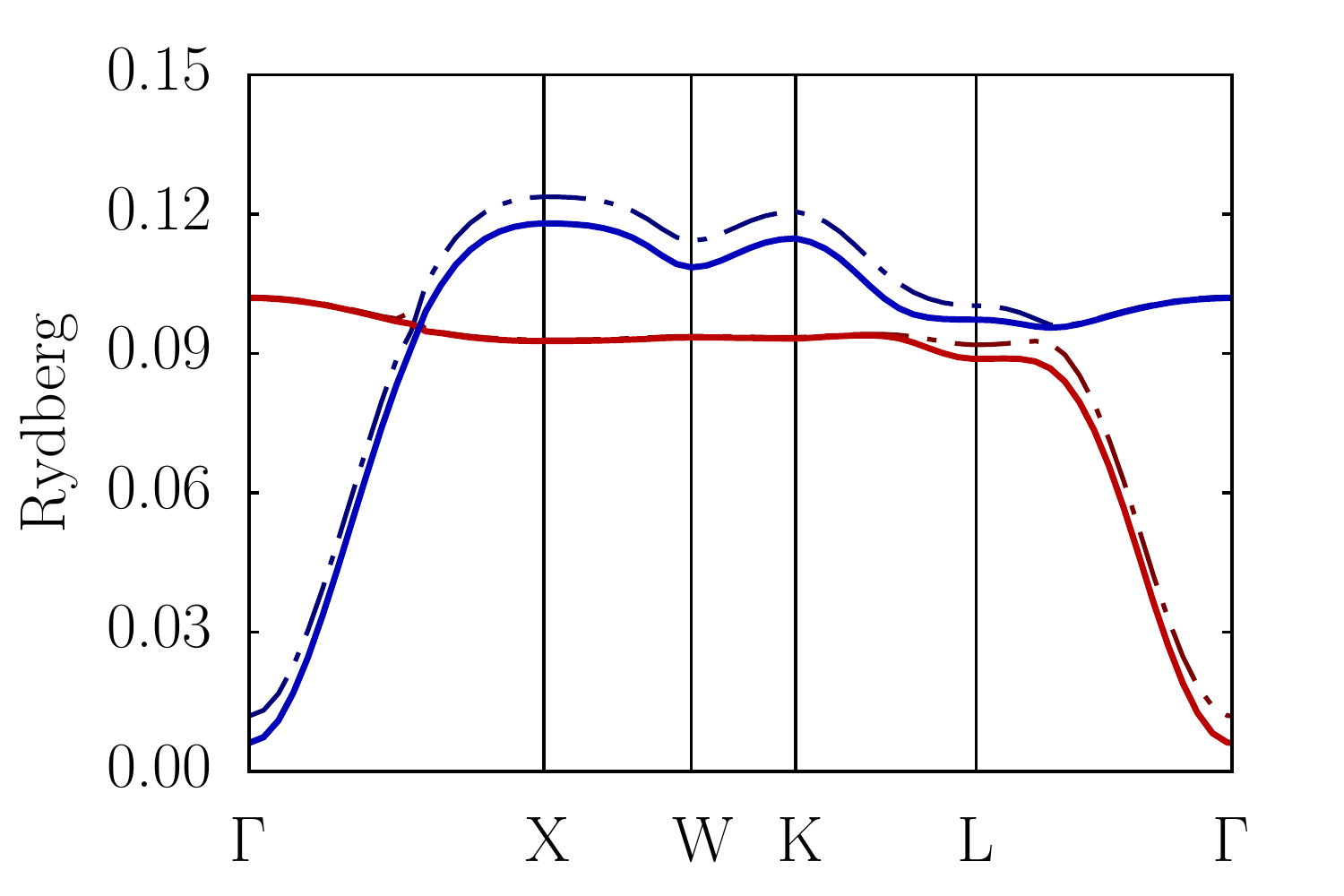}
\put(0,60) {$(c)$}
\end{overpic}
\begin{overpic}[scale=0.5]{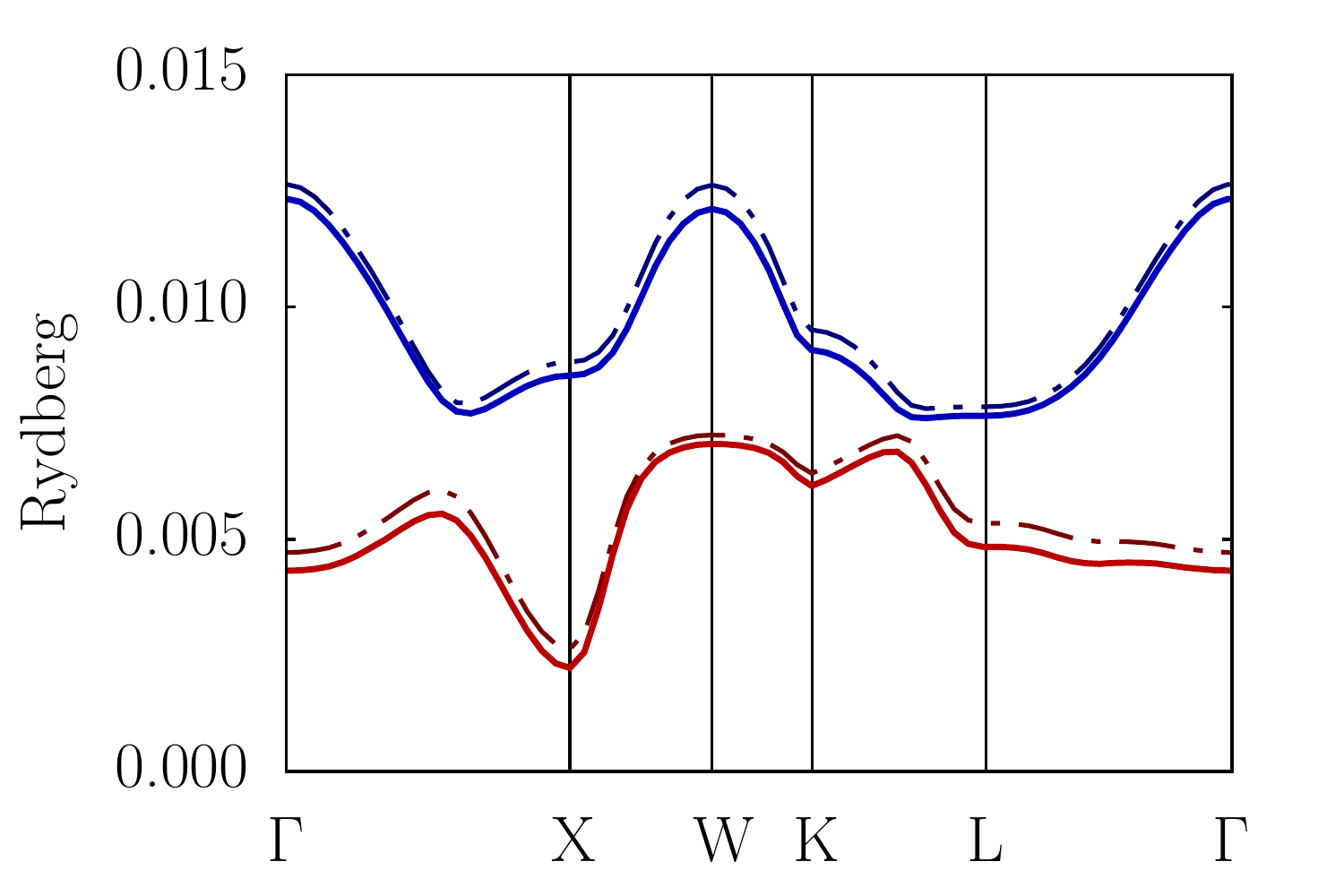}
\put(0,60) {$(d)$}
\end{overpic}
\caption{(Color online) Eigenvalues of the chemical stability matrix of Eq.~(\ref{eq:free-energy-final-result}) along special $k$ directions in the FCC Brillouin zone (solid curves) for (a) CuAgAu at 750 K, (b) NiPdPt at 1100 K, (c) RhPdAg at 5000 K, and (d) CoNiCu at 400 K. The eigenvalues represent the quadratic coefficient of the energy cost of a concentration wave with wave-vector $k$. Dash-dot curves include an Onsager reaction field. The nature of the eigenvectors is discussed in Section~\ref{sec:chemical-pol}.}
\label{fig:chemical-stab}
\end{figure}

\definecolor{highl}{rgb}{1.00,0.9843,0.8}
\begin{table}

\begin{tabular}{cc|cc|cccc}
Alloy& & T & $k$ & $\delta F^{(1)}$ & $\delta \bar{c}_1$ & $\delta \bar{c}_2$ & $ \delta \bar{c}_3$ \\
\hline
CuAgAu & & 750 & $\Gamma$ &13.744 &  -0.535875 &   0.801443 &  -0.265568 \\
CuAgAu & & 750 & $\Gamma$ & 45.722 & -0.616039 &  -0.156062 &    0.772101 \\
CuAgAu & & 750 & X &14.664 & -0.517672 &   0.805656 &    -0.287984  \\
\rowcolor{highl}[\tabcolsep]
CuAgAu & & 750 & X & 3.202 &  -0.631413 &   -0.132610 &  0.764023 \\

NiPdPt & & 1100 & $\Gamma$ & 21.437 &  -0.433233 &   0.8159761 &  -0.382743  \\
NiPdPt & & 1100 & $\Gamma$ & 69.545 &  -0.692081 &   -0.029150 &   0.721231 \\
NiPdPt & & 1100 & X & 21.140  &   0.440102 &  -0.815645 &    0.375543  \\
\rowcolor{highl}[\tabcolsep]
NiPdPt & & 1100 & X & 2.295 &  0.687733 &   0.037273 &  -0.725006 \\

RhPdAg & & 5000 & $\Gamma$ &  101.954 & 0.310183 &  -0.809186 &   0.499003  \\
\rowcolor{highl}[\tabcolsep]
RhPdAg & & 5000 & $\Gamma$ & 6.128  &   0.755284 &  -0.109015 & -0.646268 \\
RhPdAg & & 5000 & X &  92.652  &  0.224583 &   -0.792124 &   0.567540  \\
RhPdAg & & 5000 & X &  118.014 &0.785003 &    -0.198006 &   -0.586996 \\

CoNiCu & & 400 & $\Gamma$ &  4.324 & -0.786458 &   0.583262 &   0.203196  \\
CoNiCu & & 400 & $\Gamma$ & 12.331  &   -0.219431 & -0.571377 &  0.790808 \\
\rowcolor{highl}[\tabcolsep]
CoNiCu & & 400 & X &   2.229  &   0.465829 & 0.347820 &  -0.813649 \\
CoNiCu & & 400 & X &  8.523  & 0.670574 &  -0.738707 &    0.068132 \\

\end{tabular}

\caption{Chemical stability matrix eigenvalues (mRyd) and corresponding polarization vectors at $\Gamma$ and $X$
for the same temperatures as in Fig.~\ref{fig:chemical-stab}. Low energy fluctuations are highlighted. Temperatures have been chosen above the mean-field absolute instability point determined by $\bar{\Psi}_{\alpha\beta}(k;T)$. The fluctuations presented are finite but the formalism is only valid in the infinitesimal limit.}
\label{tbl:conc-polarizations}
\end{table}

In Table~\ref{tbl:pair-pot} we present the effective pair interaction of Eq.~(\ref{eq:pair-pot-all}) for the first two shells. Onsager corrections to the pair parameters are presented in Table~\ref{tbl:onsager-field-matrix}. Negative pair interactions are considered favorable. The largest pair interactions are between Cu-Au on neighboring sites (favorable) as well as Cu-Cu (unfavorable) or Au-Au (unfavorable). Therefore we can expect that a concentration wave which places Cu and Au on alternate planes will be the most favorable excitation. This is clear from Fig.~\ref{fig:chemical-stab}(a) and the highlighted row in Table~\ref{tbl:conc-polarizations}. The lowest energy fluctuation is at wave-vector at $k=$X and the corresponding chemical polarization favors opposing changes in the site concentrations of Cu and Au. The components of the chemical polarization vector are not commensurate with each other and there is no reason to expect this to be the case in the limit of infinitesimal fluctuations. The same polarization mode at the $\Gamma$-point results in a high-energy excitation because it corresponds to formation of unfavorable Cu-Cu and Au-Au clusters. The second, alternate polarization mode, as seen in Table~\ref{tbl:conc-polarizations}, sets opposing concentration variations of Ag relative to Cu or Au. The resulting band is nearly flat (c.f. Fig.~\ref{fig:chemical-stab}). From the pair potentials in Table~\ref{tbl:pair-pot} we see Cu-Ag and Ag-Au energies nearly cancel and Ag-Ag has low pair cost. Therefore there is little to no pair energy cost for redistributing Ag atoms in a system where each site is equally likely to be occupied by Cu or Au. There is still, however, an entropy cost to segregating Ag from Cu and Au atoms. The sister alloy NiPtPd mimics almost all these computational trends. We see that when a few of the pair interactions are dominant, as for CuAgAu, we can sensibly interpret the chemical stabilities of concentration waves. An isothermal section at 350 $^\circ$C of the Co-Ag-Au experimental phase diagram reveals a miscibility gap along the Cu-Ag border, multiple ordered compounds along the Cu-Au border, and another large miscibility gap along the Ag-Au border.\cite{prince_CuAgAu} While it is difficult to make a comparison, these appear to be in qualitative agreement with the sign of the largest pair potentials in Table~\ref{tbl:pair-pot}. The binary alloy Ni-Pd is miscible to as low as -200 $^\circ$C,\cite{nash_NiPd} Ni-Pt forms ordered compounds as high as 620 $^\circ$C,\cite{NiPdPt-latt} and Pd-Pt is miscible until 720 $^\circ$C.\cite{okamoto_PdPt} Again, comparison is difficult, but the formation of ordered compounds in Ni-Pt in experiment agrees well with the large, favorable pair interaction for Ni-Pt in Table~\ref{tbl:pair-pot}. However, our temperature scale of T$_c$ = 270 K is depressed from that found for the experimental binary alloys. This difference can be attributed to attempting to compare a ternary to a set of binaries as well as the lack of inclusion of charge-effects and to DFT error in general. Further, our theory is a first-order expansion of the grand potential as a function of inverse temperature $\beta$ (c.f. Eq.~(\ref{eq:omega-expansion}) and Fig.~\ref{fig:toy-model-sro}). Thus we expect the best results for high-temperatures and weakly-correlated systems. In particular, we expect better comparison to experiment of the short-range order parameters calculated at high T. At the moment this data is not available for the systems considered so far.

In RhPdAg we see from the pair parameters (c.f. Table~\ref{tbl:pair-pot}) a strong favorability to formation of Rh-Rh and Ag-Ag clusters. Therefore the low-energy fluctuation is a concentration wave with wave-vector at $\Gamma$ and a polarization mode that causes the change in site occupancy of Rh and Ag to be opposite (c.f. Table~\ref{tbl:conc-polarizations}). There is an unusual topology here: Traversing a complete circuit in k-space along the path depicted in Fig.~\ref{fig:chemical-stab} leads to one polarization mode transforming into another. Lastly, for CoNiCu, we see the pair interaction energies in Table~\ref{tbl:pair-pot} are suppressed compared to the previous examples and that no few pairs are dominant. The resulting chemical stability graph in Fig.~\ref{fig:chemical-stab}(d) has a reduced energy scale and displays more structure than the other cases. 

\section{Conclusion}

In this paper we derived a multicomponent generalization of the $S^{(2)}$ theory of binary alloys. In particular we derived an expression for the change of free-energy for any fluctuation in site occupancies. Due to translational invariance of the underlying alloy we examined these fluctuations in a basis of concentration waves. This free-energy expression showed the reciprocal connection between the magnitude of short-range order and the free-energy cost of fluctuations. The same expression also clearly splits the change in free-energy as due to a site disorder induced entropy effect and electronic effects that drive favorable atomic pairing. We also clarified the ambiguities inherent in defining chemical polarizations for multi-component alloys and described one procedure for defining these in a sensible, host-invariant manner. We further showed how to map on to an effective pair interaction model and how this can also be done in a host-invariant manner. To make these concepts clear we analyzed four representative ternary alloys: CuAgAu, NiPdPt, RhPdAg, and CoNiCu in the band-only approximation. Despite our choice of ternary alloys, the theory presents no difficulties in being applied to higher-component alloys.

We are currently developing codes to implement our linear response theory including all charge-related terms for the multicomponent case. Our goal is to apply the generalized $S^{(2)}$ theory to high-entropy alloys in order to assess their phase stability. For this purpose one of the authors has written scripts that enable high-throughput calculation of alloys for different choice of transition metals, lattice constant, structure (FCC, BCC, or HCP) and range of concentrations. We also plan to make more careful comparisons of the short-range order parameter for specific high-entropy alloys at high temperatures, the limit in which our theory becomes increasingly accurate. 

This work was supported by the Materials Sciences \& Engineering Division of the Office of Basic Energy Sciences, U.S. Department of Energy.

\begin{table}

\begin{floatrow}
\begin{tabular}{c|cccc}
$\tilde{\Lambda}_{\alpha\beta}$&  & Cu & Ag & Au \\
\hline
Cu & & 1.392 &  0.291 &  -1.678 \\
Ag & & 0.291 &  0.068 & -0.357\\
Au & & -1.678 & -0.357  & 2.029 \\
\end{tabular}

\begin{tabular}{c|cccc}
$\tilde{\Lambda}_{\alpha\beta}$&  & Ni & Pd & Pt \\
\hline
Ni & &    3.212 &   0.029 & -3.232  \\
Pd & & 0.029 &  0.005 & -0.034 \\
Pt & & -3.232 & -0.034 &  3.256 \\ 
\end{tabular}
\end{floatrow}
\vspace{1mm}

\begin{floatrow}
\begin{tabular}{c|cccc}
$\tilde{\Lambda}_{\alpha\beta}$&  & Rh & Pd & Ag \\
\hline
Rh & &   3.286 & -0.459 &  -2.819  \\
Pd & &  -0.459 &  0.193 &  0.265\\
Ag & &  -2.819 & 0.265 &   2.546 \\ 
\end{tabular}

\begin{tabular}{c|cccc}
$\tilde{\Lambda}_{\alpha\beta}$&  & Co & Ni & Cu \\
\hline
Co & &  0.314& - 0.092 &  -0.221  \\
Ni & &  -0.092 &  0.131 & 0.038\\
Cu & &  -0.221 &  -0.038 &  0.259 \\
\end{tabular}
\end{floatrow}

\caption{Onsager reaction field matrix (mRyd) (c.f. Section \ref{sec:onsager}) using a host-invariant basis (c.f. Section \ref{sec:effective-params}) at the temperatures indicated
in Fig.~\ref{fig:chemical-stab}.}
\label{tbl:onsager-field-matrix}
\end{table}

\appendix

\section{Lattice Fourier transform\label{sub:Lattice-Fourier-transform}}

All lattice Fourier transforms are according to the relations
\begin{align*}
f(k)=&\frac{1}{\sqrt{N}}\underset{i}{\sum}e^{-ik\cdot R_{i}}f_{i} &&& f_{i}=&\frac{1}{\sqrt{N}}\underset{i}{\sum}e^{ik\cdot R_{i}}f(k) & \\
A(k)=&\underset{i}{\sum}e^{-ik\cdot R_{i}}A_{i0} &&& A_{ij}=&\frac{1}{N}\underset{k}{\sum}e^{ik\cdot(R_{i}-R_{j})}A(k) &
\end{align*}
for a system with $N$ Bravais sites and translationally invariant $A_{ij}$. To simplify the derivation and notation we only consider crystals with single atom per basis throughout the Appendix. Then $\mathfrak{i} \rightarrow i$.

\section{Variation of CPA ansatz\label{sub:CPA-Variation}}

Before taking a variation of the CPA Ansatz in Eq.~(\ref{eq:CPA-condition}), we put it in a more desirable form using CPA $X_{i\mu}$ matrices. To see this, first note that by definition 
\begin{align*}
(\mathbb{\tau}^{i\mu})^{-1} = \Delta_{i\mu}I_{i} + (\mathbb{\tau}^c)^{-1}
\end{align*}
as matrices in site- and angular momentum indices and where $(I_i)_{kL;lL'}=\delta_{ki}\delta_{li}\delta_{LL'}$ is nonzero only in the $(i,L)\times(i,L')$ subblock. Multiplying on left by $\mathbb{\tau}^{i\mu}$ and right by $\mathbb{\tau}^c$ and considering the  $(i,L)\times(i,L')$ sublock:
\begin{align*}
\mathbb{\tau}^c_{ii} = \mathbb{\tau}^{i\mu}_{ii} \Delta_{i\mu} \mathbb{\tau}^c_{ii} +  \mathbb{\tau}^{i\mu}_{ii}.
\end{align*}
Substituting Eq.~(\ref{eq:cpa-definitions-D}) or Eq.~(\ref{eq:cpa-definitions-Dbar}) finds $\mathbb{\tau}^{i\mu}_{ii} = D_{i\mu} \mathbb{\tau}^c_{ii} = \mathbb{\tau}^c_{ii} \bar{D}_{i\mu}$. Plugging either of these relations for $\mathbb{\tau}^{i\mu}$ in Eq.~(\ref{eq:CPA-condition}) gives $1=\sum_\mu \bar{c}_{i\mu} D_{i\mu}$. This can be changed to 
\begin{multline*}
0=\sum_{\mu} \bar{c}_{i\mu} (D_{i\mu}^{-1}-1)D_{i\mu} \\ = \sum_\mu \bar{c}_{i\mu} ( \mathbb{\tau}^c_{ii} \Delta_{i\mu} )D_{i\mu} =  \mathbb{\tau}^c_{ii} \sum_\mu \bar{c}_{i\mu} X_{i\mu}.
\end{multline*}
Hence the CPA condition is equivalent to $0=\sum_\mu \bar{c}_{i\mu} X_{i\mu}$.

A variation on this CPA condition is 
\begin{align}
0= & \:\delta(\,\sum\limits _{\mu}\bar{c}_{i\mu}X_{i\mu}\,)=\sum\limits _{\mu}(\delta\bar{c}_{i\mu}X_{i\mu}+\bar{c}_{i\mu}\delta X_{i\mu}).  \label{eq:cpa-var-init}
\end{align}
Using Eq.~(\ref{eq:cpa-definitions-D}) and the relation $\delta M^{-1}=-M(\delta M)M$,
\begin{multline*}
\delta X_{i\mu}= \,\delta[\Delta_{i\mu}^{-1}+\tau^c_{ii}]^{-1}= \\ \bar{D}_{i\mu}(\delta t_{i\mu}^{-1}-\delta t_{ic}^{-1})D_{i\mu}-X_{i\mu}\delta\tau^c_{ii}X_{i\mu}.
\end{multline*}
We may set $X_{i\mu}\rightarrow X_{\mu},D_{i\mu}\rightarrow D_{\mu}$,
etc. because we are expanding about a homogenous reference medium.
By definition of SPO $\tau^c$ in Eq.~(\ref{eq:spo-definition}), $\delta\tau^c_{ii}=\sum_j \tau^c_{ij} \delta t_{jc}^{-1} \tau^c_{ji}$. Its lattice Fourier transform is the convolution integral
\begin{align*}
\frac{1}{\sqrt{N}}\underset{i}{\sum}\delta\tau^c_{ii}e^{-ik\cdot R_{i}}= & -\frac{1}{N}\underset{q}{\sum}\tau^c(q)\delta t_{c}^{-1}(k)\tau^c(q-k)
\end{align*}
for $k,q$ in the Brillouin zone. Thus in $k$-space Eq.~(\ref{eq:cpa-var-init}) becomes
\begin{multline*}
0= \sum\limits _{\mu}\bigg(\delta\bar{c}_{\mu}(k)X_{\mu}+\bar{c}_{\mu}\bar{D}_{\mu}(\delta t_{\mu}^{-1}(k)-\delta t_{c}^{-1}(k))D_{\mu}+ \\ 
\bar{c}_{\mu}X_{\mu}\left\{ \frac{1}{\Omega_{\text{BZ}}}\int dq\:\tau^c(q)\delta t_{c}^{-1}(k)\tau^c(q-k)\right\} X_{\mu}\bigg).
\end{multline*}
A simplification can be made using the identity
\begin{multline*}
\sum\limits _{\mu}\bar{c}_{\mu}X_{\mu}\tau^c_{00}\delta t_{c}^{-1}(k)\tau^c_{00}X_{\mu}= \\ -\delta t_{c}^{-1}(k)+\sum\limits _{\mu}\bar{c}_{\mu}\bar{D}_{\mu}\delta t_{c}^{-1}(k)D_{\mu},
\end{multline*}
which takes advantage of $\sum_{\mu}\bar{c}_{\mu}D_{\mu}=\sum_{\mu}\bar{c}_{\mu}\bar{D}_{\mu}=1$. Therefore
\begin{multline*}
0= \sum\limits _{\mu}\bigg(\delta\bar{c}_{\mu}(k)X_{\mu}+\bar{c}_{\mu}\bar{D}_{\mu}\delta t_{\mu}^{-1}(k)D_{\mu}-\delta t_{c}^{-1}(k)+\\
\bar{c}_{\mu}X_{\mu}\left\{ \frac{1}{\Omega_{\text{BZ}}}\int dq\:\Delta\tau^c(q)\delta t_{c}^{-1}(k)\Delta\tau^c(q-k)\right\} X_{\mu}\bigg)
\end{multline*}
for $\Delta\tau^c(q)=\tau^c(q)-\tau^c_{00}.$ This may be interpreted as
a supermatrix equation in the product space of angular momentum (i.e. $L\times L$) 
to be solved for $\delta t_{c}^{-1}(k)$. Using definitions in Eqs.~(\ref{eq:superD-definition})-(\ref{eq:superC-definition})
we write the compact
\begin{align*}
\left[\mathbb{I}-\mathbb{X}\mathbb{C}(k)\right]\delta t_{c}^{-1}(k)= & \sum\limits _{\mu}\left[X_{\mu}\delta\bar{c}_{\mu}(k)+\bar{c}_{\mu}\mathbb{\bar{D}_{\mu}}\delta t_{\mu}^{-1}(k)\right].
\end{align*}
Dividing by the chemical potential variation $\delta\nu_{0\gamma}$
gives Eq.~(\ref{eq:master-cpa-variation}) 
\begin{multline*}
\left[\mathbb{I}-\mathbb{X}\mathbb{C}(k)\right]\frac{\delta t_{c}^{-1}}{\delta\nu_{\gamma}}(k)= \\ \stackrel[\mu=1]{n-1}{\sum}(X_{\mu}-X_{n})\beta \bar{\Psi}_{\mu\gamma}(k)+\sum\limits _{\mu}\bar{c}_{\mu}\mathbb{\bar{D}_{\mu}}\frac{\delta t_{\mu}^{-1}}{\delta\nu_{\gamma}}(k).
\end{multline*}

\section{Variation of potential\label{sub:Potential-Variation}}

First we prove an ancillary relation. We may interpret $R_{i\mu;L}(r)$, $V_{i\mu}(r)$, and $\mathcal{J}_{L}(r):=j_\ell(r) Y_{\ell m}(r)$
of Eq.~(\ref{eq:reg-sol-definition}) as diagonal matrices over an infinite-dimensional vector space with basis elements $r \in \mathbb{R}^3$. Then Eq.~(\ref{eq:reg-sol-definition}) is
\begin{multline}
R_{i\mu;L} = \mathcal{J}_{L} + G_0 V_{i\mu} \mathcal{J}_{L} +  G_0 V_{i\mu} G_0 V_{i\mu} \mathcal{J}_L + \cdots
\\ = \mathcal{J}_L + G^\text{ss}_{i\mu} V_{i\mu} \mathcal{J}_L, \label{eq:R-ss-definition}
\end{multline}
where superscript `ss' stands for ``single-site". The variation of Eq.~(\ref{eq:R-ss-definition}) is
\begin{align}
\delta R_{i\mu;L} \mkern-20mu & \nonumber \\
=& \,G_0 \delta V_{i\mu} \mathcal{J}_L + G_0 \delta V_{i\mu} G_0 V_{i\mu} \mathcal{J}_L + G_0 V_{i\mu} G_0  \delta V_{i\mu} \mathcal{J}_L  \nonumber \\ 
& \mkern30mu + G_0 \delta V_{i\mu} G_0  \delta V_{i\mu} G_0  \delta V_{i\mu} \mathcal{J}_L + \cdots \nonumber \\
=& \,(G_0 + G_0 V_{i\mu} G_0 + \cdots) \delta V_{i\mu} (\mathcal{J}_L + G_0 V_{i\mu} \mathcal{J}_L + \cdots) \nonumber
\\ =& \,G^\text{ss}_{i\mu} \delta V_{i\mu} R_{i\mu;L}  \label{eq:delta-R}
\end{align}
In this space Eq.~(\ref{eq:tmatrix-definition}) is $t_{i\mu;LL'} = \langle \mathcal{J}_L^* | V_{i\mu} R_{i\mu;L'} \rangle $. 
Its variation is
\begin{align*}
\delta t_{i\mu;LL'} =& \,\langle \mathcal{J}_L^* | \delta V_{i\mu} R_{i\mu;L'} \rangle +  \langle \mathcal{J}_L^* | V_{i\mu} \delta R_{i\mu;L'} \rangle 
\\ =& \,\langle \mathcal{J}_L^* | \delta V_{i\mu} R_{i\mu;L'} \rangle + \langle \mathcal{J}_L^* | V_{i\mu} G^\text{ss}_{i\mu} \delta V_{i\mu} R_{i\mu;L'} \rangle
\\ =& \,\langle \mathcal{J}_L^* +  (G^\text{ss}_{i\mu})^\dag V_{i\mu} \mathcal{J}_L^*   |  \delta V_{i\mu} R_{i\mu;L'} \rangle
\\ =& \,\langle R_{i\mu;L}^*   |  \delta V_{i\mu} R_{i\mu;L'} \rangle
 \\ =& \int dr R_{i\mu;L}(r) R_{i\mu;L'}(r) \delta V_{i\mu}(r)
\end{align*}
since $G^\text{ss}_{i\mu}(r,r')$ is a symmetric in $r,r'$.\cite{0953-8984-16-36-011}
And therefore
\begin{multline}
\delta t_{i\mu;LL'}^{-1} = \\ -\int dr \sum_{L_1 L_2} t_{i\mu;LL_1}^{-1} R_{i\mu;L_1}(r) R_{i\mu;L_2}(r)  \delta V_{i\mu}(r) t_{i\mu;L_2 L'}^{-1}
\\ = -\int dr Z_{i\mu;L}(r) Z_{i\mu;L'}(r) \delta V_{i\mu}(r) \label{eq:deltat-from-zz}
\end{multline}
because $R_{i\mu;L} = \sum_{L'}  t_{i\mu;LL'} Z_{i\mu;L'} $ and $t_{i\mu;L'L} = t_{i\mu;LL'}$.\cite{zabloudil,0953-8984-16-36-011} This establishes the direct connection between site potential variation $\delta V_{i\mu}$(r) and the associated scattering $T$ matrix variation $\delta t_{i\mu}^{-1}$. 

The self-consistent site potentials which ensure the CPA
grand potential in Eq.~(\ref{eq:cpa-grand-potential}) is variational with respect to each electron density $\rho_{i\mu}(r)$
is given in Eq.~(\ref{eq:cpa-potential}). On varying Eq.~(\ref{eq:cpa-potential}) 
\begin{multline*}
\delta V_{i\mu}(r)= \frac{dV_{\text{xc}}}{d\rho}(\rho_{\mu}(r))\delta\rho_{i\mu}(r) +\\ e^{2}\int_{V_{i}}dr'\frac{\delta\rho_{_{i\mu}}(r')}{|r-r'|} 
+e^{2}\sum_{j\neq i}\int_{V_{j}}dr'\frac{\delta\overline{\rho}_{j}(r')-\delta\bar{Z}_{j}\delta(r')}{|r+R_{i}-R_{j}-r'|}.
\end{multline*}
Here $V_{\text{xc}}(\rho)$ is a univariate function of 
of $\rho$. The explicit variation of the average charge density is 
\begin{multline*}
\delta\overline{\rho}_{i}(r)-\delta\bar{Z}_{i}\delta(r)= \\ \sum_{\mu}\delta \bar{c}_{i\mu}[\rho_{\mu}(r)-Z_{\mu}\delta(r)]+\sum_{\mu}\bar{c}_{\mu}\delta\rho_{i\mu}(r).
\end{multline*}
In terms of the basis $f_n(r)$ defined in Section~\ref{sec:linear-response}; we write$\int dr'\delta\rho_{i\mu}(r') = \delta\rho_{i\mu}^{1}$. Now we make the approximation that $|r+R_{i}-R_{j}-r'|\rightarrow|R_{i}-R_{j}|$.
This is reasonable for
well-separated cells. Performing the $\int_{V_{j}}dr'(\:\cdot\:)$
integral,
\begin{multline}
\delta V_{i\mu}(r)= \frac{dV_{\text{xc}}}{d\rho}(\rho_{\mu}(r))\delta\rho_{i\mu}(r)+ 
e^{2}\int_{V_{i}}dr'\frac{\delta\rho_{_{i\mu}}(r')}{|r-r'|}\\+e^{2}\sum_{j\neq i}\frac{\sum_{\gamma}(Q_{\gamma}\delta \bar{c}_{j\gamma}+\bar{c}_{\gamma}\delta\rho_{j\gamma}^{1})}{|R_{i}-R_{j}|}.\label{eq:init-cpa-potential-variation}
\end{multline}
where $Q_{\gamma}$ is defined in Eq.~(\ref{eq:site-charge-definition}).
The Fourier transform of the last term in Eq.~(\ref{eq:init-cpa-potential-variation}) is
\begin{multline}
M(k)\left[\sum_{\gamma=1}^{n-1}(Q_{\gamma}-Q_{n})\delta \bar{c}_{\gamma}(k)+\sum_{\gamma=1}^{n}\bar{c}_{\gamma}\delta\rho_{\gamma}^{1}(k)\right] \\
 =: M(k)\delta P(k). \label{eq:pol-definition}
\end{multline}
with $M(k)$ defined in Eq.~(\ref{eq:M-definition}). In terms of the basis $f_n(r)$ we can expand $\delta\rho_{i\mu}(r)=\sum_{n}f_{n}(r)\delta\rho_{i\mu}^{n}$.
This allows one to separate the volume integral in Eq.~(\ref{eq:init-cpa-potential-variation}) from the unknown $\delta\rho_{i\mu}^{n}$.
The complete variation of the potential in $k$-space is then
\begin{multline}
\delta V_{\mu}(k;r)= \sum_{n}\bigg[\frac{dV_{\text{xc}}}{d\rho}(\rho_{\mu}(r))f_{n}(r)+ \\ \left(\int_{V_{0}}dr'\frac{e^{2}}{|r-r'|}f_{n}(r')\right)\bigg]\delta\rho_{\mu}^{n}(k)+M(k)\delta P(k). \label{eq:deltaV-mu-k-r}
\end{multline}
Using definitions in Eqs.~(\ref{eq:F-definition})-(\ref{eq:U-definition}) and Eq.~(\ref{eq:deltat-from-zz}),
\begin{align*}
\delta t_{\mu}^{-1}(k)= & \sum_{n}U_{\mu}^{n}\delta\rho_{\mu}^{n}(k)+F_{\mu}^{1}M(k)\delta P(k).
\end{align*}
On dividing by $\delta\nu_{0\gamma}$ we derive Eq.~(\ref{eq:master-pot-variation})
\begin{align*}
\frac{\delta t_{\mu}^{-1}}{\delta\nu_{\gamma}}(k)= & \sum_{n}U_{\mu}^{n}\Phi_{\mu\gamma}^{n}(k)+F_{\mu}^{1}M(k)\frac{\delta P}{\delta\nu_{\gamma}}(k).
\end{align*}
And from the definition of $\delta P(k)$ in Eq.~(\ref{eq:pol-definition}) we get Eq.~(\ref{eq:master-pol-variation})
\begin{align*}
\frac{\delta P}{\delta\nu_{\gamma}}(k)= & \sum_{\sigma=1}^{n-1}(Q_{\sigma}-Q_n)\beta \bar{\Psi}_{\sigma\gamma}(k)+\sum_{\sigma} \bar{c}_{\sigma}\Phi_{\sigma\gamma}^{1}(k).
\end{align*}

\section{Variation of grand potential\label{sub:Grand-Potential-Variation}}

Within the CPA approximation the electronic grand potential is given
by Eq.~(\ref{eq:cpa-grand-potential}) as carefully derived by Johnson et
al.\cite{PhysRevB.41.9701} $N_{c}(\epsilon$) is the Lloyd formula in Eq.~(\ref{eq:lloyd-formula}).
Consider the change of the grand potential as concentrations $\{\bar{c}_{i\mu}\}$
are varied relative to the $n^{\text{th}}$ (or host) species. This is 
\begin{multline*}
\frac{\delta\Omega_{\text{elec.}}}{\delta \bar{c}_{i\mu}}\bigg|_{\bar{c}_{j\gamma}\neq \bar{c}_{i\mu}}=  \,\frac{\partial\Omega_{\text{elec.}}}{\partial \bar{c}_{i\mu}}\bigg|_{\bar{c}_{j\gamma}\neq \bar{c}_{i\mu},\:\rho_{j\gamma}} \\ +\sum_{j\gamma}\int_{V_{j}}dr\frac{\partial\Omega_{\text{elec.}}}{\partial\rho_{j\gamma}(r)}\bigg|_{\bar{c}_{k\nu},\,\rho_{k\nu}\neq\rho_{j\gamma}}\frac{\partial\rho_{j\gamma}(r)}{\partial \bar{c}_{i\mu}}\bigg|_{\bar{c}_{k\nu}\neq \bar{c}_{i\mu}}.
\end{multline*}
As discussed by Johnson et al.\cite{PhysRevB.41.9701}, $\partial \Omega_\text{elec.}/\partial \rho_{i\mu}(r)|_{\bar{c}_{j\nu}} = 0$
when site potentials $V_{i\mu}(r)$ are defined as in Eq.~(\ref{eq:cpa-potential}). This is one of the key variational properties of the electronic grand potential. Therefore we only need
take the explicit partial:
\begin{align*}
\frac{\delta\Omega_{\text{elec.}}}{\delta \bar{c}_{i\mu}} =& \, \left(\frac{\delta\Omega}{\delta\bar{c}_{i\mu}}\right)_\text{kin.} 
+ \left(\frac{\delta\Omega}{\delta\bar{c}_{i\mu}}\right)_\text{intra.}
+ \left(\frac{\delta\Omega}{\delta\bar{c}_{i\mu}}\right)_\text{inter.}
\end{align*}
{ \allowdisplaybreaks
\begin{align*}
\left(\frac{\delta\Omega}{\delta\bar{c}_{i\mu}}\right)_\text{kin.}  =& 
\\ & \mkern-75mu +\frac{1}{\pi}\int d\epsilon f(\epsilon-\mu)\,\text{Im}\left[\,\log||\bar{D}_{i\mu}^{-1}||-\log||\bar{D}_{in}^{-1}||\,\right]  
 \\ &\mkern-75mu -\frac{1}{\pi}\int d\epsilon f(\epsilon-\mu)\,\text{Im}\left[\,\log||\alpha_{i\mu} t_{i\mu}^{-1} ||-\log||\alpha_{in} t_{in}^{-1}||\,\right]
 \\  & \mkern-75mu -\int_{V_i} dr \left[\,\rho_{i\mu}(r)V_{i\mu}(r)-\rho_{in}(r)V_{in}(r)\,\right] \\ \\
  \left(\frac{\delta\Omega}{\delta\bar{c}_{i\mu}}\right)_\text{intra.} =& \, \frac{e^{2}}{2}\int_{V_{i}}dr\int_{V_{i}}dr'\frac{1}{|r-r'|} \times \\
  & \mkern-105mu \bigg\{ \rho_{i\mu}(r')\left[\rho_{i\mu}(r)-2Z_{\mu}\delta(r)\right]
-\rho_{in}(r')\left[\rho_{in}(r)-2Z_{n}\delta(r)\right]\bigg\}
\\ & \mkern-75mu +\int_{V_{i}}dr\left[\,\rho_{i\mu}(r)\epsilon_{\text{xc}}(\rho_{i\mu}(r))-\rho_{in}(r)\epsilon_{\text{xc}}(\rho_{in}(r))\,\right] \\ 
\\ \left(\frac{\delta\Omega}{\delta\bar{c}_{i\mu}}\right)_\text{inter.} =& \\
 & \mkern-90mu e^{2}\sum_{j\neq i}\int_{V_{i}}dr\int_{V_{j}}dr'\frac{1}{|r+R_{i}-R_{j}-r'|}\,\times\\
 & \mkern-90mu \left[\rho_{i\mu}(r)-\rho_{in}(r)-(Z_{\mu}-Z_{n})\delta(r)\right]\left[\bar{\rho}_{j}(r')-\bar{Z}_{j}\delta(r')\right]
 \\[-10pt]
\end{align*}
for site average electron density $\bar{\rho}_{j}(r)=\sum_{\gamma}\bar{c}_{j\gamma}\rho_{j\gamma}(r)$
and atomic number $\bar{Z}_{j}=\sum_{\gamma}\bar{c}_{j\gamma}Z_{\gamma}$.
} 

Now consider the variation of $\delta\Omega/\delta\bar{c}_{i\mu}$ itself. We also consider this in three pieces:
\begin{align}
\delta\left( \frac{\delta\Omega}{\delta\bar{c}_{i\mu}}\right) = T_\text{MS} + T_\text{SS} + T_\text{Q}. \label{eq:tms-tss-tq}
\end{align}
$T_\text{MS}$ includes any terms containing $D_{i\mu}$; $T_\text{SS}$ any terms including $\alpha_{i\mu}$ or $t_{i\mu}$; and $T_\text{Q}$ any remaining terms.

We have
\begin{multline}
T_\text{MS}= \\
 -\frac{1}{\pi}\int d\epsilon\frac{\partial f}{\partial\epsilon}(\epsilon-\mu)\text{Im}\left[\,\log||\bar{D}_{i\mu}^{-1}||-\log||\bar{D}_{in}^{-1}||\,\right]\delta\mu\\
+ \frac{1}{\pi}\int d\epsilon f(\epsilon-\mu)\text{Im}\left[\text{Tr}\left\{ \bar{D}_{i\mu}\delta\bar{D}_{i\mu}^{-1}-\bar{D}_{in}\delta\bar{D}_{in}^{-1}\right\} \right]. \label{TMS-definition}
\end{multline}
To evaluate this we need $\text{Tr}(\bar{D}_{i\mu}\delta\bar{D}_{i\mu}^{-1})$.
This is
\begin{align}
\text{Tr}(\bar{D}_{i\mu}\delta\bar{D}_{i\mu}^{-1})= & \,\text{Tr}(\bar{D}_{i\mu}[\delta\Delta_{i\mu}\tau^c_{ii}-\Delta_{i\mu}\sum_{j}\tau^c_{ij}\delta t_{jc}^{-1}\tau^c_{ji}]\,). \label{eq:TMS-trace}
\end{align}
Consider the on-site $i=j$ terms separately. These are
\begin{align*}
 \text{Tr}(\bar{D}_{i\mu}(\delta t_{i\mu}^{-1}-\delta t_{ic}^{-1})\tau^c_{ii}-\bar{D}_{i\mu}\Delta_{i\mu}\tau^c_{ii}\delta t_{ic}^{-1}\tau^c_{ii})&=\\
& \mkern-300mu =\text{Tr}[\bar{D}_{i\mu}\delta t_{i\mu}^{-1}\tau^c_{ii}]-\text{Tr}[\bar{D}_{i\mu}(1+\Delta_{i\mu}\tau^c_{ii})\delta t_{ic}^{-1}\tau^c_{ii}] \\
& \mkern-300mu =\text{Tr}[\bar{D}_{i\mu}\delta t_{i\mu}^{-1}\tau^c_{ii}]-\text{Tr}[\delta t_{ic}^{-1}\tau^c_{ii}].
\end{align*}
The second term is independent of $\mu$ and therefore cancels with
the corresponding term from the host in Eq.~(\ref{TMS-definition}). While
\begin{align}
\text{Tr}[\bar{D}_{i\mu}\delta t_{i\mu}^{-1}\tau^c_{ii}]&= 
 \text{Tr}[\delta t_{i\mu}^{-1}\tau^c_{ii}\bar{D}_{i\mu}] =\text{Tr}[\delta t_{i\mu}^{-1} D_{i\mu}\tau^c_{ii}] \nonumber \\
& \mkern-80mu = -\sum_{L_{1}L_{2}}\int drZ_{i\mu;L_{1}}(r)Z_{i\mu;L_{2}}(r)(\tau^{i\mu}_{ii})_{L_{2}L_{1}}\delta V_{i\mu}(r) \label{TMS-zzterm}
\end{align}
using Eq.~(\ref{eq:deltat-from-zz}) and $\tau^{i\mu}_{ii} = D_{i\mu} \tau^c_{ii}$ proved in Appendix~\ref{sub:Potential-Variation}. 
Eq.~(\ref{TMS-zzterm}) can be recognized as a major subexpression in the charge-density $\rho_{i\mu}(r)$ expressed using Eq.~(\ref{eq:green-function}) and Eq.~(\ref{eq:charge-from-green}). 
Now consider the off-site $i\neq j$ terms in Eq.~(\ref{eq:TMS-trace}), including subtraction for host in Eq.~(\ref{TMS-definition}).
This is
\begin{multline*}
-\text{Tr}[ (\bar{D}_{i\mu}\Delta_{i\mu}-\bar{D}_{in}\Delta_{in} )\sum_{j\neq i}\tau^c_{ij}\delta t_{jc}^{-1}\tau^c_{ji}\,]= \\
 -\text{Tr}[ (X_{i\mu}-X_{in})\sum_{j\neq i}\tau^c_{ij}\delta t_{jc}^{-1}\tau^c_{ji}\,] .
\end{multline*}
Prior literature\cite{PhysRevB.50.1450} expresses this as
\begin{align*}
\bar{D}_{i\mu}\Delta_{i\mu}-\bar{D}_{in}\Delta_{in}= & -(\bar{D}_{i\mu}-\bar{D}_{in})(\tau^c_{ii})^{-1}.
\end{align*}
Altogether Eq.~(\ref{TMS-definition}) becomes
\begin{align}
T_\text{MS} &= \nonumber \\
 & \mkern-30mu -\frac{1}{\pi}\int d\epsilon\frac{\partial f}{\partial\epsilon}(\epsilon-\mu)\text{Im}\left[\,\log||\bar{D}_{i\mu}^{-1}||-\log||\bar{D}_{in}^{-1}||\,\right]\delta\mu \nonumber
 \\ & \mkern-30mu -\frac{1}{\pi}\int d\epsilon f(\epsilon-\mu) \bigg\{ \text{Im}\text{Tr}[\,(X_{i\mu}-X_{in})\sum_{j\neq i}\tau_{ij}\delta t_{jc}^{-1}\tau_{ji}\,] \nonumber
 \\ & \mkern-15mu + \text{Im} \sum_{LL'}\int drZ_{i\mu;L}(r)Z_{i\mu;L'}(r)(\tau^{i\mu}_{ii})_{LL'}\delta V_{i\mu}(r) \nonumber
  \\ & \mkern-15mu - \text{Im} \sum_{LL'}\int drZ_{in;L}(r)Z_{in;L'}(r)(\tau^{in}_{ii})_{LL'}\delta V_{in}(r) \bigg\}. \label{TMS-final}
\end{align}

The $T_\text{SS}$ piece in Eq.~(\ref{eq:tms-tss-tq}) is
\begin{align}
T_\text{SS} &= \frac{1}{\pi}\int d\epsilon\frac{\partial f}{\partial\epsilon}(\epsilon-\mu) \times \nonumber \\
& \quad  \text{Im}\left[\,\log||\alpha_{i\mu} t_{i\mu}^{-1}||-\log||\alpha_{in} t_{in}^{-1}||\,\right]\delta\mu \nonumber \\
& \mkern-20mu - \frac{1}{\pi}\int d\epsilon f(\epsilon-\mu)\text{Im}\left[\text{Tr}\left\{ \alpha_{i\mu}^{-1}\delta \alpha_{i\mu}-\alpha_{in}^{-1} \delta\alpha_{in} \right\} \right] \nonumber \\
& \mkern-20mu - \frac{1}{\pi}\int d\epsilon f(\epsilon-\mu)\text{Im}\left[\text{Tr}\left\{ t_{i\mu}\delta t_{i\mu}^{-1}-t_{in}\delta t_{in}^{-1}\right\} \right]. \label{TSS-definition}
\end{align}
Before continuing, we establish basic relations of the $\alpha_{i\mu}$ matrix. An alternative definition\cite{0953-8984-16-36-011} is
\begin{align}
\alpha_{i\mu;LL'} = \delta_{LL'} + \int dr \mathcal{H}_L(r) V_{i\mu}(r) R_{i\mu;L'}(r) \label{eq:alpha-definition}
\end{align}
for $\mathcal{H}(E;r)_L = -i \sqrt{E} h_\ell(\sqrt{E}r) Y_{\ell m}(r)$ and spherical Hankel of the first kind $h_\ell(r)$. Also let $H_{i\mu;L}(E;r)$ be the solution of $(-\nabla^2+V_{i\mu}(r))\psi=E\psi$ with boundary
condition $H_{i\mu;L}(r) = \mathcal{H}_L(r)$ for $r \notin V_i$. As in Appendix \ref{sub:Potential-Variation}, we may view $R_{i\mu}(r)$, $\mathcal{H}_L(r)$, and $H_{i\mu;L}(r)$ as diagonal matrices over an infinite-dimensional vector space with basis elements $r \in \mathbb{R}^3$. In this space, 
\begin{align}
H_{i\mu;L} = \sum_{L'} \alpha_{i\mu;LL'}^{-1}(\mathcal{H}_{L'} + G^\text{ss}_{i\mu} V_{i\mu} \mathcal{H}_{L'}), \label{eq:zeller-irreg}
\end{align}
as proved by Zeller.\cite{0953-8984-16-36-011} Therefore, using Eq.~(\ref{eq:alpha-definition}), Eq.~(\ref{eq:delta-R}), and Eq.~(\ref{eq:zeller-irreg});
\begin{align}
\delta \alpha_{i\mu;LL'} &= \langle \mathcal{H}_L^* | \delta V_{i\mu} R_{i\mu;L'} \rangle + \langle \mathcal{H}_L^* | V_{i\mu} \delta R_{i\mu;L'} \rangle \nonumber  \\
& \mkern-30mu = \langle \mathcal{H}_L^* | \delta V_{i\mu} R_{i\mu;L'} \rangle + \langle \mathcal{H}_L^* | V_{i\mu} G^\text{ss}_{i\mu} \delta V_{i\mu} R_{i\mu;L'} \rangle \nonumber \\
& \mkern-30mu = \langle \mathcal{H}_L^* + (G^\text{ss}_{i\mu})^\dag V_{i\mu} \mathcal{H}_L^* | \delta V_{i\mu} R_{i\mu;L'} \rangle \nonumber \\
& \mkern-30mu = \sum_{L''} \langle \alpha_{i\mu;LL''}^* H_{i\mu;L''}^* | \delta V_{i\mu} R_{i\mu;L'} \rangle \nonumber \\
& \mkern-30mu = \sum_{L''} \alpha_{i\mu;LL''} \int dr  H_{i\mu;L''}(r) R_{i\mu;L'}(r) \delta V_{i\mu}(r) \label{eq:delta-alpha}.
\end{align}
This gives a major term in Eq.~(\ref{TSS-definition});
\begin{align}
\text{Tr}\{\alpha_{i\mu}^{-1} \delta \alpha_{i\mu}\} = \sum_L \int dr  H_{i\mu;L}(r) R_{i\mu;L}(r) \delta V_{i\mu}(r). \label{eq:a-times-deltaa}
\end{align}
But this contains a well-known expression for single-site Green function $G^\text{ss}_{i\mu}(r,r) = \sum_L H_{i\mu;L}(r) R_{i\mu;L}(r)$; as shown in Appendix A of Zeller.\cite{0953-8984-16-36-011} On the other hand, using Eq.~(\ref{eq:green-function}) with $\tau^\text{ss}_{i\mu} := t_{i\mu}$;
\begin{multline}
G^\text{ss}_{i\mu}(r,r) = \sum_{LL'} Z_{i\mu;L}(r) t_{i\mu;LL'} Z_{i\mu;L'}(r) \\ - \sum_L Z_{i\mu;L}(r) J_{i\mu;L}(r). \label{eq:gss-from-tmatrix}
\end{multline}
The other major term in Eq.~(\ref{TSS-definition}) is 
\begin{multline}
\text{Tr}\{t_{i\mu} \delta t_{i\mu}^{-1} \} = \\ -\int dr \sum_{LL'} t_{i\mu;LL'} Z_{i\mu;L'}(r) Z_{i\mu;L}(r) \delta V_{i\mu}(r) \label{eq:t-times-delta-t}
\end{multline}
by Eq.~(\ref{eq:deltat-from-zz}). Inserting Eqs.~(\ref{eq:a-times-deltaa})-(\ref{eq:t-times-delta-t}) into Eq.~(\ref{TSS-definition});
\begin{align}
T_\text{SS} &= \frac{1}{\pi}\int d\epsilon\frac{\partial f}{\partial\epsilon}(\epsilon-\mu) \times \nonumber \\
& \quad  \text{Im}\left[\,\log||\alpha_{i\mu} t_{i\mu}^{-1}||-\log||\alpha_{in} t_{in}^{-1}||\,\right]\delta\mu \nonumber \\
& \mkern-20mu + \frac{1}{\pi}\int d\epsilon f(\epsilon-\mu)  \text{Im} \int dr\, \delta V_{i\mu}(r) \times \nonumber \\
& \quad  \sum_L[ Z_{i\mu;L}(r) J_{i\mu;L}(r) -Z_{in;L}(r) J_{in;L}(r) ]. \label{TSS-final}
\end{align}
On combining Eq.~(\ref{TMS-final}) and Eq.~(\ref{TSS-final}) and identifying the expression for charge density from Eq.~(\ref{eq:green-function}) and Eq.~(\ref{eq:charge-from-green}); 
\begin{align}
T_\text{MS}+T_\text{SS} &=  \nonumber \\
 & \mkern-50mu -\frac{1}{\pi}\int d\epsilon\frac{\partial f}{\partial\epsilon}(\epsilon-\mu)\text{Im}\left[\,\log||\bar{D}_{i\mu}^{-1}||-\log||\bar{D}_{in}^{-1}||\,\right]\delta\mu \nonumber
 \\ & \mkern-50mu -\frac{1}{\pi}\int d\epsilon f(\epsilon-\mu) \bigg\{ \text{Im}\text{Tr}[\,(X_{i\mu}-X_{in})\sum_{j\neq i}\tau_{ij}\delta t_{jc}^{-1}\tau_{ji}\,] \nonumber \\
& \mkern-50mu + \frac{1}{\pi}\int d\epsilon\frac{\partial f}{\partial\epsilon}(\epsilon-\mu) \times \nonumber \\
& \quad  \text{Im}\left[\,\log||\alpha_{i\mu} t_{i\mu}^{-1}||-\log||\alpha_{in} t_{in}^{-1}||\,\right]\delta\mu \nonumber \\
& \mkern-50mu +\int dr [\rho_{i\mu}(r) \delta V_{i\mu}(r) - \rho_{in}(r) \delta V_{in}(r)]. \label{eq:tms-tss-final}
\end{align}

The variation of the charge term $T_\text{Q}$ in Eq.~(\ref{eq:tms-tss-tq}) is straightforward:
{ \allowdisplaybreaks
\begin{align*}
T_\text{Q} = \mkern-20mu &\\
& -\int_{V_{i}}dr\bigg[\delta\rho_{i\mu}(r)V_{i\mu}(r)+\rho_{i\mu}\delta V_{i\mu}(r)\\
& \mkern50mu -\rho_{in}(r)\delta V_{in}(r)-\delta\rho_{in}(r)V_{in}(r)\bigg]\\
& + \int_{V_{i}}dr\left[\delta\rho_{i\mu(r)}\epsilon_{\text{xc}}(\rho_{i\mu}(r))+\rho_{i\mu}(r)\frac{\delta\epsilon_{\text{xc}}}{\delta\rho}(\rho_{i\mu}(r))\delta\rho_{i\mu}(r)\right]\\
& -\int_{V_{i}}dr\left[\delta\rho_{in(r)}\epsilon_{\text{xc}}(\rho_{in}(r))+\rho_{in}(r)\frac{\delta\epsilon_{\text{xc}}}{\delta\rho}(\rho_{in}(r))\delta\rho_{in}(r)\right]\\
&+ \,\int_{V_{i}}dr\int_{V_{i}}dr'\frac{e^{2}}{|r-r'|}\left[(\rho_{i\mu}(r)-Z_{\mu}\delta(r))\delta\rho_{i\mu}(r')\right]\\
&-  \,\int_{V_{i}}dr\int_{V_{i}}dr'\frac{e^{2}}{|r-r'|}\left[(\rho_{in}(r)-Z_{n}\delta(r))\delta\rho_{in}(r')\right]\\
&+  \sum_{j\neq i}\int_{V_{i}}dr\int_{V_{j}}dr'\frac{e^{2}}{|r+R_{i}-R_{j}-r'|} \times \\ 
& \mkern50mu \left\{ [\delta\rho_{i\mu}(r)-\delta\rho_{in}(r)][(\bar{\rho}_{j}(r')-\bar{Z}_{j}\delta(r')]\right\} \\
&+  \sum_{j\neq i}\int_{V_{i}}dr\int_{V_{j}}dr'\frac{e^{2}}{|r+R_{i}-R_{j}-r'|} \times \\ 
 & \mkern50mu \bigg\{ [\rho_{i\mu}(r)-\rho_{in}(r)-(Z_{\mu}-Z_{n})\delta(r)] \times \\
 & \mkern100mu [\delta\bar{\rho}_{j}(r')-\delta\bar{Z}_{j}\delta(r')]\bigg\} .
\end{align*} }
Most of these terms can be identified as the self-consistent CPA potential given in Eq.~(\ref{eq:cpa-potential}).
A major cancellation then results in
\begin{align}
T_\text{Q} =& - \int_{V_{i}}dr\left[\rho_{i\mu}\delta V_{i\mu}(r)-\rho_{in}(r)\delta V_{in}(r)\right] \nonumber \\
& \mkern-40mu +\sum_{j\neq i}\int_{V_{i}}dr\int_{V_{j}}dr'\frac{e^{2}}{|r+R_{i}-R_{j}-r'|}\times \nonumber \\
& \mkern-30mu  [\rho_{i\mu}(r)-\rho_{in}(r)-(Z_{\mu}-Z_{n})\delta(r)][\delta\bar{\rho}_{j}(r')-\delta\bar{Z}_{j}\delta(r')]. 
\raisetag{-5pt}
\label{eq:tq-final}
\end{align}
Adding Eq.~(\ref{eq:tms-tss-final}) and Eq.~(\ref{eq:tq-final}) resolves Eq.~(\ref{eq:tms-tss-tq}) as
\begin{align*}
  \delta\left(\frac{\partial\Omega_{\text{elec.}}}{\partial c_{i\mu}}\right)= \mkern-120mu &\\
& -\frac{1}{\pi}\int d\epsilon\frac{\partial f}{\partial\epsilon}(\epsilon-\mu)\text{Im}\left[\,\log||\bar{D}_{\mu}^{-1}||-\log||\bar{D}_{n}^{-1}||\,\right]\delta\mu\\
& + \frac{1}{\pi}\int d\epsilon\frac{\partial f}{\partial\epsilon}(\epsilon-\mu) \times \nonumber \\
& \mkern50mu  \text{Im}\left[\,\log||\alpha_{\mu} t_{\mu}^{-1}||-\log||\alpha_{n} t_{n}^{-1}||\,\right]\delta\mu \nonumber \\
 & -\frac{1}{\pi}\int d\epsilon f(\epsilon-\mu)\text{Im}\text{Tr}\left[\,(X_{\mu}-X_{n})\sum_{j\neq i}\tau_{ij}\delta t_{jc}^{-1}\tau_{ji}\,\right]\\
 & +\sum_{j\neq i}\int_{V_{i}}dr\int_{V_{j}}dr'\frac{e^{2}}{|r+R_{i}-R_{j}-r'|}\times \\
  & \left\{ [\rho_{\mu}(r)-\rho_{n}(r)-(Z_{\mu}-Z_{n})\delta(r)][\delta\bar{\rho}_{j}(r')-\delta\bar{Z}_{j}\delta(r')]\right\} .
\end{align*}
We have at this stage dropped unnecessary site indices $i$. We now wish to Fourier transform. As usual, we make the approximation $|r+R_{i}-R_{j}-r'|\rightarrow|R_{i}-R_{j}|$.
The transform of the first and second term vanishes if we restrict ourselves
to finite $k$. The transform of the fourth term is given by Eq.~(\ref{eq:pol-definition}).
Using the definitions in Eq.~(\ref{eq:pol-definition}) and Eq.~(\ref{eq:superC-definition});
\begin{multline*}
  \frac{1}{\sqrt{{\scriptstyle N}}}\sum_{i}e^{-k\cdot R_{i}}\delta\left(\frac{\partial\Omega_{\text{elec.}}}{\partial c_{i\mu}}\right)=\quad\quad\\
  -\frac{1}{\pi}\int d\epsilon f(\epsilon-\mu)\text{Im}\text{Tr}\left[\,(X_{\mu}-X_{n})\mathbb{C}(k)\delta t_{c}^{-1}(k)\right]\\
 +(Q_{\mu}-Q_{n})M(k)\delta P(k).
\end{multline*}
Dividing
by chemical potential change $\delta\nu_{0\gamma}$ gives
\begin{multline}
\frac{1}{\delta\nu_{\gamma}}  \sum_{i}e^{-k\cdot R_{i}}\delta\left(\frac{\partial\Omega_{\text{elec.}}}{\partial c_{i\mu}}\right)=\\
  -\frac{1}{\pi}\int d\epsilon f(\epsilon-\mu)\text{Im}\text{Tr}\left[\,(X_{\mu}-X_{n})\mathbb{C}(k)\frac{\delta t_{c}^{-1}}{\delta\nu_{\gamma}}(k)\right]\\
  +(Q_{\mu}-Q_{n})M(k)\frac{\delta P}{\delta\nu_{\gamma}}(k). \label{eq:omega-var-fourier-final}
\end{multline}

\section{Variation of site concentrations\label{sub:Concentration-Variation}}

The optimal variational parameters $\{\bar{c}_{i\mu}\}$ are fixed by Eq.~(\ref{eq:optimal-site-concentrations}).
The variation of the first term about the homogenous reference is
\begin{multline*}
\delta\left(\,\beta^{-1}\log\frac{\bar{c}_{i\mu}}{\bar{c}_{in}}\right) = \\ \beta^{-1}\sum_{\gamma=1}^{n-1}\left(\frac{\delta_{\mu\gamma}}{\bar{c}_{\mu}}+\frac{1}{\bar{c}_{n}}\right)\delta\bar{c}_{i\gamma} 
=\sum_{\gamma=1}^{n-1}\ \beta^{-1}C_{\mu\gamma}^{-1}\delta\bar{c}_{i\gamma} 
\end{multline*}
for $C_{\mu\gamma}$ defined in Eq.~(\ref{eq:cmat-definition}).
The variation of Eq.~(\ref{eq:optimal-site-concentrations}) is therefore
\begin{align*}
0= \sum_{\gamma=1}^{n-1}\ \beta^{-1}C_{\mu\gamma}^{-1}\delta\bar{c}_{i\gamma}-\delta\nu_{i\mu}
+\delta\left(\frac{\partial\langle\Omega_\text{elec.}\rangle_{0}}{\partial c_{i\mu}}\right).
\end{align*}
On dividing by $\delta\nu_{0\sigma}$ and Fourier transforming, one
gets
\begin{multline*}
0= \sum_{\gamma=1}^{n-1}\ C_{\mu\gamma}^{-1}\bar{\Psi}_{\gamma\sigma}(k)-\delta_{\mu\sigma}\\
+\frac{1}{\delta\nu_{\sigma}}\sum_{i}e^{-k\cdot R_{i}}\delta\left(\frac{\partial\langle\Omega_\text{elec.}\rangle_{0}}{\partial c_{i\mu}}\right).
\end{multline*}
Substituting Eq.~(\ref{eq:omega-var-fourier-final})
on the variation of the grand potential we have
\begin{multline*}
\sum_{\gamma=1}^{n-1}\ C_{\mu\gamma}^{-1}\bar{\Psi}_{\gamma\sigma}(k)= \\
 \delta_{\mu\sigma}+\frac{1}{\pi}\int d\epsilon f(\epsilon-\mu)\text{Im}\text{Tr}\left[\,(X_{\mu}-X_{n})\mathbb{C}(k)\frac{\delta t_{c}^{-1}}{\delta\nu_{\sigma}}(k)\right]\\
 -(Q_{\mu}-Q_{n})M(k)\frac{\delta P}{\delta\nu_{\sigma}}(k).
\end{multline*}
Multiplying through by $\beta C$ gives Eq.~(\ref{eq:master-conc-variation})
\begin{multline*}
\beta \bar{\Psi}_{\mu\gamma}(k)=  \beta C_{\mu\gamma}+\sum_{\sigma=1}^{n-1}\beta C_{\mu\sigma}\bigg\{ \\
 \frac{1}{\pi}\int d\epsilon f(\epsilon-\mu)\text{Im}\text{Tr}\left[\,(X_{\sigma}-X_{n})\mathbb{C}(k)\frac{\delta t_{c}^{-1}}{\delta\nu_{\gamma}}(k)\right]\\
-(Q_{\sigma}-Q_{n})M(k)\frac{\delta P}{\delta\nu_{\gamma}}(k)\bigg\} .
\end{multline*}

\section{Variation of charge density\label{sub:Charge-Variation}}

The site electron density $\rho_{i\mu}(r)$ is given by Eq.~(\ref{eq:charge-from-green}) with $G=G_{i\mu}(\epsilon;r,r')$ the site impurity Green function. The variation may be decomposed into three contributions:
\begin{align}
\delta\rho_{i\mu}(r)=  \,\delta\rho_{i\mu}(r)\bigg|_{\delta V_{i\mu}(r)}+\delta\rho_{i\mu}(r)\bigg|_{\delta t_{jc}}+\delta\rho_{i\mu}(r)\bigg|_{\delta\mu}. \raisetag{-5pt} \label{eq:charge-response-decomp}
\end{align}
These may be expressed using Eq.~(\ref{eq:charge-from-green})  as
\begin{align*}
\delta \rho_{i\mu}(r)\bigg|_{\delta V_{i\mu}(r)} =&  -\frac{1}{\pi}\int d\epsilon\,f(\epsilon-\mu) \text{ Im }\delta G_{i\mu}(\epsilon;r,r)\bigg|_{\delta V_{i\mu}(r)} \\
\delta\rho_{i\mu}(r)\bigg|_{\delta t_{jc}} =&  
  -\frac{1}{\pi}\int d\epsilon\,f(\epsilon-\mu) \times \\ 
  & \mkern-25mu \text{ Im } \sum_{LL'}\bigg[Z_{\mu;L}(r)\delta(D_{i\mu}\tau^c_{ii})_{LL'}\bigg|_{\delta t_{jc}}Z_{\mu;L'}(r)\bigg]\\
 \delta\rho_{i\mu}(r)\bigg|_{\delta\mu} =&\frac{1}{\pi}\int d\epsilon\,\frac{\partial f}{\partial\epsilon}(\epsilon-\mu)\text{ Im }G_{\mu}(\epsilon;r,r)\delta\mu
\end{align*}
where we use $Z_{i\mu}\rightarrow Z_{\mu}$, etc. when expanding about
a homogenuous medium. We know by the Born series expansion of the
impurity Green function
\begin{multline*}
G_{i\mu}=  \,G_{0}+G_{0}(\mathcal{V}_{i\mu}+\delta \mathcal{V}_{i\mu})G_{0}+\\
G_{0}(V_{i\mu}+\delta \mathcal{V}_{i\mu})G_{0}(\mathcal{V}_{i\mu}+\delta \mathcal{V}_{i\mu})G_{0}+\cdots
\end{multline*}
for $\mathcal{V}_{i\mu}$ the full potential for CPA medium with embedded impurity
$\mu$ at the $i^{\text{th}}$ site. And therefore
\begin{multline}
\delta G_{i\mu}=  \,(G_{0}+G_{0}\mathcal{V}_{i\mu}G_{0}+\cdots)\delta \mathcal{V}_{i\mu} \times \\
(G_{0}+G_{0}\mathcal{V}_{i\mu}G_{0}+\cdots)=G_{i\mu}\delta \mathcal{V}_{i\mu}G_{i\mu}. \label{eq:deltaG-born}
\end{multline}
Using Eq.~(\ref{eq:deltaG-born}); the first term in Eq.~(\ref{eq:charge-response-decomp}) is
\begin{multline*}
\,\delta\rho_{i\mu}(r)\bigg|_{\delta V_{i\mu}(r)}=  -\frac{1}{\pi}\int d\epsilon\,f(\epsilon-\mu) \times \\ \text{ Im }\int_{V_{i}}dr'G_{\mu}(\epsilon;r,r')\delta V_{i\mu}(r')G_{\mu}(\epsilon;r',r).
\end{multline*}
Taking the Fourier transform and substituting Eq.~(\ref{eq:deltaV-mu-k-r});
\begin{multline*}
  \delta\rho_{\mu}(k;r)\bigg|_{\delta V_{\mu}(k;r)}=\\
 -\frac{1}{\pi}\int d\epsilon\,f(\epsilon-\mu)\text{ Im }\int_{V_{0}}dr'G_{\mu}(\epsilon;r,r') \times\\
  \bigg\{\sum_{n}\left[\frac{dV_{\text{xc}}}{d\rho}(\rho_{\mu}(r'))f_{n}(r')+\left(\int_{V_{0}}dr''\frac{e^{2}}{|r'-r''|}f_{n}(r'')\right)\right]
  \\ \times \delta\rho_{\mu}^{n}(k) +M(k)\delta P(k)\bigg\} G_{\mu}(\epsilon;r',r).
\end{multline*}
Integrating both sides by $\int dr\,f_{m}(r)(\,\cdot\,)$;
\begin{align}
\delta\rho_{\mu}^{m}(k)\bigg|_{\delta V_{\mu}(k;r)}=\sum_{n}A_{\mu}^{mn}\delta\rho_{\mu}^{n}(k)+B_{\mu}^{m}M(k)\delta P(k)
 \label{eq:delta-rho-vimu}
\end{align}
using definitions in Eqs.~(\ref{eq:A-definition})-(\ref{eq:B-definition}).
Now we focus on the second term of Eq.~(\ref{eq:charge-response-decomp}). This requires
\begin{multline}
\delta(D_{i\mu}\tau_{ii}^c)\bigg|_{\delta t_{jc}}= \\
 -D_{i\mu}(-\sum_{j}\tau_{ij}^c\delta t_{jc}^{-1}\tau_{ji}^c\Delta_{i\mu}-\tau_{ii}^c\delta t_{ic}^{-1})D_{i\mu}\tau_{ii}^c\\
 -D_{i\mu}\sum_{j}\tau_{ij}^c\delta t_{jc}^{-1}\tau_{ji}^c. \label{eq:delta-D-tau}
\end{multline}
The $i=j$ terms in Eq.~(\ref{eq:delta-D-tau}) vanish;
\begin{multline*}
  -D_{i\mu}(-\tau_{ii}^c\delta t_{ic}^{-1}\tau_{ii}^c\Delta_{i\mu}-\tau_{ii}^c\delta t_{ic}^{-1})D_{i\mu}\tau_{ii}^c\\
  -D_{i\mu}\tau_{ii}^c\delta t_{ic}^{-1}\tau_{ii}^c=0.
\end{multline*}
While the remaining terms $i \neq j$ in Eq.~(\ref{eq:delta-D-tau}) are
\begin{multline*}
\delta(D_{i\mu}\tau_{ii}^c)\bigg|_{\delta t_{jc}}= \\
 D_{i\mu}\sum_{j\neq i}\tau_{ij}^c\delta t_{jc}^{-1}\tau_{ji}^c\Delta_{i\mu}D_{i\mu}\tau_{ii}^c-D_{i\mu}\sum_{j\neq i}\tau_{ij}^c\delta t_{jc}^{-1}\tau_{ji}^c\\
=  -D_{i\mu}\sum_{j\neq i}\tau_{ij}^c\delta t_{jc}^{-1}\tau_{ji}^c\bar{D}_{i\mu}.
\end{multline*}
using Eqs.~(\ref{eq:cpa-definitions-Dbar})-(\ref{eq:cpa-definitions-D}). Therefore we have lattice Fourier transform
\begin{multline*}
\delta\rho_{\mu}(k;r)\bigg|_{\delta t_{c}(k)}= 
 -\frac{1}{\pi}\int d\epsilon\,f(\epsilon-\mu)\times \\
 \text{ Im}\sum_{LL'}\left[Z_{\mu;L}(r)(-\mathcal{\mathbb{D}}_{\mu}\mathbb{C}(k)\delta t_{c}^{-1}(k))_{LL'}Z_{\mu;L'}(r)\right].
\end{multline*}
The Fourier transform of $\delta\rho_{i\mu}(r)\bigg|_{\delta\mu}$vanishes
for finite $k$. 
On integrating both sides by $\int dr\,f_{m}(r)(\,\cdot\,)$;
\begin{multline}
\delta\rho_{\mu}^{m}(k)\bigg|_{\delta t_{c}(k)}= \\
 -\frac{1}{\pi}\int d\epsilon\,f(\epsilon-\mu)\text{ Im}\sum_{LL'}F_{\mu;LL'}^{m}(\mathcal{\mathbb{D}}_{\mu}\mathbb{C}(k)\delta t_{c}^{-1}(k))_{LL'}. \label{eq:delta-rho-tc}
\end{multline}
Therefore, combining Eq.~(\ref{eq:delta-rho-vimu}) and Eq.~(\ref{eq:delta-rho-tc});
\begin{multline*}
\delta\rho_{\mu}^{m}(k)=  \sum_{n}A_{\mu}^{mn}\delta\rho_{\mu}^{n}(k)+B_{\mu}^{m}M(k)\delta P(k) \\
-\frac{1}{\pi}\int d\epsilon\,f(\epsilon-\mu)\text{ Im}\sum_{LL'}F_{\mu;LL'}^{m}(\mathcal{\mathbb{D}}_{\mu}\mathbb{C}(k)\delta t_{c}^{-1}(k))_{LL'}.
\end{multline*}
Dividing by $\delta\nu_{0\gamma}$ gives Eq.~(\ref{eq:master-charge-variation})
\begin{multline*}
\Phi_{\mu\nu}^{m}(k)= \sum_{n}A_{\mu}^{mn}\Phi_{\mu\nu}^{n}(k)+B_{\mu}^{m}M(k)\frac{\delta P}{\delta\nu_{\gamma}}(k)\\
-\frac{1}{\pi}\int d\epsilon\,f(\epsilon-\mu)\text{ Im}\sum_{LL'}F_{\mu;LL'}^{m}(\mathcal{\mathbb{D}}_{\mu}\mathbb{C}(k)\frac{\delta t_{c}^{-1}}{\delta\nu_{\gamma}}(k))_{LL'}.
\end{multline*}

\bibliography{s2formalism}

\end{document}